\documentclass[11pt]{article}
\usepackage[utf8]{inputenc}
\usepackage{jheppub}
\usepackage{graphicx}
\usepackage{caption}
\usepackage{subcaption}
\usepackage{hyperref}
\usepackage[T1]{fontenc}

\newcommand{\be}{\begin{equation}}
\newcommand{\ee}{\end{equation}}
\newcommand{\bea}{\begin{eqnarray}}
\newcommand{\eea}{\end{eqnarray}}

\newcommand{\nn}{\nonumber\\}
\newcommand\underrel[3][]{\mathrel{\mathop{#3}\limits_{%
      \ifx c#1\relax\mathclap{#2}\else#2\fi}}}

\def\q{{\bf q}}

\def\CA{\mathcal{A}}
\def\CB{\mathcal{B}}

\def\CL{\mathcal{L}}

\def\CN{\mathcal{N}}

\def\CO{\mathcal{O}}

\def\CW{\mathcal{W}}

\def\qfr{\mathfrak{q}}
\def\wfr{\mathfrak{w}}

\def\lgb{\lambda_{\scriptscriptstyle GB}}

\def\gammagb{\gamma_{\scriptscriptstyle GB}}
\def\ggb{\gamma_{\scriptscriptstyle GB}}
\def\re{\mbox{Re}}
\def\im{\mbox{Im}}

\title{Hydrodynamic dispersion relations at finite coupling}

\author[a]{Sa\v{s}o Grozdanov,}
\author[b]{Andrei O.~Starinets,}
\author[c]{and Petar Tadi\'{c}}

\affiliation[a]{University of Ljubljana, Faculty of Mathematics and Physics, Jadranska ulica 19, SI-1000 Ljubljana, Slovenia}

\affiliation[b]{Rudolf Peierls Centre for Theoretical Physics, Clarendon Lab,  Oxford, OX1 3PU, UK}

\affiliation[c]{School of Mathematics, Trinity College Dublin, Dublin, D02 W272, Ireland}

\emailAdd{saso.grozdanov@fmf.uni-lj.si}
\emailAdd{andrei.starinets@physics.ox.ac.uk}
\emailAdd{tadicp@maths.tcd.ie}

\abstract{By using holographic methods, the radii of convergence of the hydrodynamic shear and sound dispersion relations were previously computed in the ${\cal N} = 4$ supersymmetric Yang-Mills theory at infinite 't Hooft coupling and infinite number of colours. Here, we extend this analysis to the domain of large but finite 't Hooft  coupling. To leading order in the perturbative expansion, we find that the radii grow with  increasing inverse coupling, contrary to naive expectations. However, when the equations of motion are solved using a qualitative non-perturbative resummation, the dependence on the coupling becomes piecewise continuous and the initial growth is followed by a decrease. The piecewise  nature of the dependence is related to the dynamics of branch point singularities of the energy-momentum tensor finite-temperature two-point functions in the complex plane of spatial momentum squared. We repeat the study using the Einstein-Gauss-Bonnet gravity as a model where the equations can be solved fully non-perturbatively, and find the expected decrease of the radii of convergence with the effective inverse coupling which is also piecewise continuous. Finally, we provide arguments in favour of the non-perturbative approach and show that the presence of non-perturbative modes in the quasinormal spectrum can be indirectly inferred from the analysis of perturbative critical points.}

\preprint{OUTP-21-11P}

\begin{document}
\maketitle
\flushbottom

\section{Introduction}
In the hydrodynamic regime, quantum field theory is expected to contain collective excitations such as sound waves \cite{landau-6,Kovtun:2012rj}. These hydrodynamic modes are characterised in momentum space by their gapless dispersion relations $\omega =\omega (\bf{q})$, where $\omega$ is  the frequency of the mode and ${\bf q}$ is  its wave-vector. In the simplest case of a relativistic neutral isotropic fluid, two hydrodynamic modes  known as  shear and sound modes have dispersion relations
\begin{align}
&\qquad \omega_{\rm \tiny shear} ({\bf q}^2) = -i D {\bf q}^2 + \cdots\,, \label{PuiseuxShear1} \\
& \qquad \omega_{\rm \tiny sound} ({\bf q}^2) =  \pm v_s ( {\bf q}^2)^{\frac{1}{2}} - i \frac{\Gamma}{2}  {\bf q}^2 +\cdots\,.\label{PuiseuxSound1}
\end{align}
These modes arise as  linearised fluctuations of an equilibrium state and describe transverse momentum (shear) and longitudinal energy-momentum  (sound) transfer. The coefficients of the series such as the speed of sound $v_s$, the transverse momentum diffusion constant $D=\eta/s T$ and the sound attenuation constant $\Gamma =(\zeta + 4\eta/3)/sT$, where $\eta$ and $\zeta$ are, respectively, shear and bulk viscosities and $s$ is the equilibrium entropy density at temperature $T$,  are determined by the underlying microscopic quantum field theory  \cite{Kovtun:2012rj}. In the following, it will be convenient to use  the  frequency $\wfr = \omega/2\pi T$ and the spatial momentum  ${\bf \qfr} = {\bf q}/2\pi T$ normalised by the Matsubara frequency.

Recently, in the context of exploring the domain of applicability of hydrodynamics, the radii of convergence of the series \eqref{PuiseuxShear1}, \eqref{PuiseuxSound1} have been investigated in some strongly interacting quantum field theories by using  their dual gravitational descriptions in refs.~\cite{Withers:2018srf,Grozdanov:2019kge,Grozdanov:2019uhi,Abbasi:2020ykq,Jansen:2020hfd,Grozdanov:2020koi} and by using field theory methods in refs.~\cite{Choi:2020tdj,Baggioli:2020loj}. In particular, by promoting ${\bf \qfr}^2$ to a complex variable and analysing critical points of the associated spectral curves, in  refs.~\cite{Grozdanov:2019kge,Grozdanov:2019uhi}, it was found that for  the $\mathcal{N}=4$ supersymmetric $SU(N_c)$ Yang-Mills theory (SYM)
in the limit of infinite number of colours $N_c \to \infty$ and infinite 't Hooft coupling $\lambda=g_{YM}^2 N_c\to \infty$, the radii of convergence $R=| {\bf \qfr}^2|$ of the hydrodynamic series  $\wfr = \wfr (\qfr^2)$ in the complex $\qfr^2$-plane are given by 
\begin{align}
R_{\rm \tiny shear}^{\tiny (\infty)} &\approx 2.22\, ,\label{RN4ShearInfa} \\
R_{\rm \tiny sound}^{\tiny (\infty)} &= 2\,. \label{RN4SoundInfa}
\end{align} 
The physical reason behind the breakdown of  the convergence of  hydrodynamic series is the presence of the gapped non-hydrodynamic degrees of freedom 
whose spectra ``cross levels'' with the hydrodynamic  degrees of freedom at some (generically complex) value of $\qfr^2$.

Our main goal in this paper is to find the  't Hooft coupling constant corrections to the infinite coupling results \eqref{RN4ShearInfa}, \eqref{RN4SoundInfa}, similar to the coupling constant corrections to the entropy \cite{Gubser:1998nz,Pawelczyk:1998pb}, shear viscosity \cite{Buchel:2004di,Buchel:2008sh} and other transport coefficients (see ref.~\cite{Grozdanov:2014kva} and references therein)  computed for the $\mathcal{N}=4$ SYM theory earlier. Our methods are discussed in detail in section \ref{section-cp-res}, and in appendices \ref{app-critical-points-radius} and \ref{darboux}.

Naively, one may expect that the radius of convergence $R(\lambda)$ decreases with the coupling decreasing from its infinite value. Indeed, schematically \cite{Grozdanov:2016vgg}, while at infinite coupling the characteristic spectral distance $\nu^{\tiny (\infty)}$ (set by the location of quasinormal modes in the dual gravity theory \cite{Horowitz:1999jd,Starinets:2002br,Kovtun:2005ev}) is coupling-independent, its counterpart $ \nu^{\tiny (0)}$ at small coupling (set by the eigenvalues of a suitable  linearised collision operator \cite{Arnold:2000dr,Arnold:2003zc}, \cite{Huot:2006ys}) is coupling-dependent and parametrically small, hence,
\begin{eqnarray}
&\,& R^{\tiny (\infty)} \sim \nu^{\tiny (\infty)}/T\,  \sim \, 1\, ,\label{RN4ShearInfaq} \\
&\,&R^{\tiny (0)}  \sim \nu^{\tiny (0)}/T \, \sim\,  \lambda^2 \ln{\lambda^{-1}} \ll 1\,, \label{RN4SoundInfaq}
\end{eqnarray} 
where eq.~\eqref{RN4ShearInfaq}  is clearly consistent with the results \eqref{RN4ShearInfa}, \eqref{RN4SoundInfa}. 

\begin{figure*}[th]
\centering
\includegraphics[width=0.48\textwidth]{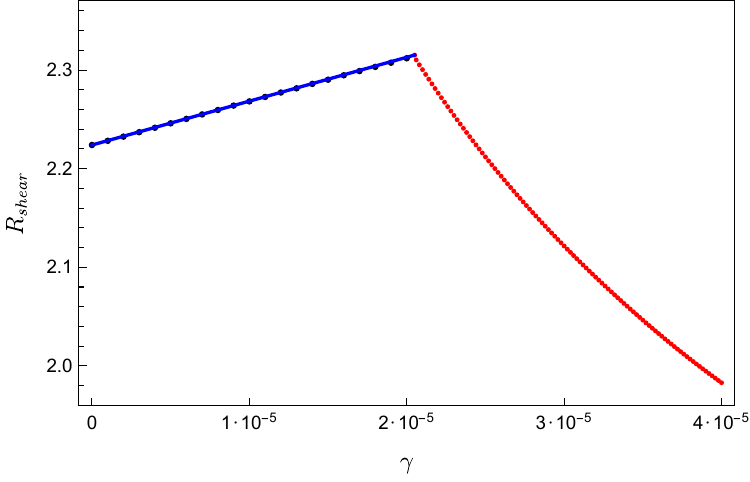}
\hspace{0.2cm}
\includegraphics[width=0.48\textwidth]{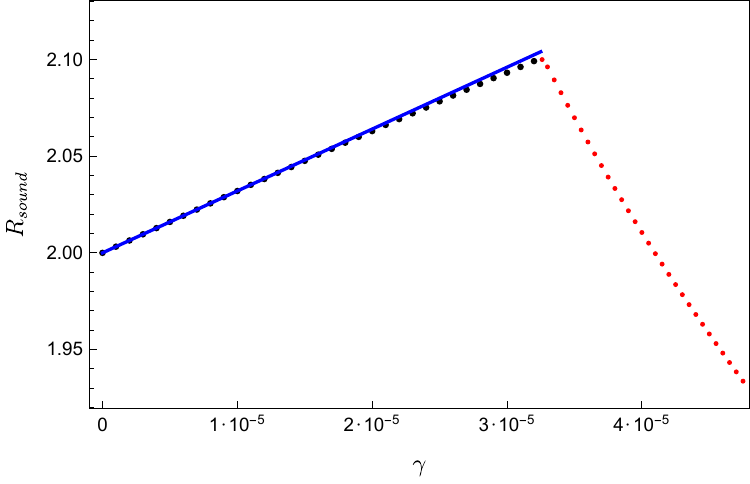}
\caption{{\small Radii of convergence $R_{\rm shear}$ and $R_{\rm sound}$ of the hydrodynamic shear (left panel) 
and sound (right panel) modes in the $\CN =4$ SYM theory as a function of the coupling $\gamma \propto \lambda^{-3/2}$. Solid blue  curves correspond to the perturbative results of eqs.~\eqref{RN4Shear}, \eqref{RN4Sound}, black dots are the non-perturbative results. The red curves are
 determined by the level-crossings of the hydrodynamic modes with the modes  not present in the perturbative spectrum. 
}}
\label{fig:radius_coupling_dependence_n4_main_intro}
\end{figure*}

However, these expectations are shattered by a concrete calculation. Using perturbative methods only, in section \ref{sec:N4} we find instead that  at large coupling the radius of convergence increases with the coupling decreasing from its infinite value, namely,
\begin{align}
R_{\rm shear} (\lambda) &= R_{\rm shear}^{\tiny (\infty)}  \left( 1 + 674.15 \,   \lambda^{-3/2} + \cdots \right)\,,\label{RN4Shear} \\
R_{\rm sound} (\lambda) &= R_{\rm sound}^{\tiny (\infty)} \left( 1 + 481.68 \,  \lambda^{-3/2} + \cdots \right)\,,\label{RN4Sound}
\end{align}
where $R_{\rm shear}^{\tiny (\infty)} $ and $R_{\rm sound}^{\tiny (\infty)}$ are given by eqs.~\eqref{RN4ShearInfa} and \eqref{RN4SoundInfa}. This result is unexpected. Admittedly, the large numerical coefficients in eqs.~\eqref{RN4Shear} and \eqref{RN4Sound} may reflect  
the necessity of a non-perturbative ``resummation'' along the lines of ref.~\cite{Waeber:2015oka}. Indeed, applying such a resummation (discussed in detail below), we find that $R_{\rm shear} (\lambda)$ and $R_{\rm sound} (\lambda)$ become {\it decreasing} functions of the decreasing $\lambda$ after the initial growth that is well-approximated by eqs.~\eqref{RN4Shear} and \eqref{RN4Sound} (see fig.~\ref{fig:radius_coupling_dependence_n4_main_intro}). In consequence, for the $\CN = 4$ SYM theory, our analysis 
 implies that $R(\lambda)$ should be a non-monotonic and piecewise continuous function. The  dependence of  $R_{\rm shear}$ and $R_{\rm sound}$ on $\gamma \sim \lambda^{-3/2}$ shown in fig.~\ref{fig:radius_coupling_dependence_n4_main_intro}  constitutes the main result of this paper. The non-perturbative part of the curve $R_{\rm shear} (\gamma)$ (the red  segment of the curve in the right panel of fig.~\ref{fig:radius_coupling_dependence_n4_main_intro}) coincides with the boundary of validity of hydrodynamics previously discussed in ref.~\cite{Grozdanov:2016vgg}. We note that the piecewise character of this dependence is similar to the one recently observed for infinitely strongly coupled theories with finite chemical potential \cite{Jansen:2020hfd,aiks} and the Sachdev-Ye-Kitaev chain at finite coupling \cite{Choi:2020tdj}. 

To understand the non-perturbative aspects of the analysis better, in section \ref{sec:GB}, we 
compute the  radii of convergence of hydrodynamic series using the Einstein-Gauss-Bonnet gravity in five dimensions as a theoretical laboratory. There, the second-order bulk equations of motion can be solved fully non-perturbatively in the Gauss-Bonnet coupling \cite{Brigante:2007nu,Grozdanov:2016vgg,Grozdanov:2016fkt}, and thus the outcome of relevant perturbative resummations can be compared with exact results. We find  (see figs.~\ref{fig:qc-lambda-shear} and \ref{fig:radius_coupling_dependence_sound_GB}) that the radius of convergence decreases  (and the dependence is piecewise continuous)  with what can be phenomenologically identified as the direction of decreasing CFT coupling \cite{Grozdanov:2016vgg,Grozdanov:2016fkt,Grozdanov:2016zjj,Andrade:2016rln}, which is qualitatively similar to the results obtained for the $\CN = 4$ SYM theory in section \ref{sec:N4}.

The question of  the hydrodynamic series convergence at finite coupling was recently addressed in ref.~\cite{Baggioli:2020loj} for experimentally realisable fluids, and in ref.~\cite{Choi:2020tdj} for the Sachdev-Ye-Kitaev chain. The calculations of ref.~\cite{Baggioli:2020loj} are based on estimating the size of the $k$-gap \cite{Baggioli:2019jcm} and show an increasing $R$ with increasing Coulomb coupling strength, while ref.~\cite{Choi:2020tdj} finds a non-monotonic dependence, with $R$ growing towards weak coupling. Radii of convergence of hydrodynamic series in relativistic kinetic theory (in the relaxation time approximation) 
were recently studied in ref.~\cite{Heller:2020hnq}.

This paper is structured as follows. In section \ref{section-cp-res}, we  briefly review the method of critical points of spectral curves introduced in refs.~\cite{Grozdanov:2019kge,Grozdanov:2019uhi} to compute the radii of convergence, as well as the non-perturbative resummation approach for theories with higher-derivative equations of motion. In section \ref{sec:N4}, we use the dual higher-derivative gravity to compute the radii of convergence for the shear and sound modes in  the $\CN = 4$ SYM theory at large but finite `t Hooft coupling. This analysis is done perturbatively and non-perturbatively by using the ``resummed'' version of the first-order theory. In section \ref{sec:GB}, we perform similar calculations in Einstein-Gauss-Bonnet gravity to check the validity of our approach. We also demonstrate  level-crossings at higher momenta. Then, in section \ref{sec:validity}, we  discuss the validity of non-perturbative ``resummations'' used in holography. We first consider a toy algebraic example and then study in detail the shear channel of the Einstein-Gauss-Bonnet theory. These examples are used to draw plausible conclusions about  the $\CN = 4$ SYM theory and the emergence of purely relaxing gapped modes in that theory. We conclude with a discussion of open problems in section \ref{sec:disc}. Appendix \ref{app-critical-points-radius} is a short review of the methods involving critical points and quasinormal level-crossing. Appendix \ref{darboux} introduces a useful method for determining the Puiseux exponent at a critical point by analysing the coefficients of  hydrodynamic series of a dispersion relation $\wfr = \wfr (\qfr^2)$. Finally, appendices \ref{sec:appendix-N=4} and \ref{sec:appendix-GB} contain the coefficients of the differential equations used in the paper.

\section{Critical points, quasinormal level-crossings and the ``resummation''}
\label{section-cp-res}

In this section, we briefly review the methods used to obtain the main results of the paper. These methods were formulated in refs.~\cite{Grozdanov:2019kge,Grozdanov:2019uhi}, where the interested reader can find more details and examples.

\subsection{Critical points, Puiseux series  and quasinormal level-crossing}
In momentum space, the hydrodynamic dispersion relations arise from the hydrodynamic spectral curve $P_H( \qfr^2,\wfr)=0$ given by the zeros of the determinant of the matrix of linearised fluctuations around an equilibrium 
state  \cite{Grozdanov:2019kge,Grozdanov:2019uhi}. Using the symmetry of the system and applying the Newton polygon method, one can write generic expressions for the dispersion relations describing transverse momentum and longitudinal energy-momentum fluctuations  in terms of the converging Puiseux series centred at the origin:
\begin{align}
&\qquad \wfr_{\rm \tiny shear} (\qfr^2) = -i \sum_{n=1}^\infty c_n \left(\qfr^2\right)^n\,, \label{PuiseuxShear} \\
& \qquad \wfr_{\rm \tiny sound} (\qfr^2) =  - i \sum_{n=1}^\infty a_n e^{\pm \frac{i\pi n}{2}} \left(\qfr^2\right)^{n/2}\,. \label{PuiseuxSound}
\end{align}
The modes \eqref{PuiseuxShear} and \eqref{PuiseuxSound} are gapless. The coefficients $c_n$ and $a_n$ of the series  are real functions proportional to the transport coefficients. In particular, $c_1=2\pi T D$, $a_1=\pm v_s$, $a_2=-\Gamma \pi T$. In the  underlying quantum field theory, the full spectral curve $P( \qfr^2,\wfr)$, which reduces to  $P_H( \qfr^2,\wfr)$ in the hydrodynamic limit,  is proportional to the denominator of the two-point retarded correlation function of the corresponding conserved current (here, the energy-momentum tensor). The spectral curve equation $P( \qfr^2,\wfr)=0$ contains the full spectrum of  modes $\wfr = \wfr (\qfr^2)$, gapless and gapped. Identifying gapless modes with  \eqref{PuiseuxShear} and \eqref{PuiseuxSound}, one reads off the transport coefficients in terms of the quantum field theory parameters.

Each Puiseux series is an expansion around a critical point $(\qfr_{\rm c}^2,\wfr_{\rm c})$ of order $p$ which is a solution of the following set of equations:
\begin{equation}
P(\qfr_{\rm c}^2,\wfr_{\rm c}) = 0\, , ~ \partial_\wfr P(\qfr_{\rm c}^2,\wfr_{\rm c}) = 0\, , ~ \cdots ,  ~ \partial^p_\wfr P(\qfr_{\rm c}^2,\wfr_{\rm c})\neq 0\, . \label{CriticalPointsX}
\end{equation}
The order $p$ determines the number of branches of the curve at the critical point. The analytic properties of the branches can be found by using e.g.~the Newton polygon method, as explained in ref.~\cite{Grozdanov:2019uhi} and references therein. Accordingly, a critical point may constitute a branch point singularity (or worse) or be a regular point depending on the coefficients of the original complex curve. For example, a point with $p=1$ is always a regular point, as guaranteeed by the implicit function theorem, in which case a Puiseux series is the ordinary Taylor series. In terms of eq.~\eqref{CriticalPointsX}, the shear mode is a Puiseux series in  $\qfr^2$ of order $p = 1$ around the origin $(\qfr_{\rm c}^2,\wfr_{\rm c})=(0,0)$ (i.e., a Taylor series around a regular point), whereas for the sound mode, the origin is a branch point singularity generating a Puiseux series of order $p = 2$ \cite{Grozdanov:2019kge,Grozdanov:2019uhi}.

The radii of convergence of the series   \eqref{PuiseuxShear} and \eqref{PuiseuxSound}  are set by the locations of the closest to the origin singularities of the functions $\wfr_{\rm \tiny shear} (\qfr^2)$ and  $\wfr_{\rm \tiny sound} (\qfr^2)$, respectively, in the complex $\qfr^2$-plane. Critical points of the spectral curve  $P( \qfr^2,\wfr)=0$ having branch points at $\qfr^2=\qfr^2_c$ are the common source of such singularities. At the critical points with $p>1$, the equation $P(\qfr_{\rm c}^2,\wfr_{\rm c}) = 0$ has  multiple roots, and hence the hydrodynamic mode ``collides'' with
 one or more gapped modes  in the complex $\wfr$-plane. If the corresponding $\qfr^2=\qfr^2_c$ is a branch point, this is the level-crossing phenomenon\footnote{At the critical points with regular branches we have ``level-touching'' rather than ``level-crossing'', as happens e.g. for the BTZ background \cite{Grozdanov:2019kge}. See Appendix \ref{app-critical-points-radius} for details.},  albeit happening here at complex values of frequency and momentum squared. We illustrate this with simple examples in Appendix \ref{app-critical-points-radius}. In Appendix \ref{darboux}, we also introduce a method based on the Darboux theorem which allows one to compute the Puiseux exponent at a critical point closest to (but different from) the origin by analysing the coefficients of  a series  centred at the origin.

Computing a spectral curve in quantum field theory, even perturbatively, is a difficult problem. However, for strongly interacting theories with gravity dual descriptions, this task is rather straightforward. Indeed, the recipe for computing the two-point retarded correlators from dual gravity \cite{Son:2002sd} implies that the spectral curve $P(\qfr^2,\wfr)=0$  is determined by the boundary value $Z(u=0,\qfr^2,\wfr)$ of the solution $Z(u,\qfr^2,\wfr)$ to the bulk equations of motion for the fluctuations coupled  to the relevant conserved current: $P(\qfr^2,\wfr)=Z(u=0,\qfr^2,\wfr)=0$ \cite{Grozdanov:2019kge,Grozdanov:2019uhi}. Analytic properties of the spectral curve such as the location of branch point singularities are thus inherited from the properties of the bulk ODEs. In practice, the ODEs are sufficiently complicated and have to be solved numerically. Having such a solution, one first solves 
eqs.~\eqref{CriticalPointsX} to find the critical points in the complex $\qfr^2$-plane, and then determines the degree of singularity at the critical points by considering the quasinormal mode behaviour in the complex $\wfr$-plane under the monodromy $\qfr^2= |\qfr^2| e^{i\varphi}$, 
where $\varphi \in [0,2\pi]$ (see  Appendix \ref{app-critical-points-radius}). The closest to the origin (in the complex $\qfr^2$-plane) critical point exhibiting a branch point singularity sets the radius of convergence of the series \eqref{PuiseuxShear}, \eqref{PuiseuxSound}. In the $\CN = 4$ SYM theory at infinite $N_c$ and infinite `t Hooft coupling, the critical points closest to the origin in the shear and sound channels are located at
\begin{align}
\text{Shear}:&\qquad \qfr_{\rm c}^2 \approx 1.8906469 \pm 1.1711505i, \quad \wfr_{\rm c}\approx\pm 1.4436414 - 1.0692250i\,, \\
\text{Sound}:&\qquad \qfr_{\rm c}^2 = \pm 2 i , \quad \wfr_{\rm c} = \pm 1 - i\,,
\end{align}
leading to the radii of convergence \eqref{RN4ShearInfa} and \eqref{RN4SoundInfa}. In general, a multitude of critical points is expected to exist in the complex $\q^2$-plane,  representing  level-crossings among two or more branches of the spectrum. Moreover, at finite $N_c$ or at weak coupling described by kinetic theory, one may also expect other types of singularities to appear \cite{Heller:2020hnq}.

Here, we continue working in the limit $N_c\to\infty$, and extend the approach 
of refs.~\cite{Grozdanov:2019kge,Grozdanov:2019uhi} to bulk gravity theories with higher derivative terms, i.e., to the domain of large but finite `t Hooft coupling.

\subsection{Non-perturbative ``resummation''}
\label{resum-intro}
Inverse 't Hooft coupling corrections in the $\CN =4$ SYM theory arise from  higher-derivative terms in the dual type IIB  string theory  low energy effective action (see e.g. refs.~\cite{Gubser:1998nz,Pawelczyk:1998pb,Buchel:2004di,Buchel:2008sh,Grozdanov:2014kva}). In a holographic calculation of a quasinormal spectrum, the bulk equations of motion typically produce a  differential equation for the background fluctuation  $Z=Z (u, \wfr, \qfr^2)$ of the form \cite{Buchel:2004di,Buchel:2008sh,Grozdanov:2014kva} 
\begin{equation}\label{QNMPertEQxs}
\partial^2_u Z + \CA(u, \wfr, \qfr^2)  \partial_u  Z+ \CB(u, \wfr, \qfr^2)  Z = \gamma H [Z,\partial_u Z, \partial^2_u Z, \partial^3_u Z, \ldots]\, ,
\end{equation} 
 where $u$ is the radial coordinate in the bulk with $u=0$ the location of the boundary, $\gamma$ is a small  parameter proportional to the inverse coupling (e.g. $\gamma \sim \lambda^{-3/2}$ in the $\CN =4$ SYM with the `t Hooft coupling $\lambda$), and the right hand side comes from the leading higher-derivative correction to Einsten-Hilbert action (e.g. from the $R^4$ term in   type IIB supergravity). To avoid issues such as  Ostrogradsky instability  (see e.g. ref.~\cite{Grozdanov:2016fkt} and references therein), the higher-derivative terms in eq.~\eqref{QNMPertEQxs} are usually treated as perturbations of the second-order ODE,  and its left-hand-side is used to eliminate all derivatives higher than the first one from the right-hand-side, ignoring contributions of order $\gamma^2$ and higher. The resulting equation,
\begin{equation}\label{QNMPertEQxx}
\partial^2_u Z + \bar{\CA}(u, \wfr, \qfr^2,\gamma)  \partial_u  Z+ \bar{\CB} (u, \wfr, \qfr^2,\gamma)  Z = 0\, ,
\end{equation} 
is a homogeneous linear second-order  ODE whose coefficients  $\bar{\CA}$ and $\bar{\CB}$ now depend on $\gamma$. As discussed above, the spectral curve is then determined by the boundary value of the solution $Z(u,\wfr,\qfr^2,\gamma)$ to that ODE, i.e.  $P(\qfr^2, \wfr) \equiv Z(u\to0,\wfr,\qfr^2,\gamma) = 0$.
 
 In the standard approach, one looks for a perturbative solution to eq.~\eqref{QNMPertEQxx} in
  the form $Z = Z_0 + \gamma Z_1$. Similarly, the perturbative ansatz for the spectrum is $\wfr = \wfr_0 +\gamma \wfr_1$, where $\wfr_0$ is the quasinormal frequency at $\gamma=0$. Alternatively, one can solve eq.~\eqref{QNMPertEQxx} without assuming a perturbative ansatz. Such a non-perturbative solution, if it can be expanded in  series in powers of $\gamma \ll 1$, will not be fully quantitatively correct beyond  linear order in $\gamma$, since both in the original equation \eqref{QNMPertEQxs} and in the steps leading to eq.~\eqref{QNMPertEQxx} terms of order $\gamma^2$ and higher were ignored. Quantitatively, the solution only  captures the non-perturbative effects in $\gamma$ related to eq.~\eqref{QNMPertEQxx} and in this sense only partially ``resums'' the 
  contributions nonlinear in $\gamma$ in the approximation to the full solution. However, such a solution may provide a more faithful qualitative approximation to the exact solution at finite $\gamma$. Moreover, if the exact solution is non-perturbative in $\gamma \ll 1$, any perturbative ansatz would necessarily miss it completely, whereas the non-perturbative approach is capable of describing the situation qualitatively correctly. The choice of a correct ansatz is the crucial step in singular perturbation theory \cite{hinch-book}. We discuss these issues in more detail in 
  section \ref{sec:validity} and illustrate them with  simple  examples. In the context of holography, partial ``resummations'' have been used in 
refs.~\cite{Waeber:2015oka}, \cite{Grozdanov:2016vgg,Solana:2018pbk}  and criticised in ref.~\cite{Buchel:2018eax}. A crucial feature of such a ``resummation'' in holography, first pointed out in \cite{Grozdanov:2016vgg}, is that the quasinormal spectrum now contains new, non-pertubative gapped modes which seem to play an important role in describing physics at finite coupling  qualitatively correctly \cite{Grozdanov:2016vgg,Solana:2018pbk,Grozdanov:2018fic,Grozdanov:2018gfx}. We shall see in section  \ref{sec:N4} that the situation with the radii of convergence is similar: the non-perturbative ``resummation'' reverses the tendency seen in eqs.~\eqref{RN4Shear}, \eqref{RN4Sound}, making the radii to decrease (after an initial rise) with the coupling decreasing. In section \ref{sec:GB}, we compare this behaviour with that in the Einstein-Gauss-Bonnet theory, where both perturbative and non-perturbative results are available, using it as a theoretical laboratory to test our methods, and find a qualitative agreement with the $\CN = 4$ SYM case. Curiously, in section \ref{sec:validity} we find that it is in fact possible in some cases to infer the existence of non-perturbative critical points by using perturbative data. While we are able to explicitly demonstrate this in the Einstein-Gauss-Bonnet theory, for the $\CN = 4$ SYM theory, this may serve as an indicative argument that the same behaviour is plausible.

\section{Convergence of hydrodynamic series in the ${\cal N}=4$ SYM theory}\label{sec:N4}
We begin by studying the coupling dependence of the radii of convergence of the hydrodynamic  shear and sound modes of the ${\cal N}=4$ $SU(N_c)$ SYM theory in the $N_c \to \infty$ limit and at large but finite 't Hooft coupling $\lambda$. Our analysis uses its gravitational dual, namely, the type IIB supergravity with higher-derivative terms in the action. For the ${\cal N}=4$ SYM theory, the source of finite 't Hooft coupling corrections is the ten-dimensional low-energy effective action of type IIB string theory
\begin{align}
S_{IIB} = \frac{1}{2\kappa_{10}^2} \int d^{10} x \sqrt{-g} \left( R - \frac{1}{2} \left(\partial \phi\right)^2 - \frac{1}{4\cdot 5!} F_5^2 + \gamma e^{-\frac{3}{2} \phi} \CW + \ldots \right)\, ,
\label{eq:10DAct}
\end{align}
where $\gamma = \alpha'^3 \zeta(3) / 8$, with $\alpha'$ set by the length of the fundamental string, and the term $\CW$   proportional to the contractions of the four copies of the Weyl tensor,
\begin{align}
\label{eq:Wterm}
\CW = C^{\alpha\beta\gamma\delta}C_{\mu\beta\gamma\nu} C_{\alpha}^{~\rho\sigma\mu} C^{\nu}_{~\rho\sigma\delta} + \frac{1}{2} C^{\alpha\delta\beta\gamma} C_{\mu\nu\beta\gamma} C_{\alpha}^{~\rho\sigma\mu} C^\nu_{~\rho\sigma\delta}\, .
\end{align}
Considering corrections to the AdS-Schwarzschild black brane background and its fluctuations, potential  $\alpha'$ corrections to supergravity fields other than the metric and the five-form field have been argued to be irrelevant \cite{Myers:2008yi}. Moreover, as discussed in \cite{Buchel:2008ae}, for the purposes of computing the corrected quasinormal spectrum, one can use the Kaluza-Klein reduced five-dimensional action
\begin{align}
S = \frac{1}{2\kappa_5^2} \int d^5 x \sqrt{-g} \left(R  + \frac{12}{L^2} + \gamma \CW \right)\, ,
\label{eq:hd-action-x}
\end{align}
where $\CW$ is now given by eq.~(\ref{eq:Wterm}) in $5d$. The effective five-dimensional gravitational constant is related to the rank of the gauge group $SU(N_c)$ by the expression $\kappa_5 = 2\pi /N_c$. The parameter $\gamma$ is related to the value of the 't Hooft coupling constant $\lambda$ in the $\CN=4$ SYM theory via $\gamma  = \lambda^{-3/2}\zeta (3) L^6/8$.  This parameter is  dimensionless in units of $L$. Higher derivative terms in the equations of motion are treated as perturbations in $\gamma$. In the following, we shall use $\lambda$ and 
\begin{align}
\gamma  = \frac{\zeta (3)}{8 \lambda^{3/2}} \ll 1
\label{gamma-def}
\end{align}
interchangeably.

The black brane solution to the equations of motion following from the action (\ref{eq:hd-action-x}), which is dual to an equilibrium  thermal state of the CFT at temperature $T$, is given by  \cite{Gubser:1998nz,Pawelczyk:1998pb}
\begin{align}
ds^2 = \frac{(\pi T L)^2}{u} \left( - e^{A(u)} f(u) dt^2 + dx^2 +dy^2 +dz^2 \right) + e^{B(u)} \frac{L^2 du^2}{4u^2 f}\, ,
\label{eq:corrected_metric-x}
\end{align}
where $f(u) = 1 - u^2$. The radial coordinate is denoted by $u$, with the boundary  located at $u=0$ and the horizon at $u=1$. To leading order in $\gamma$, the functions $A(u)$ and $B(u)$ were found to be 
\begin{align}
A(u) =  - 15\gamma\left(5u^2+5u^4-3 u^6 \right)\, , \qquad B(u) =  15\gamma \left(5u^2 + 5 u^4 - 19 u^6 \right)\, .
\end{align}
The correction to the $\CN=4$ SYM entropy density is then given by  \cite{Gubser:1998nz}
\begin{equation}
\frac{s}{s_0} = \frac{3}{4}\left( 1 +  15 \gamma + \cdots \right)\, ,
\end{equation}
where $s_0 = 2 \pi^2 N_c^2 T^3 / 3$ is the Stefan-Boltzmann entropy density of the ideal gas of particles in the $\CN = 4$ SYM theory (i.e., in the theory at $\lambda = 0$). The metric \eqref{eq:corrected_metric-x} was also used to compute `t Hooft  coupling constant corrections to the ratio of shear viscosity $\eta$ to entropy density \cite{Buchel:2004di,Buchel:2008sh},
\begin{equation}
\frac{\eta}{s} = \frac{1}{4\pi}\left( 1 + 120 \gamma + \cdots \right)\, ,
\end{equation}
 and to all of the second-order transport coefficients of the $\CN=4$ SYM plasma.\footnote{The complete list of the coefficients can be found e.g. in refs.~\cite{Grozdanov:2014kva,Grozdanov:2016fkt}.} Computing transport coefficients and, in general, correlation functions of the energy-momentum tensor in the $\CN=4$ SYM plasma requires considering small fluctuations of the metric $g_{\mu\nu} = g_{\mu\nu}^{(0)} +  h_{\mu\nu}(u,t,x,y,z)$, where $g_{\mu\nu}^{(0)}$ is the background \eqref{eq:corrected_metric-x}. Due to translational invariance and spatial isotropy of the background, we can Fourier transform the fluctuations and choose the direction of  spatial momentum along $z$, so that
 \begin{align}
h_{\mu\nu} (u,t,z) = \int \frac{d\omega d q}{(2\pi)^2}\, e^{-i \omega t + i q z}\, h_{\mu\nu} (u, \omega, q)\,.
\end{align}
Following the recipes of ref.~\cite{Kovtun:2005ev} and choosing the radial gauge  $h_{ u \nu} = 0$, one can write down the linearised equations of motion for the three gauge-invariant linear combinations $Z_i$, $i=1,2,3$, of the modes $h_{\mu\nu}(u,\omega,q)$ in the scalar, shear and sound channels, respectively \cite{Policastro:2002se,Kovtun:2005ev,Grozdanov:2016vgg}.

The linearised equations of motion obtained following  the procedure outlined in section \ref{section-cp-res} are given in the three channels by \cite{Grozdanov:2016vgg}
\begin{equation}
\partial^2_u Z_i + {\cal A}_{(i)}(u, \wfr, \qfr^2, \gamma) \partial_u Z_i +  {\cal B}_{(i)}(u, \wfr, \qfr^2, \gamma) Z_i =0\, ,
\label{fluct-eq-main-x-n4}
\end{equation}
 where ${\cal A}_{(i)}(u, \wfr, \qfr^2, \gamma) = {\cal A}_{(i)}^{(0)}(u, \wfr, \qfr^2) + \gamma {\cal A}_{(i)}^{(1)}(u, \wfr, \qfr^2)$ and ${\cal B}_{(i)}(u, \wfr, \qfr^2, \gamma) = {\cal B}_{(i)}^{(0)}(u, \wfr, \qfr^2) + \gamma {\cal B}_{(i)}^{(1)}(u, \wfr, \qfr^2)$. The coefficients are given explicitly in Appendix \ref{sec:appendix-N=4}. As discussed in section \ref{section-cp-res}, using the ODEs \eqref{fluct-eq-main-x-n4}, the quasinormal spectrum  can now be computed  either perturbatively by expanding $Z = Z_0 + \gamma Z_1$ along with $\wfr = \wfr_0 + \gamma \wfr_1$ and $\qfr^2 = \qfr^2_0 + \gamma \qfr^2_1$ \cite{Stricker:2013lma}, or non-perturbatively by treating eq.~\eqref{fluct-eq-main-x-n4} as being 
 exact in the parameter $\gamma$ \cite{Waeber:2015oka}, \cite{Grozdanov:2016vgg,Solana:2018pbk}. 
 
\begin{figure}[t!]
\centering
\includegraphics[width=0.45\textwidth]{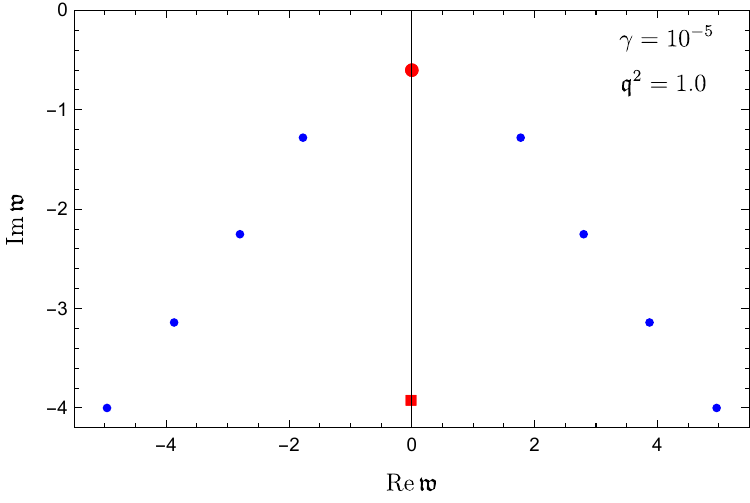}
\hspace{0.05\textwidth}
\includegraphics[width=0.45\textwidth]{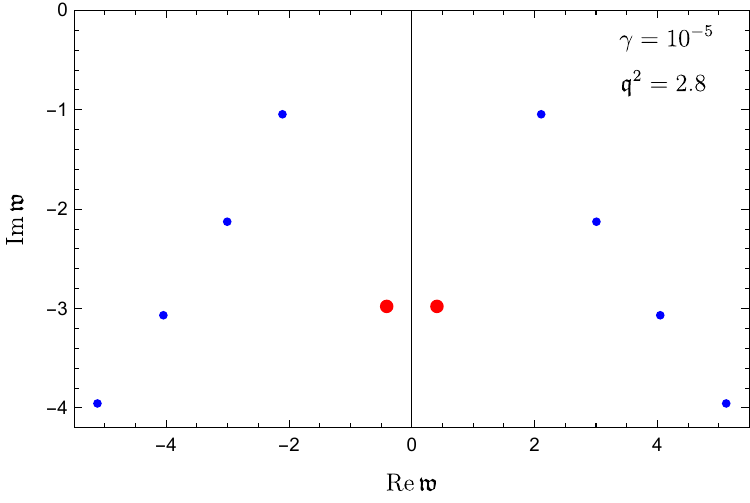}
\caption{{\small Quasinormal spectrum in the shear channel of  the $\CN=4$ SYM for  $\gamma = 1\,\cdot10^{-5}$ and $\qfr^2 = 1$ (left panel).
The hydrodynamic shear mode at $\wfr \approx -0.60064 i$ is shown by the red circle. The new feature, not seen in a perturbative calculation, is the appearance of an extra mode on the imaginary axis (shown by the red square), ascending from complex infinity with $\gamma$ increasing \cite{Grozdanov:2016vgg}. With real $\qfr^2$ increasing, the hydrodynamic mode moves down the imaginary axis, while the new non-perturbative mode moves up. They collide at $\qfr^2_* 
\approx 2.72$ and move off the axis for $\qfr^2>\qfr^2_*$ (right panel).}}
\label{shear-qnm-spectrum-example-xddx}
\end{figure}

\begin{figure}[t!]
\centering
\includegraphics[width=0.55\textwidth]{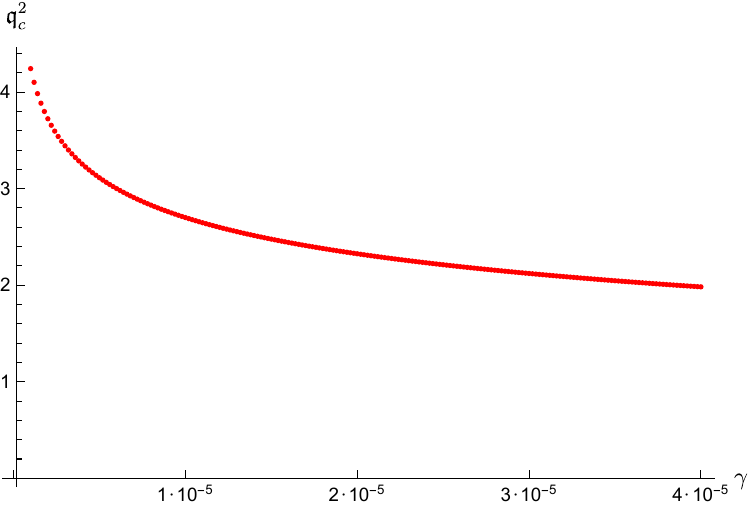}
\caption{{\small The value of the   (real) spatial momentum squared, limiting the hydrodynamic regime,
as a function of the (inverse) coupling $\gamma$ in the shear channel of the $\CN=4$ SYM \cite{Grozdanov:2016vgg}. 
Hydrodynamics has a wider range of applicability in $\qfr^2$ at smaller $\gamma$ 
(larger 't Hooft coupling).}}
\label{shear-qnm-spectrum-end-of-hydro}
\end{figure}

 As an example, the shear channel quasinormal spectrum for  $\gamma = 1\cdot 10^{-5}$ is shown in fig.~\ref{shear-qnm-spectrum-example-xddx}. Its novel feature, discussed in detail in ref.~\cite{Grozdanov:2016vgg}, is the existence of the non-perturbative (in $\gamma$) gapped modes on the imaginary axis. The highest (closest to the real axis) of those modes is shown in fig.~\ref{shear-qnm-spectrum-example-xddx} by the red square: with   real $\qfr^2$ increasing, this mode moves up the axis and at 
 $\qfr^2=\qfr^2_*$ it collides with the hydrodynamic shear mode (shown in fig.~\ref{shear-qnm-spectrum-example-xddx} by the red circle). For $\qfr^2>\qfr^2_*$, the two  modes move off the imaginary axis, effectively destroying the diffusive pole of the correlator. In ref.~\cite{Grozdanov:2016vgg}, this was interpreted as the end of  the hydrodynamic regime at sufficiently large spatial momentum (small wavelength), where the microscopic effects prevail over the collective ones. The dependence $\qfr^2_* =\qfr^2_* (\gamma)$, shown in fig.~\ref{shear-qnm-spectrum-end-of-hydro}, suggests that the domain of applicability of  the hydrodynamic description is smaller at larger $\gamma$ (i.e., at smaller `t Hooft coupling), but it seems to extend to arbitrarily large momentum in the limit of infinite coupling (at $\gamma \to 0$) \cite{Grozdanov:2016vgg}.

 To see how this qualitative picture is amended at very large but finite  't Hooft coupling, we now consider the radius of convergence of the hydrodynamic shear and sound dispersion series in this theory, which requires us to solve eq.~\eqref{CriticalPointsX} and look for critical points with $p=2$. With $P(\qfr^2, \wfr)$ given by $P(\qfr^2, \wfr)=Z(\wfr,\qfr^2) \equiv Z(u=0,\wfr,\qfr^2)$ for any channel (we omit the index ``i'' labeling the channel), we are therefore looking for solutions $\wfr = \wfr_{\rm c}$ and $\qfr^2=\qfr_{\rm c}^2$ to the system 
\begin{equation}\label{CritN4-1}
Z(\wfr,\qfr^2) = 0\, , \qquad \partial_\wfr Z(\wfr,\qfr^2) = 0\, ,
\end{equation}
with $\partial^2_\wfr Z(\wfr,\qfr^2) \neq 0 $. In the non-perturbative approach (in $\gamma$), critical points follow directly from eq.~\eqref{CritN4-1}, where $Z(\wfr,\qfr^2)$ can be found by constructing a Frobenius series solution $Z(u,\wfr,\qfr^2)$ 
to eq.~\eqref{fluct-eq-main-x-n4} around the horizon in the standard way \cite{Horowitz:1999jd,Starinets:2002br,Kovtun:2005ev}, and setting $u=0$. For $\gamma=0$, this is the same procedure as the one used in refs.~\cite{Grozdanov:2019kge,Grozdanov:2019uhi}.

\begin{figure}[t!]
\centering
\includegraphics[width=0.55\textwidth]{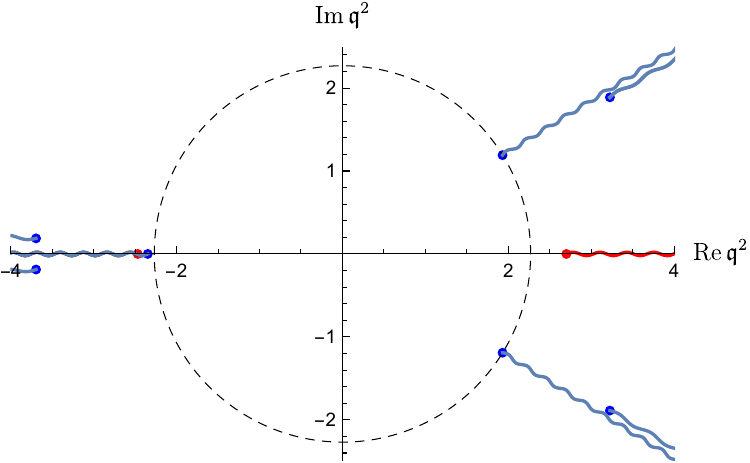}
\caption{{\small The closest to the origin branch points (see Table \ref{tab:table1x}) of the ${\cal N}=4$ SYM shear mode dispersion relation $\wfr=\wfr(\qfr^2)$ in the complex $\qfr^2$-plane at 
$\gamma =1\cdot 10^{-5}$.  The critical points seen in perturbation theory and their branch cuts are shown by blue colour, the two non-perturbative critical points on the real axis are shown in red. The radius of convergence at $\gamma =1\cdot 10^{-5}$, $R_{\rm shear} \approx 2.27$, is determined by the pair of 
branch points at $\qfr^2 \approx 1.93 \pm  1.19 i$.  We note that it is the sole inclusion of the non-perturbative red critical point on the positive real $\qfr^2$-axis that leads to the non-perturbative result of fig.~\ref{shear-qnm-spectrum-end-of-hydro}.}
}
\label{shear-qnm-spectrum-example-xooq}
\end{figure}

To find the critical points perturbatively, we expand equations \eqref{CritN4-1} as 
\begin{equation}\label{exp-4}
\begin{aligned}
Z_0 (\wfr,\qfr^2)+ \gamma Z_1 (\wfr,\qfr^2) &=0\, ,\\
\partial_\wfr Z_0 (\wfr,\qfr^2) + \gamma \partial_\wfr Z_1 (\wfr,\qfr^2) &=0\, ,
\end{aligned}
\end{equation}
and also expand 
\begin{equation}
\label{exp-5}
\wfr_{\rm c} = \wfr_{{\rm c},0} +\gamma \wfr_{{\rm c},1}\, , \qquad \qfr^2_{\rm c} = \qfr_{{\rm c},0}^2 +\gamma \qfr_{{\rm c},1}^2\, .
\end{equation}
Together, these expansions yield a system of equations at $\CO(\gamma^0)$:
\begin{equation}
\label{exp-6-0}
\begin{aligned}
Z_0(\wfr_{\rm c,0}, \qfr^2_{\rm c,0}) &= 0\, , \\
\partial_\wfr Z_0(\wfr_{\rm c,0}, \qfr^2_{\rm c, 0}) &= 0\, ,
\end{aligned}
\end{equation}
and a system at $\CO(\gamma)$:
\begin{equation}
\label{exp-6}
\begin{aligned}
Z_1 (\wfr_{\rm c,0},\qfr_{\rm c,0}^2) +  \partial_{\qfr^2} Z_0 (\wfr_{\rm c,0}, \qfr^2_{\rm c,0}) \, \qfr^2_{\rm c,1} &=0\, ,\\
\partial_\wfr Z_1 (\wfr_{\rm c,0}, \qfr^2_{\rm c,0}) + \partial^2_\wfr Z_0 (\wfr_{\rm c,0}, \qfr^2_{\rm c,0})  \,  \wfr_{\rm c,1} + \partial_\wfr \partial_{\qfr^2} Z_0 (\wfr_{\rm c,0}, \qfr^2_{\rm c,0})  \,   \qfr_{\rm c,1}^2  &= 0\, .
\end{aligned}
\end{equation} 
Eqs.~\eqref{exp-6-0} and \eqref{exp-6} are sufficient to find $\wfr_{\rm c,0}$, $\wfr_{\rm c,1}$, $\qfr^2_{\rm c,0}$ and $\qfr_{\rm c,1}^2$. More explicitly, eq. \eqref{exp-6} allows us to express 
\begin{equation}
\begin{aligned}
\qfr_{\rm c,1}^2 &= - 
\frac{  Z_1 (\wfr_{\rm c,0},\qfr_{\rm c,0}^2) }{ \partial_{\qfr^2} Z_0 (\wfr_{\rm c,0}, \qfr^2_{\rm c,0}) }\, ,
\\
\wfr_{\rm c,1} &=  \frac{1}{ \partial^2_\wfr Z_0 (\wfr_{\rm c,0}, \qfr^2_{\rm c,0}) }  \left(  \frac{ Z_1 (\wfr_{\rm c,0},\qfr_{\rm c,0}^2)  \partial_\wfr \partial_{\qfr^2} Z_0 (\wfr_{\rm c,0}, \qfr^2_{\rm c,0}) }{ \partial_{\qfr^2} Z_0 (\wfr_{\rm c,0}, \qfr^2_{\rm c,0})  }  -  \partial_\wfr Z_1 (\wfr_{\rm c,0}, \qfr^2_{\rm c,0}) \right)\, . 
\end{aligned}
\label{corr-1}
\end{equation}
As before, the function $Z_0 (\wfr,\qfr^2)$ can be obtained as the boundary value of the Frobenius solution to eq.~\eqref{fluct-eq-main-x-n4} with $\gamma=0$ and  $Z_1 (\wfr,\qfr^2)$ as the boundary value of the Frobenius solution to the corresponding inhomogeneous equation.

\subsection{Shear channel}
\label{N=4-shear}
At infinite `t Hooft coupling, the shear mode dispersion relation $\wfr=\wfr (\qfr^2)$ has numerous 
branch point singularities \cite{Grozdanov:2019kge,Grozdanov:2019uhi}. At finite coupling, we expect those singularities, now parametrised by $\gamma \propto \lambda^{-3/2}$,  to move in the complex $\qfr^2$-plane with $\gamma$ varying. As discussed in section \ref{resum-intro}, one can compute  relevant corrections by using either the ``conservative'' perturbative or the non-perturbative approach.

\begin{table}[h!]
  \begin{center}
    \begin{tabular}{l|c|c|c|c} 
      \# & $\qfr_c^2$ (non-pert.) & $|\qfr_c^2|$ (non-pert.) & $\qfr_c^2$ (pert.) & 
      $|\qfr_c^2|$ (pert.)\\
      \hline
      1 & $1.93027 \pm  1.19123 i$  & $2.268$ & $1.93147\pm 1.18977i$ & $2.269$ \\
      2 & $-2.34715$ &  $2.347$ & $-2.35128$ & $2.351$\\
      3 & $-2.46848$ &  $2.469$ &  n/a  & n/a  \\
      4 & $2.70094$ & 2.701 & n/a & n/a \\
      5 & $-3.69434\pm 0.18770 i  $ & $3.699$ & $ -3.63611 \pm 0.03644 i  $ & $3.528$ \\
      6 & $3.22474 \pm 1.88845 i$&$3.737$ & $3.31204 \pm 1.84293 i$ &$3.790$ \\
    \end{tabular}
    \caption{\small{The six closest to the origin (in the complex $\qfr^2$-plane) critical points 
    for $\gamma = 1\cdot 10^{-5}$.}}
    \label{tab:table1x}
  \end{center}
\end{table}
\begin{table}[h!]
  \begin{center}
    \begin{tabular}{l|c|c} 
      \# & $\wfr_c$ (non-perturbative) & $\wfr_c$ (perturbative)  \\
      \hline
      1 & $\pm 1.47755 - 1.05400 i $ & $\pm 1.48035 - 1.05562 i $\\
  2 & $-1.73447 i$ &  $-1.71258 i$\\
  3 & $-2.60274 i$ & n/a\\
  4 & $-2.93397 i  $ & n/a \\
  5 & $\pm 1.53654 - 2.71700 i$ &$ \pm 1.72062 - 2.98177 i$\\
  6 & $\pm 2.54972 - 1.99267 i$ &$ \pm 2.65553 - 1.99615  i$\\
    \end{tabular}
    \caption{\small{The  six closest to the origin (in the complex $\wfr$-plane) critical points for $\gamma = 1\cdot 10^{-5}$.}}
    \label{tab:table2}
  \end{center}
\end{table}

\subsubsection{Perturbative calculation}
For a perturbative calculation of the  coupling constant correction to the radius of convergence in the 
shear channel of the $\CN = 4$ SYM theory, we use eq.~\eqref{fluct-eq-main-x-n4} with $i=2$. To first order in  $\gamma \propto \lambda^{-3/2} $, from eqs.~\eqref{corr-1} we find the following first set of critical points closest to the origin in the complex $\qfr^2$-plane:
\begin{eqnarray}
\label{crit-pointn4-1x}
&\,&\qfr_{\rm c}^2 \approx1.89065  \pm 1.17115  i +\gamma(4081.99 \pm 1862.06 i)\,,\\
&\,&\wfr_{\rm c} \approx\pm 1.44364 - 1.06923  i + \gamma(\pm 3671.27  + 1360.52 i) 
\label{crit-pointn4-2x}\,.
\end{eqnarray}
The value of $\qfr_{\rm c}^2$ in eq.~\eqref{crit-pointn4-1x} gives the convergence radius $|\qfr_{\rm c}^2|$ quoted in eq.~\eqref{RN4Shear}. The radius increases with $\gamma$ increasing (i.e. with the coupling $\lambda$ decreasing from its infinite value). The numerical coefficients multiplying the parameter $\gamma$ in eqs.~\eqref{crit-pointn4-1x}, \eqref{crit-pointn4-2x} are large: the perturbative terms give small corrections to the $\gamma=0$ result only for $\gamma 
\lesssim 10^{-4} - 10^{-5}$. 

The next closest to the origin critical point (i.e.~the critical point with larger value of $|\qfr^2|$ than \eqref{crit-pointn4-1x})
 is located on the negative real axis of $\qfr^2$:
\begin{eqnarray}
\label{crit-pointn4-1xx}
&\,&    \qfr_{\rm c,1}^2 \approx-2.37737  + \gamma \, 2608.88     \,,\\
&\,&   \wfr_{\rm c,1} \approx -1.64659 i - \gamma\, 6599.64 i    
\label{crit-pointn4-2xx}
\,.
\end{eqnarray}
At $\gamma =0$, this critical point plays no role in determining the radius of convergence. At finite $\gamma$, the point \eqref{crit-pointn4-1xx} moves closer to the origin with $\gamma$ increasing, whereas the pair of points \eqref{crit-pointn4-1x} moves away from it. At $\gamma = \gamma_* \approx  2.172 \cdot 10^{-5}$, the critical point \eqref{crit-pointn4-1xx} formally becomes dominant (closest to the origin), changing the situation qualitatively. We view 
this as  an indication of the breakdown of linear (in $\gamma$) approximation . 

The next  two sets of critical points with yet larger values of $|\qfr^2|$ are
\begin{eqnarray}
\label{crit-pointn4-1}
&\,&   \qfr_{\rm c,2}^2 \approx-3.11051 \mp 0.81050i +\gamma(-52560.3  \pm 77406.3 i)     \,,\\
&\,&   \wfr_{\rm c,2}\approx\pm 1.41043 -2.87086 i + \gamma(\pm 31019.2  -11091.7i)  \,, \\
&\,&    \qfr_{\rm c,3}^2\approx2.90684  \pm 1.66612 i +\gamma(40520.1  \pm 17681.1i)     \,,\\
&\,&   \wfr_{\rm c,3}\approx\pm 2.38819  -2.13154 i + \gamma(\pm 26733.9  +13539.1i) \label{crit-pointn4-1a} \,.
\end{eqnarray}
Notice again the large numerical coefficients multiplying the perturbative parameter $\gamma$ in eqs.~\eqref{crit-pointn4-1}--\eqref{crit-pointn4-1a}. For illustration, several perturbative closest
 to the origin critical points in the complex $\qfr^2$-plane for $\gamma =1\cdot 10^{-5}$ are shown in fig.~\ref{shear-qnm-spectrum-example-xooq}  in blue colour.

\subsubsection{Non-perturbative calculation}
\begin{figure}[t!]
\centering
\includegraphics[width=0.45\textwidth]{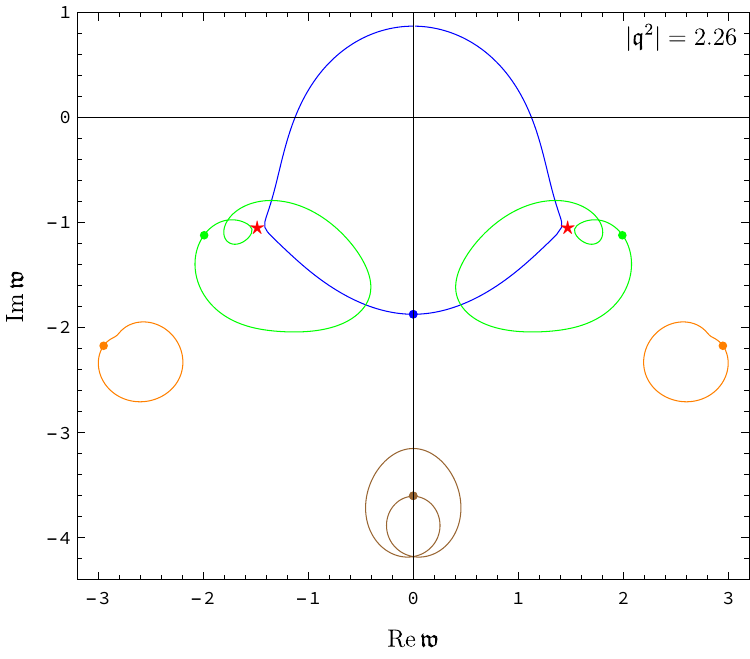}
\hspace{0.05\textwidth}
\includegraphics[width=0.45\textwidth]{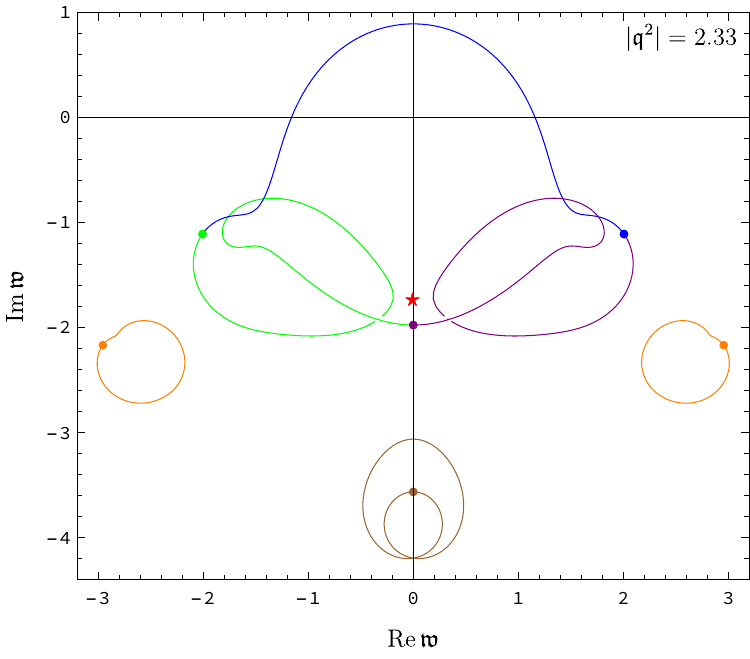}\\
\includegraphics[width=0.45\textwidth]{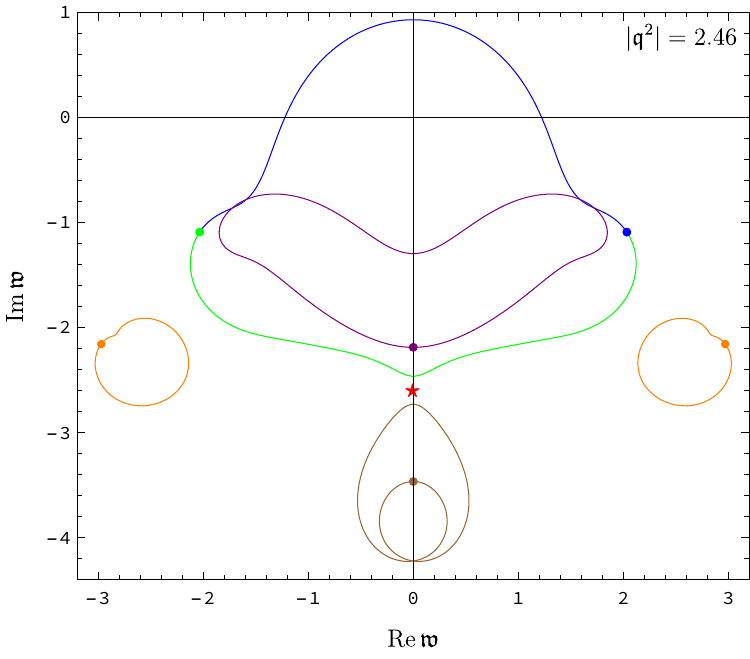}
\hspace{0.05\textwidth}
\includegraphics[width=0.45\textwidth]{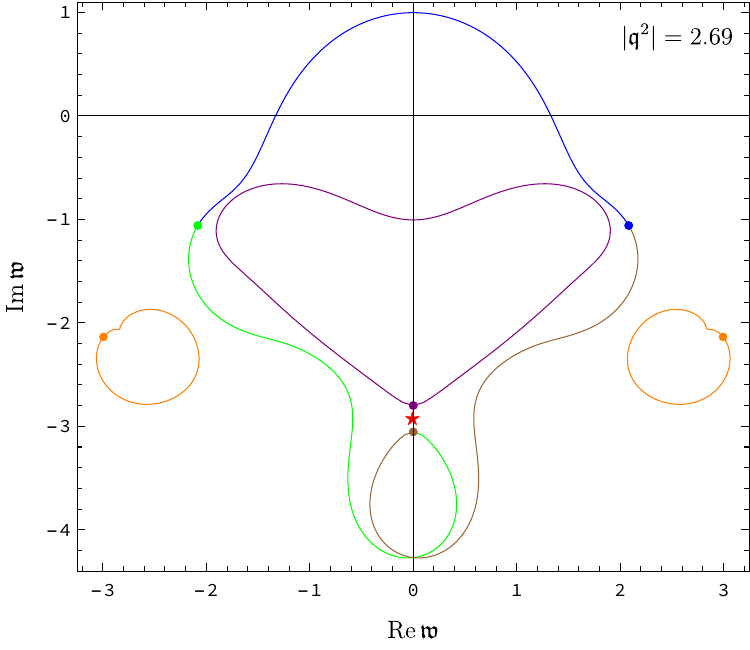}
\caption{{\small Quasinormal spectrum in the shear channel at $\gamma = 1\cdot 10^{-5}$, 
computed non-perturbatively in $\gamma$. The trajectories are plotted for complex values of the spatial momentum squared, $\qfr^2= |\qfr^2|e^{i \varphi}$, where phase $\varphi$ is varied from 0 to $2\pi$.  The positions of quasinormal modes at $\varphi=0$ are shown by dots. The positions of the  two critical points (closest to the origin in the complex $\qfr^2$-plane) are shown by red stars. The first pair of critical points corresponds to the collision of trajectories of the two top gapped modes (shown in green) 
with the hydrodynamic shear mode trajectory (shown in blue) at  $\qfr_c^2 \approx 1.93027 \pm  1.19123 i$ and $\wfr_c \approx \pm 1.47755 - 1.05400 i$. The corresponding value  $|\qfr^2|\approx 2.27$ sets the radius of convergence of the hydrodynamic mode (top left panel). The second critical point arises from the collision on the imaginary axis between the parts of the common curve involving the three top modes including the shear mode (top right panel). 
The two plots in the bottom row show the third and the fourth critical points. Both points arise on the imaginary axis of complex frequency from the collision involving  the new, non-perturbative mode in the quasinormal spectrum.}}
\label{shear-qnm-spectrum-example-1}
\end{figure}

\begin{figure}[t!]
\centering
\includegraphics[width=0.4\textwidth]{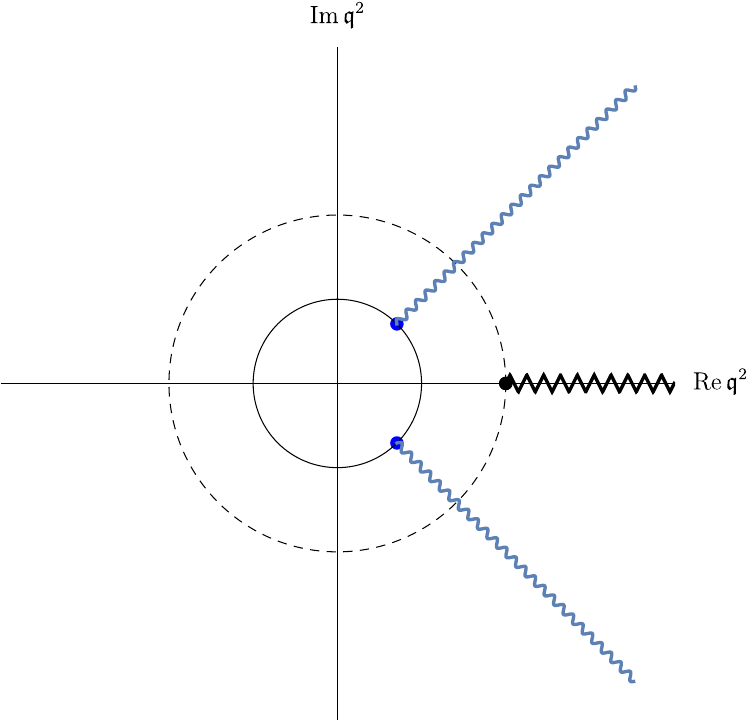}\hspace{0.05\textwidth}
\includegraphics[width=0.4\textwidth]{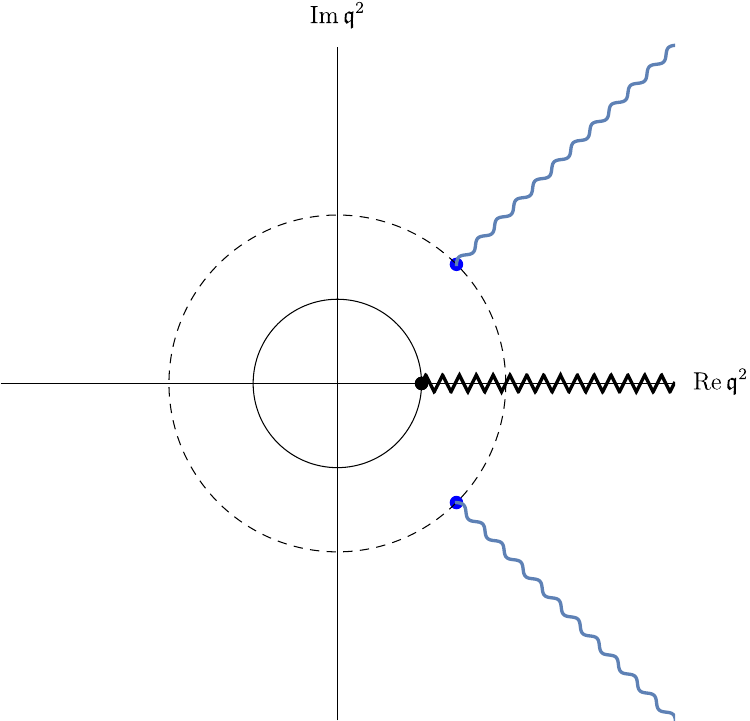}
\caption{{\small The closest to the origin branch points of the shear mode 
dispersion relation $\wfr=\wfr(\qfr^2)$ in the complex $\qfr^2$-plane (shown schematically). The radius of convergence is determined by the pair of branch points (left panel) or the branch point on the real axis (right panel).
}
}
\label{shear-qnm-spectrum-example-x}
\end{figure}

\begin{figure}[t!]
\centering
\includegraphics[width=0.45\textwidth]{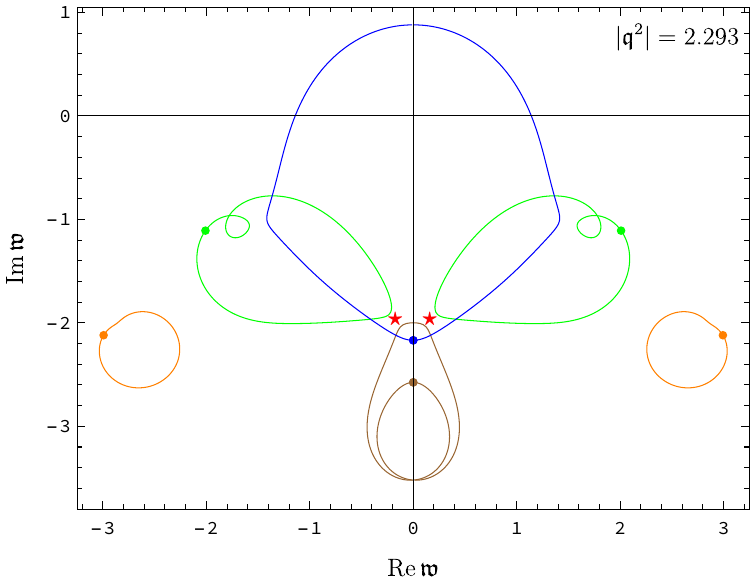}
\hspace{0.05\textwidth}
\includegraphics[width=0.45\textwidth]{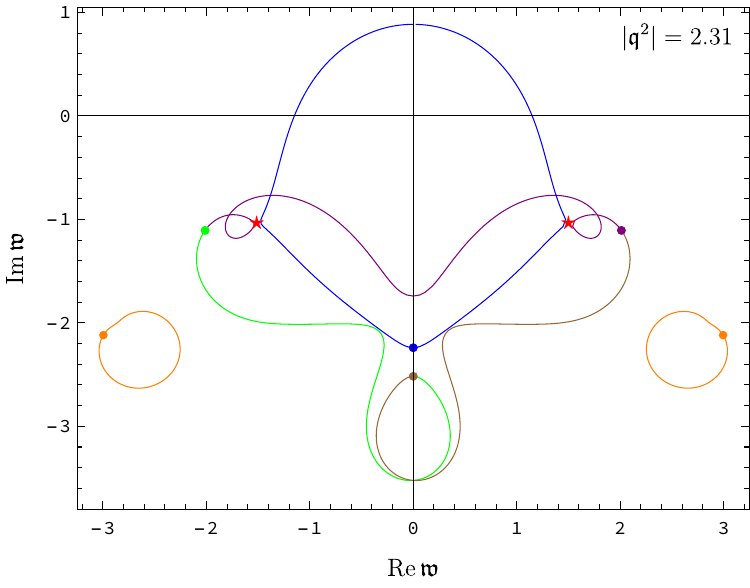}\\
\includegraphics[width=0.45\textwidth]{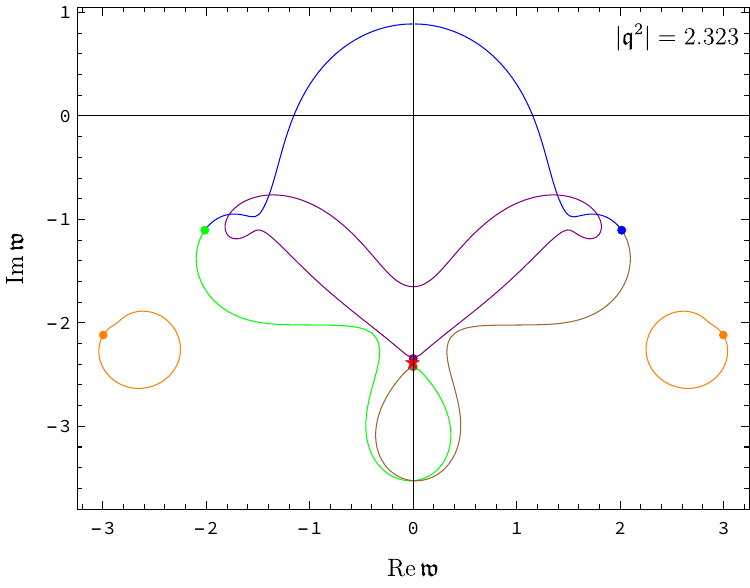}
\hspace{0.05\textwidth}
\caption{{\small Quasinormal spectrum in the shear channel at $\gamma = 2\cdot 10^{-5}$, 
computed non-perturbatively in $\gamma$. The level-crossing  occurring at the smallest value of $|\qfr^2|$ (top left panel) does not affect the hydrodynamic mode (shown in blue colour). The first (smallest in $|\qfr^2|$) critical point of the shear mode arises from the level-crossing with the top gapped modes (top right panel). This point sets the radius of convergence of the hydrodynamic series. The critical point with an even higher value of $|\qfr^2|$ is shown in the bottom panel.}}
\label{shear-qnm-spectrum-example-2}
\end{figure}

\begin{figure}[t!]
\centering
\includegraphics[width=0.45\textwidth]{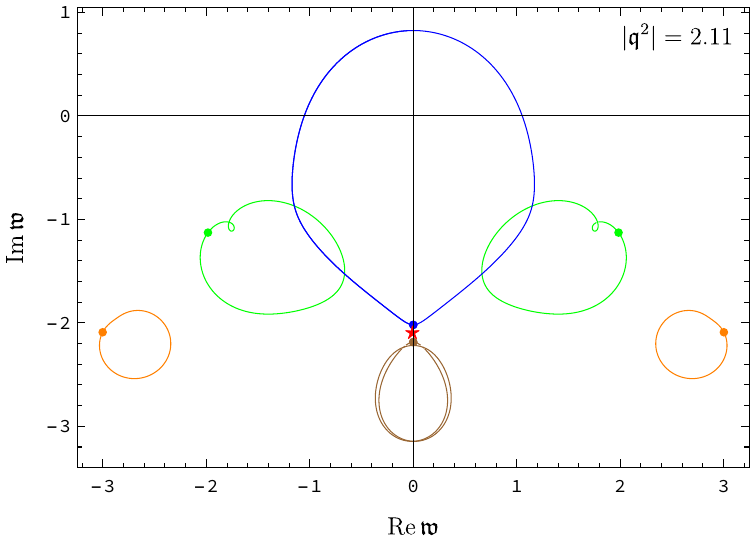}
\hspace{0.05\textwidth}
\includegraphics[width=0.45\textwidth]{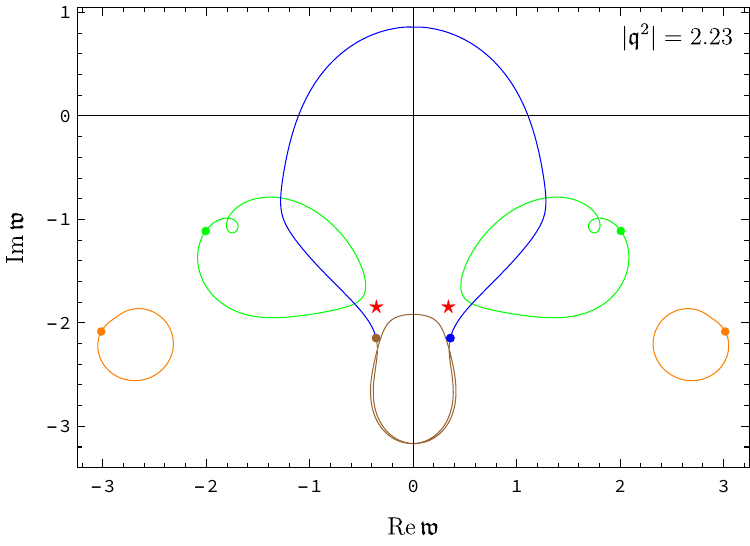}\\
\caption{{\small Quasinormal spectrum in the shear channel at $\gamma = 3\cdot 10^{-5}$, 
computed non-perturbatively in $\gamma$. The radius of convergence is set by the level-crossing between the shear mode and the non-perturbative mode on the imaginary axis (left panel). This critical point coincides with the endpoint of the hydrodynamic regime discussed in ref.~\cite{Grozdanov:2016vgg}. An example of critical points with a larger $|\qfr^2|$ is shown in the right panel.}}
\label{shear-qnm-spectrum-example-3}
\end{figure}

For a non-perturbative calculation, we solve eqs.~\eqref{fluct-eq-main-x-n4} and \eqref{CritN4-1} numerically 
without assuming $\gamma$ to be small. We observe three qualitatively different scenarios of quasinormal modes' behaviour, and illustrate them by showing the modes at $\gamma = 1\cdot10^{-5}$, $\gamma= 2\cdot10^{-5}$ 
and $\gamma = 3\cdot10^{-5}$, respectively (see figs.~\ref{shear-qnm-spectrum-example-1}, \ref{shear-qnm-spectrum-example-2}, \ref{shear-qnm-spectrum-example-3}):

a) At $\gamma = 1\cdot10^{-5}$, the top modes in the spectrum are shown in fig.~\ref{shear-qnm-spectrum-example-1} in the complex plane of $\wfr$,  for complex values of the spatial 
momentum squared $\qfr^2= |\qfr^2|e^{i \varphi}$, where the phase $\varphi$ is varied from 0 to $2\pi$. 
The figure shows how the critical points (we show the first four points closest to the origin) arise from the collision of quasinormal modes trajectories as the phase $\varphi$ varies. The closest to the origin (in the complex $\qfr^2$-plane; see fig.~\ref{shear-qnm-spectrum-example-xooq}) pair of critical points sets the radius of convergence 
$R_{\rm shear}=|\qfr^2_c| \approx 2.27$ of the hydrodynamic series (in the complex $\wfr$-plane, this point 
is shown in the top left panel in fig.~\ref{shear-qnm-spectrum-example-1}). The location of the six closest to the origin critical points at $\gamma = 1\cdot 10^{-5}$ is given in Tables \ref{tab:table1x} and  \ref{tab:table2}, where a comparison between perturbative and non-perturbative results is also made. For  $\gamma = 1\cdot 10^{-5}$, the location of the critical points is in a reasonably good agreement with the perturbative results \eqref{crit-pointn4-1x}, \eqref{crit-pointn4-2x} and  \eqref{crit-pointn4-1xx}, \eqref{crit-pointn4-2xx}, except when the collision of modes involves the  mode on the imaginary axis which has no perturbative analogue.
b) At $\gamma = 2\cdot10^{-5}$, the top modes in the spectrum are shown in fig.~\ref{shear-qnm-spectrum-example-2}. Here, the first level-crossing (i.e., the level-crossing with the minimal value of $|\qfr^2|$) occurs between the two top gapped modes and the non-perturbative mode on the imaginary axis, as shown in the top left panel of  fig.~\ref{shear-qnm-spectrum-example-2}. However, the shear mode is not affected by this crossing: its first non-analyticity 
still arises as a result of the collision with the top two gapped modes as shown in the top right panel of  fig.~\ref{shear-qnm-spectrum-example-2}.   This collision sets the radius of convergence of the shear mode at 
$R_{\rm shear}(\gamma) =|\qfr^2_c|\approx 2.31$. In the complex $\qfr^2$-plane, the position of the branch point singularities is thus qualitatively the same as at $\gamma = 1\cdot10^{-5}$ 
(see fig.~\ref{shear-qnm-spectrum-example-x}, left panel), but the radius of convergence $R_{\rm shear}(\gamma)$ increases with $\gamma$ increasing (i.e. with the coupling $\lambda$ decreasing).

c) Finally, at $\gamma = 3\cdot10^{-5}$, the top modes in the spectrum are shown in fig.~\ref{shear-qnm-spectrum-example-3}. In this case, the situation is qualitatively different.  Now the first level-crossing occurs at a {\it real} value of $\qfr^2_c \approx 2.12157$ and at $\wfr_c\approx -2.096 i$, which is a result of the collision between the shear mode and the non-perturbative mode on the imaginary axis (top left panel in fig.~\ref{shear-qnm-spectrum-example-3}). This is the collision of the type shown in fig.~\ref{shear-qnm-spectrum-example-xddx} and interpreted in 
ref.~\cite{Grozdanov:2016vgg}, where it was discovered, as the end point $\qfr_*^2(\gamma)$ of the hydrodynamic regime (in the sense that for real $\qfr^2>\qfr_*^2$ the hydrodynamic purely imaginary shear mode does not exist). The radius of convergence in the $\qfr^2$-plane is set by the corresponding value of $|\qfr^2_*|=|\qfr^2_c|\approx 2.12157$ (see fig.~\ref{shear-qnm-spectrum-example-x}, right panel, where this situation is shown schematically). Thus, at $\gamma = 3\cdot10^{-5}$, the radius of convergence is determined by the non-perturbative mode. In this regime, the convergence radius decreases with $\gamma$ increasing.

The three examples considered above illustrate the general situation fully. For infinitesimally small $\gamma$, the radius of convergence of the shear mode  dispersion relation is an increasing function of $\gamma$. In the complex $\qfr^2$-plane, the obstacle to convergence is the pair of critical points, as shown in fig.~\ref{shear-qnm-spectrum-example-x} (left panel). These points move away from the origin with $\gamma$ increasing. At $\gamma = \gamma_* \approx 2.05 \cdot 10^{-5}$, the picture changes qualitatively, as the transition between the  regimes a) and c) occurs. Now the new critical point arising from the level-crossing with the non-perturbative mode is located closer to the origin in the complex $\qfr^2$-plane than the previous pair of critical points (fig.~\ref{shear-qnm-spectrum-example-x}, right panel). This new critical point is located on the positive real axis of $\qfr^2$ and corresponds to the end point of hydrodynamic regime as discussed in ref.~\cite{Grozdanov:2016vgg}. This point moves closer to the origin with $\gamma$ increasing. The dependence of the radius of convergence on $\gamma$ is thus 
a piecewise  continuous function\footnote{Curiously, the same type of a piecewise smooth dependence is observed when considering the radius of convergence as a function of chemical potential \cite{aiks,Jansen:2020hfd} and as a function of the coupling in the Sachdev-Ye-Kitaev chain \cite{Choi:2020tdj}.}, shown
 in fig.~\ref{fig:radius_coupling_dependence_n4_main_intro} (left panel), which is the main result of this section. The results of the perturbative and non-perturbative calculations for the closest to the origin critical points (the ones seen in perturbation theory) are in good agreement, as can be observed from Table \ref{tab:table1x}.

We emphasise that the non-perturbative effects discussed in this section are at best qualitative, since terms of 
order $\gamma^2$ and higher will inevitably modify them. We believe these effects are qualitatively correct as they fit well with various physical expectations \cite{Waeber:2015oka,Grozdanov:2016vgg,Solana:2018pbk,Grozdanov:2018gfx}. In particular, the existence of the non-perturbative critical point makes the radius of convergence decrease with the `t Hooft coupling decreasing from its infinite value.  Admittedly, an alternative conservative point of view would simply limit any considerations by the range of $\gamma$ up to $\gamma \lesssim 2 \cdot 10^{-5}$, beyond which the perturbative treatment becomes unreliable.

\subsection{Sound channel}
For the sound channel, the analysis follows the strategy used in the previous section and 
in refs.~\cite{Grozdanov:2019kge,Grozdanov:2019uhi} very closely. The relevant equation of motion is eq.~\eqref{fluct-eq-main-x-n4} with $i=3$.

\subsubsection{Perturbative calculation}
Solving the equations  perturbatively to linear order  $\gamma$, we find the 
following correction to the location of the closest to the origin critical point
\begin{eqnarray}
&\,& \qfr_{\rm c}^2 \approx \pm 2i +\gamma(166.844  \pm 3201.39 i )\,, \label{crit-point-soundn4} \\
&\,& \wfr_{\rm c} \approx\pm 1 - i + \gamma(\pm 2948.55  +1459.36 i)\,.
\label{crit-point-soundn4a}
\end{eqnarray}
Eq.~\eqref{crit-point-soundn4}  gives the radius of convergence $R_{\rm sound}=|\qfr_c^2|$ stated in eq.~\eqref{RN4Sound}. The next critical point is given by
\begin{eqnarray}
\label{crit-point-soundn4-1}
&\,& \qfr_{\rm c,1}^2 \approx -0.01681 \pm 3.12967i +\gamma(9108.90  \pm 36862.6 i)\,,\\
&\,& \wfr_{\rm c,1}\approx\pm 1.90135  - 2.04492i + \gamma(\pm 24615.9   +12589.5  i)\,.
\end{eqnarray}
As in the case of the shear mode, the coefficients of the perturbative expansion are quite large: the correction becomes comparable to the $\gamma=0$ result already  for $\gamma \sim 10^{-4} - 10^{-5}$.

\subsubsection{Non-perturbative calculation}
\begin{figure}[t!]
\centering
\includegraphics[width=0.45\textwidth]{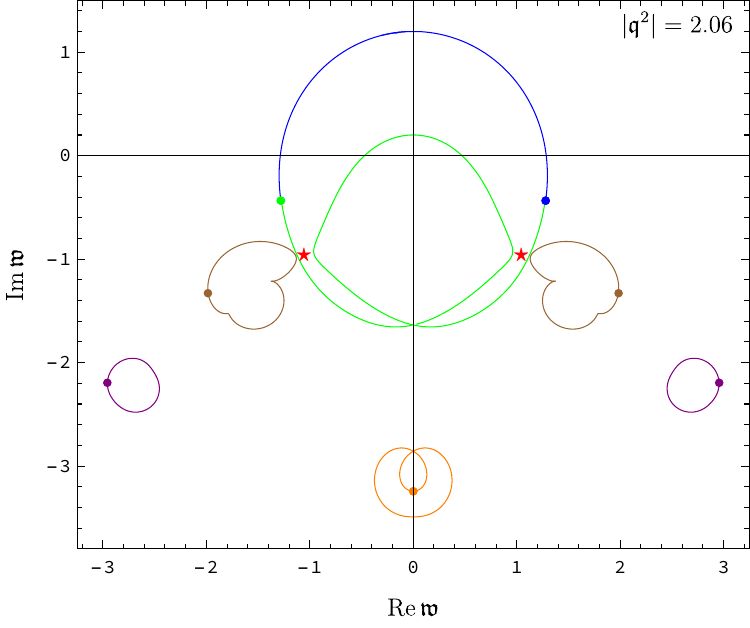}
\hspace{0.05\textwidth}
\includegraphics[width=0.45\textwidth]{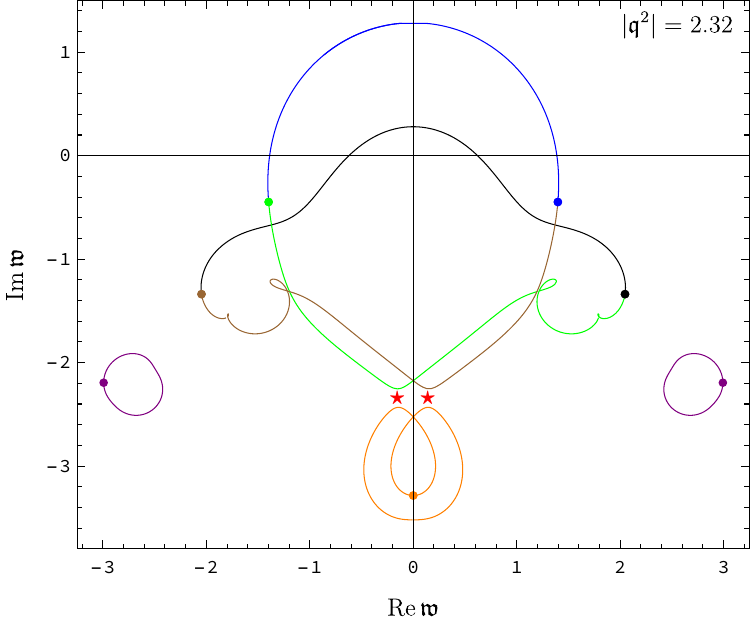}
\hspace{0.05\textwidth}
\caption{{\small Quasinormal spectrum in the sound channel at $\gamma = 2\cdot 10^{-5}$, computed non-perturbatively in $\gamma$. The positions of  critical points are shown by red stars. 
}}
\label{sound-qnm-spectrum-example-1}
\end{figure}

\begin{figure}[t!]
\centering
\includegraphics[width=0.45\textwidth]{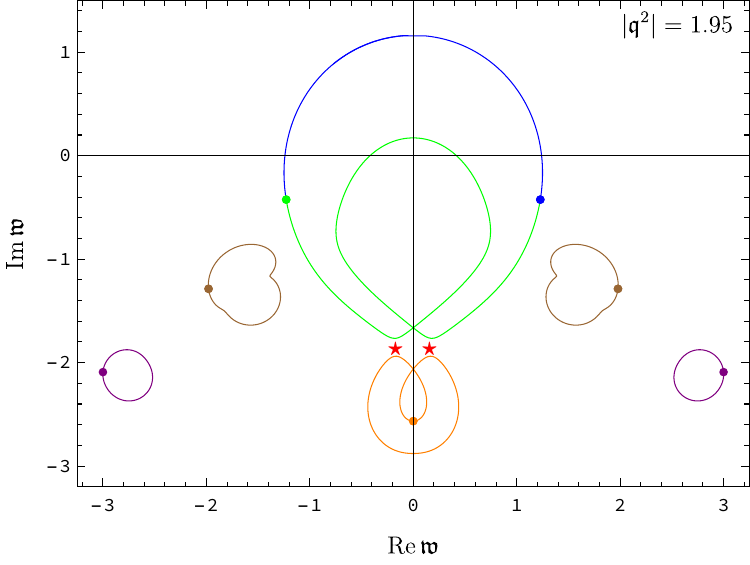}
\hspace{0.05\textwidth}
\includegraphics[width=0.45\textwidth]{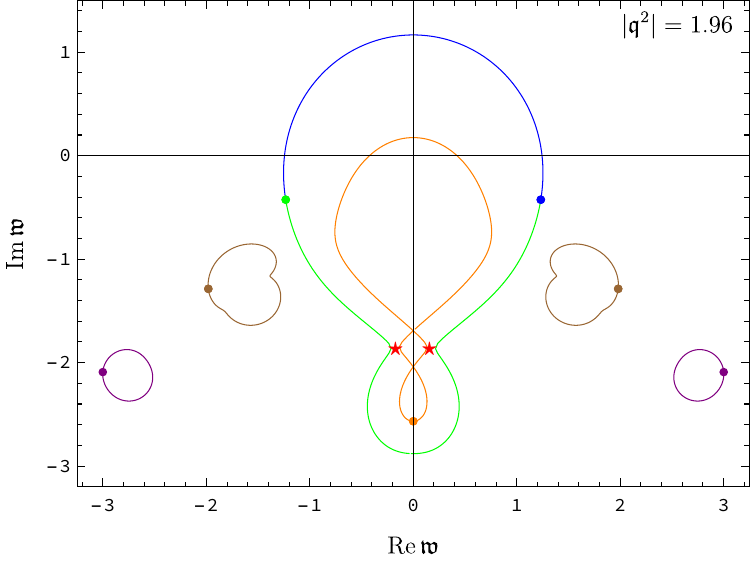}\\
\includegraphics[width=0.45\textwidth]{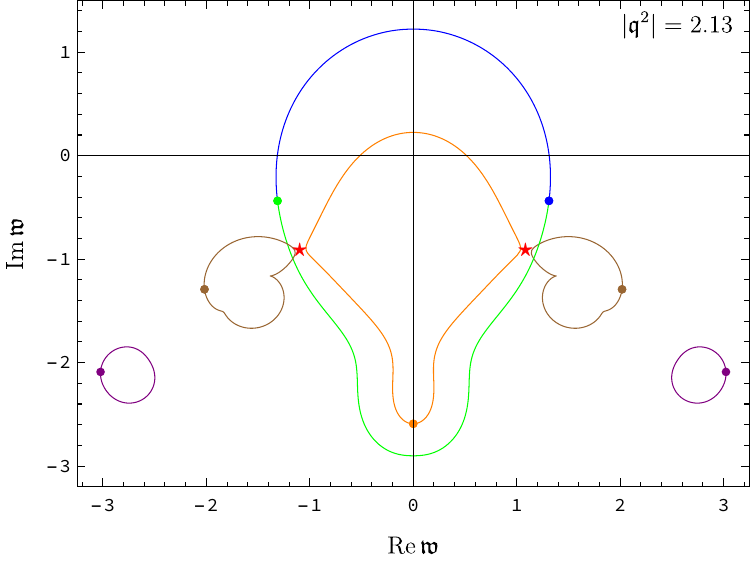}
\hspace{0.05\textwidth}
\caption{{\small Quasinormal spectrum in the sound channel at $\gamma = 4.5\cdot 10^{-5}$, computed non-perturbatively in $\gamma$. }}
\label{sound-qnm-spectrum-example-2}
\end{figure}

Non-perturbative treatment implies solving the equation of motion \eqref{fluct-eq-main-x-n4} without 
assuming $\gamma$ to be small. As in the shear mode case, there are essentially two qualitatively different scenarios of the distribution of critical points. We illustrate them by showing  trajectories of quasinormal modes in the complex $\wfr$-plane at complex $\qfr^2=|\qfr^2| e^{i\varphi}$, $\varphi\in [0,2\pi]$,  for $\gamma = 2\cdot 10^{-5}$ and $\gamma = 4.5 \cdot 10^{-5}$ in figs.~\ref{sound-qnm-spectrum-example-1} and \ref{sound-qnm-spectrum-example-2}, respectively.

In fig.~\ref{sound-qnm-spectrum-example-1}, left panel, the critical point limiting the radius of convergence of the sound mode's dispersion relation arises from the level-crossing between that mode and the top gapped modes (this situation is qualitatively the same as at $\gamma=0$ \cite{Grozdanov:2019kge,Grozdanov:2019uhi}). The pair of critical points closest to the origin in the complex $\qfr^2$-plane is well approximated by the perturbative expression \eqref{crit-point-soundn4-1} for $\gamma \lesssim 3\cdot 10^{-5}$ (fig.~\ref{fig:radius_coupling_dependence_n4_main_intro}, right panel). The radius of convergence, $R_{\rm sound} = |\qfr_{\rm c}^2|$, increases with $\gamma$ (see the blue curve in fig.~\ref{fig:radius_coupling_dependence_n4_main_intro}, right panel).     

For $\gamma > \gamma_* \approx 3.225 \cdot 10^{-5}$, the situation changes qualitatively, as illustrated in fig.~\ref{sound-qnm-spectrum-example-2}. Now the sound mode first collides with the non-perturbative mode (in fig.~\ref{sound-qnm-spectrum-example-2}, i.e. at $\gamma = 4.5 \cdot 10^{-5}$, this happens  at $\qfr^2_c = 0.9568 \pm 1.7083i$, $\wfr_c \approx  \mp 0.16513 -1.8681i$, corresponding to $R_{\rm sound} = |\qfr_{\rm c}^2|\approx 1.958$). In this regime, the radius of convergence, $R_{\rm sound} = |\qfr_{\rm c}^2|$, decreases with $\gamma$ (see the red 
curve in fig.~\ref{fig:radius_coupling_dependence_n4_main_intro}, right panel). This dependence is very similar, although not identical, to the one observed in  fig.~\ref{fig:radius_coupling_dependence_n4_main_intro} (left panel) for the shear channel. 

As before, in a conservative approach we would limit ourselves to the blue part of the curve  in the right panel of fig.~\ref{fig:radius_coupling_dependence_n4_main_intro}, which ends in the region where perturbation theory becomes unreliable. We believe, however, that the red part of the curve, although not  quantitatively precise, reflects the dependence of the radius of convergence on coupling qualitatively correctly. The piecewise smooth dependence on the coupling, shown in fig.~\ref{fig:radius_coupling_dependence_n4_main_intro} (right panel), is likely to persist even with $\gamma^2$ and higher terms in the action taken into account quantitatively correctly, since it corresponds to the discrete change of  ``status'' of the closest to the origin critical point, even though the critical points move continuously in the complex plane with varying coupling. 


\section{Convergence of hydrodynamic series in the Einstein-Gauss-Bonnet theory}
\label{sec:GB}
We now consider  the radii of convergence of  gapless quasinormal modes in five-dimensional Einstein-Gauss-Bonnet gravity. Although this theory may not have a healthy QFT dual \cite{Camanho:2014apa}, it is nevertheless a very useful theoretical laboratory for relevant bulk calculations in higher-derivative gravity: by design, its equations of motion  are second-order in derivatives and can thus be solved fully non-perturbatively in terms of the higher-derivative coupling. 

 The Einstein-Gauss-Bonnet action in $5d$ is
\begin{align}
\label{eq:GBaction}
S_{\scriptscriptstyle GB} = \frac{1}{2\kappa_5^2} \int d^5 x \sqrt{-g} \left[ R  + \frac{12}{L^2} + \frac{l^2_{\scriptscriptstyle GB}}{2} \left( R^2 - 4 R_{\mu\nu} R^{\mu\nu} + R_{\mu\nu\rho\sigma} R^{\mu\nu\rho\sigma} \right) \right]\,,
\end{align}
where the scale $l^2_{\scriptscriptstyle GB}$ of the higher-derivative term can be chosen to be set by a cosmological constant, $l^2_{\scriptscriptstyle GB} = \lgb L^2$, where $\lgb$ is a dimensionless parameter. 
 
The black brane metric solution of Einstein-Gauss-Bonnet equations of motion can be found analytically and is given by\footnote{Exact solutions and thermodynamics of black branes and black holes in Einstein-Gauss-Bonnet gravity were considered in \cite{Cai:2001dz} (see also \cite{Nojiri:2001aj,Cho:2002hq,Neupane:2002bf,Neupane:2003vz}).}
\begin{align}
ds^2 = - f(r) N^2_{\scriptscriptstyle GB} dt^2 + \frac{1}{f(r)} dr^2 + \frac{r^2}{L^2} \left(dx^2 + dy^2 +dz^2 \right)\,,
\label{eq:BB}
\end{align}
where
\begin{align}
f(r) = \frac{r^2}{L^2} \frac{1}{2\lgb} \left[1 - \sqrt{1-4\lgb \left(1 - \frac{r^4_0}{r^4} \right) } \right]
\label{eq:BBf}
\end{align}
and the constant $N_{\scriptscriptstyle GB}$ can be chosen to normalise the speed of light at the boundary to $c=1$:
\begin{align}
N_{\scriptscriptstyle GB}^2 = \frac{1}{2} \left(1+\sqrt{1-4\lgb} \right)\,.
\label{eq:NGBDef}
\end{align}
The position of the horizon is at $r = r_0$. The Hawking temperature corresponding to the solution \eqref{eq:BB} is given by
\begin{align}
T =  \frac{N_{\scriptscriptstyle GB} r_0}{\pi L^2} = \frac{r_0\sqrt{ 1+\gammagb}}{\sqrt{2} \pi L^2 }\,,
\label{eq:GBTemperature}
\end{align}
where we introduced the notation $\gammagb \equiv \sqrt{1-4\lgb}$. We shall use $\lgb$ and $\gammagb$ interchangeably in the following. The range $\lgb < 0 $ corresponds to $\gammagb \in (1,\infty)$ and the interval $\lgb \in (0,1/4]$ maps onto $\gammagb \in [0,1)$, with $\lgb =0$ corresponding to  $\gammagb=1$. 

To compute the quasinormal mode spectrum, we again consider linearised metric perturbations. In analogy with our discussion in section \ref{sec:N4}, we arrive at three gauge-invariant equations of motion for the scalar, shear and sound channels ($i=1,2,3$, respectively):
\begin{equation}
\partial^2_u Z_i + {\cal A}_{(i)}(u, \wfr, \qfr^2, \lgb) \partial_u Z_i +  {\cal B}_{(i)}(u, \wfr, \qfr^2, \lgb) Z_i =0\, ,
\label{fluct-eq-main-n}
\end{equation}
where the coefficients ${\cal A}_{(i)}$ and ${\cal B}_{(i)}$ are given in Appendix \ref{sec:appendix-GB}. All relevant details regarding the theory and the derivation of these equations can be found in ref.~\cite{Grozdanov:2016fkt}.

\subsection{Shear channel} 

The shear channel spectrum is determined by eq.~\eqref{fluct-eq-main-n} with $i=2$. As in the case of the $\CN = 4$ theory, we compare calculations done perturbatively and non-perturbatively  in the Gauss-Bonnet coupling $\lgb$.
\begin{figure*}[t]
\centering
\includegraphics[width=0.30\textwidth]{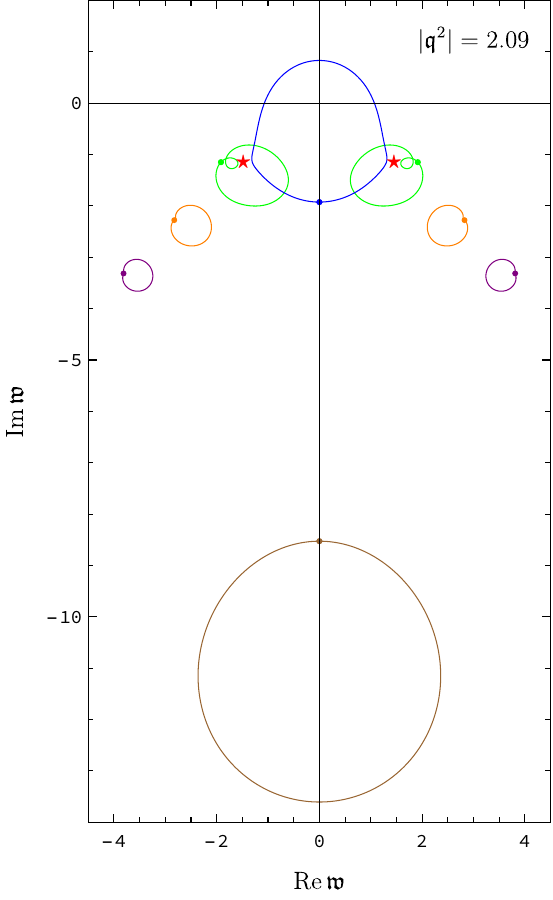}
\hspace{0.01\textwidth}
\includegraphics[width=0.30\textwidth]{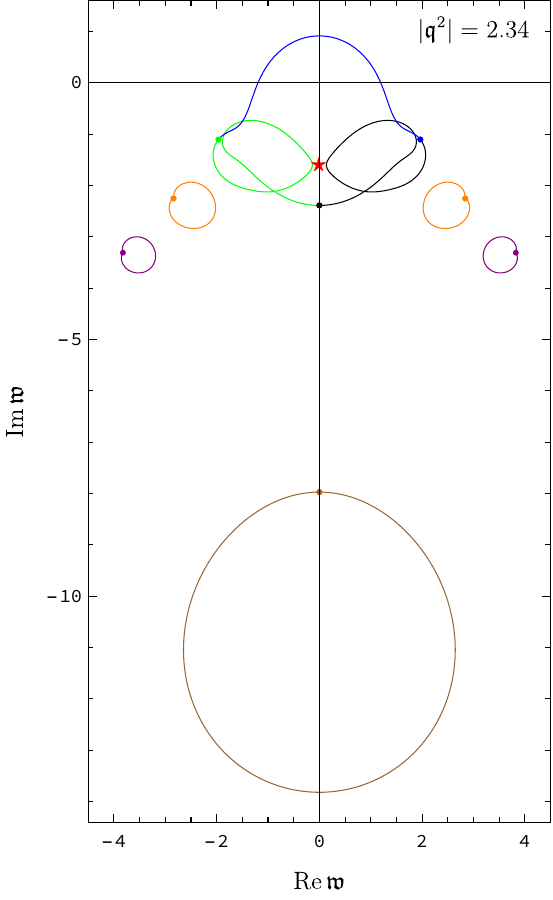}
\hspace{0.01\textwidth}
\includegraphics[width=0.30\textwidth]{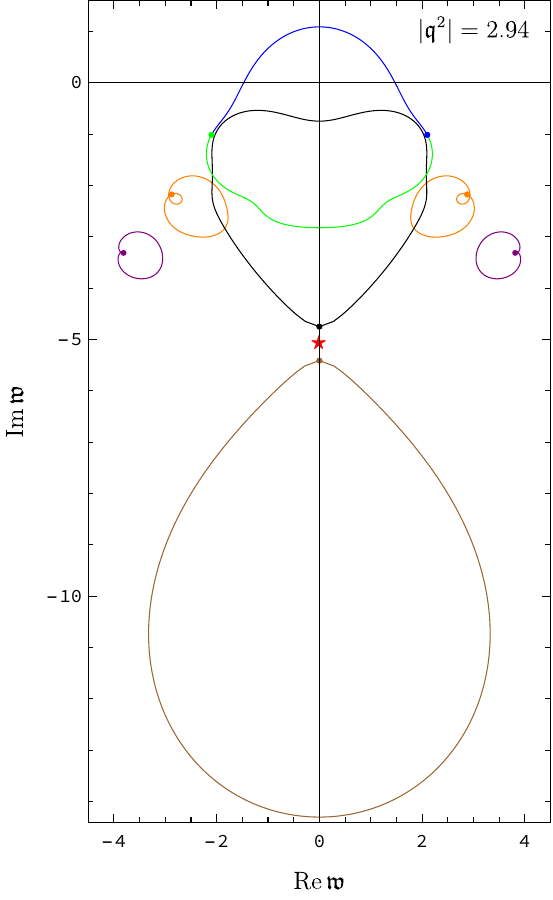}
\caption{
{\small  Quasinormal spectrum in the shear channel of Gauss-Bonnet theory, computed non-perturbatively in $\lgb$, at $\lgb=-0.01$. The trajectories are plotted for complex values of the spatial momentum squared, $\qfr^2= |\qfr^2|e^{i \varphi}$, where phase $\varphi$ is varied from 0 to $2\pi$.  The positions of quasinormal modes at $\varphi=0$ are shown by dots. The positions of the  critical points  are shown by red stars. 
}}
\label{GB-shear-1}
\end{figure*}
\begin{figure*}[t]
\centering
\includegraphics[width=0.45\textwidth]{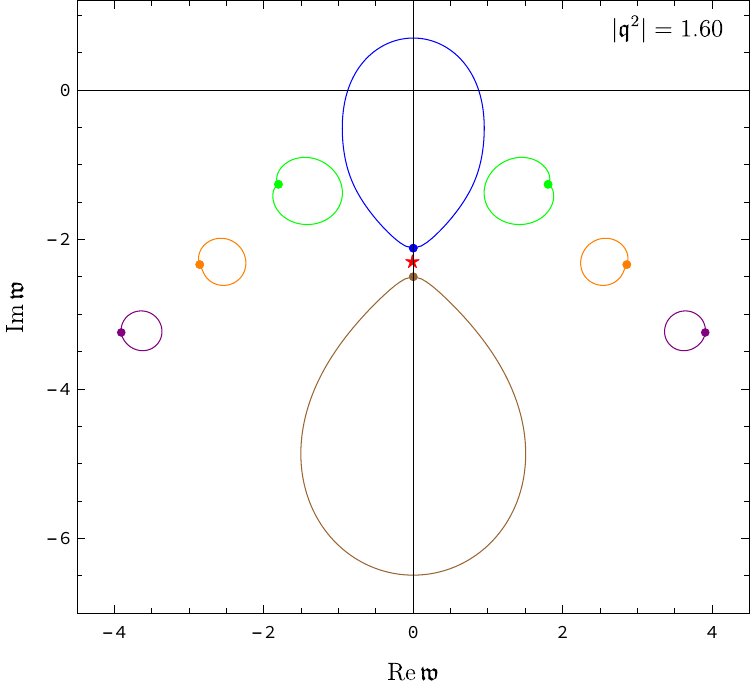}
\hspace{0.05\textwidth}
\includegraphics[width=0.45\textwidth]{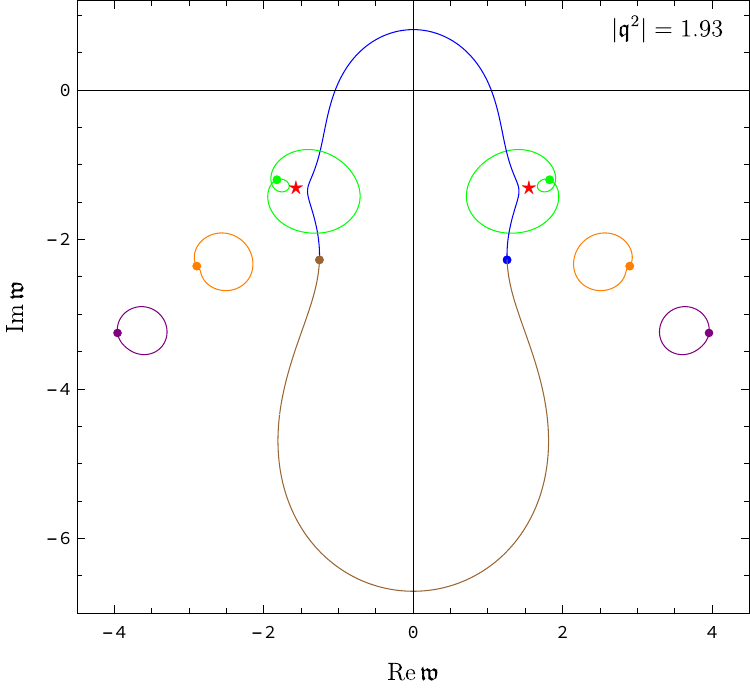}
\\
\includegraphics[width=0.45\textwidth]{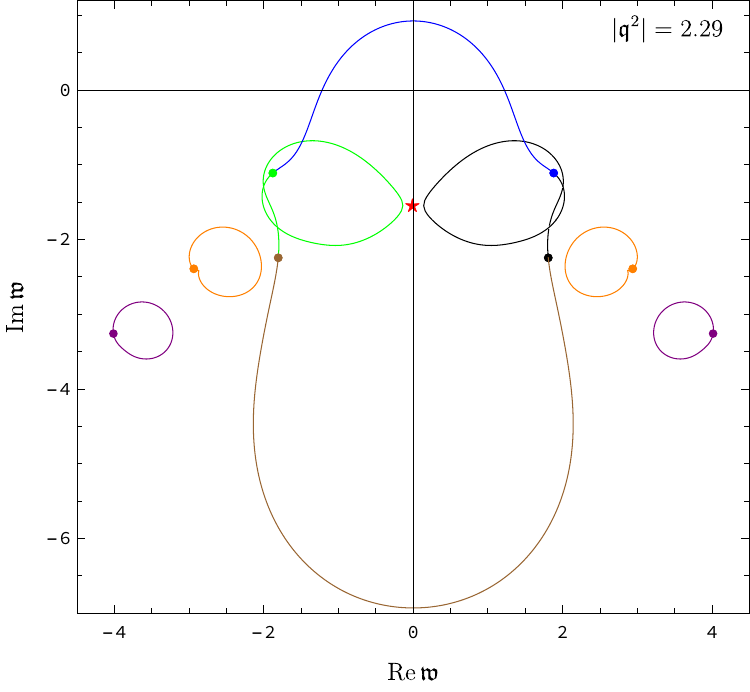}
\caption{
{\small Quasinormal spectrum in the shear channel of Gauss-Bonnet theory, computed non-perturbatively in $\lgb$, at $\lgb=-0.03$. 
}}
\label{GB-shear-2}
\end{figure*}

\subsubsection{Perturbative calculation}
Solving  eqs.~\eqref{corr-1} (with the eq.~\eqref{fluct-eq-main-n}, $i=2$, as the underlying equation of motion)  perturbatively in $\lgb$, we find the critical point closest to the origin in the 
complex $\qfr^2$-plane:
\begin{eqnarray} 
\label{crit-point}
&\,& \qfr_{\rm c}^2  \approx 1.89065  \pm 1.17115 i + \lgb (-2.01742 \pm 22.5317 i)  + \mathcal{O}(\lgb^2)\,,\\
&\,& \wfr_{\rm c}  \approx\pm 1.44364 - 1.06923 i + \lgb (\mp 1.69340  + 8.39996 i)  +  \mathcal{O}(\lgb^2)\,.
\end{eqnarray}
The  radius of convergence for  the shear hydrodynamic mode is then given by $R_{\rm shear}=|\qfr_c^2|$:
\begin{equation}\label{RshearGB}
R_{\rm shear}(\lgb) \approx  2.22 + 22.6 \, \lgb + \CO(\lgb^2)\,.
\end{equation}
The coefficient in front of $\lgb$ is a positive number, which is similar to the   $\CN = 4$ case. However, the ``physical'' regime of the hypothetical field theory dual to Gauss-Bonnet gravity corresponds to  $\lgb < 0$ \cite{Grozdanov:2016vgg,Grozdanov:2016fkt,Grozdanov:2016zjj,Andrade:2016rln,DiNunno:2017obv}. In this case, the radius of convergence decreases with $|\lgb|$ increasing, the trend opposite to the one found for the $\CN =4 $ SYM theory in section \ref{N=4-shear}.
\begin{figure*}[th]
\centering
\includegraphics[width=0.55\textwidth]{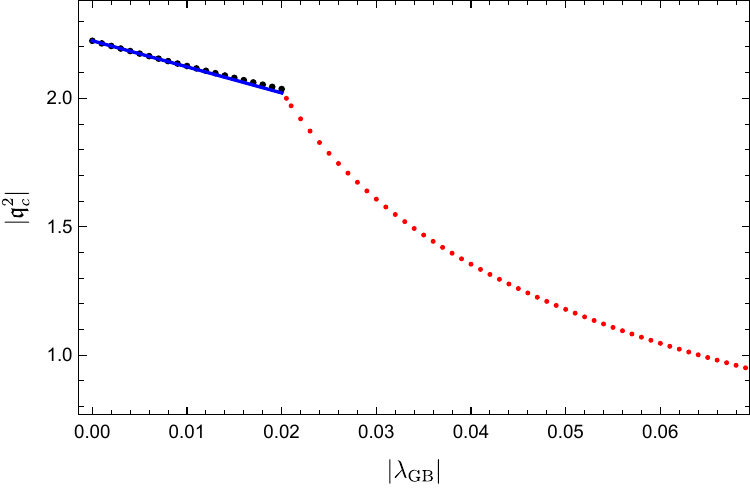}
\caption{{\small Radius of convergence of the hydrodynamic shear mode in the Einstein-Gauss-Bonnet theory as a function of 
the higher-derivative coupling $\lgb$ ($\lgb <0$). The blue line is the perturbative result \eqref{RshearGB}. Black dots on top of the blue
 line denote the radius of convergence computed non-perturbatively in $\lgb$ (see  fig.~\ref{GB-shear-1}, top left panel). The red dots correspond to the critical point involving the shear mode and the non-perturbative mode (see  fig.~\ref{GB-shear-2}, top left panel). The transition between the two regimes occurs at $\lgb \approx -0.0198$.}}
\label{fig:qc-lambda-shear}
\end{figure*}

We also compute perturbative $\lgb$ corrections to the next three higher critical points, finding
\begin{eqnarray}
\label{crit-point1}
&\,& \qfr_{\rm c,1}^2 \approx -2.37737  -2.59245  \, \lgb + \mathcal{O}(\lgb^2)\,,\\
&\,& \wfr_{\rm c,1} \approx -1.64659 -3.44719  \, \lgb +  \mathcal{O}(\lgb^2)\,,
\end{eqnarray}
for the first point,   
\begin{eqnarray}
\label{crit-point2}
&\,& \qfr_{\rm c,2}^2\approx -3.11051\mp 0.81050 i +\lgb (-1.70074\pm 5.98722 i)  + \mathcal{O}(\lgb^2)\,,\\
&\,& \wfr_{\rm c,2}\approx \pm 1.41043-2.87086  i + \lgb (\mp 0.31643-4.26447  i) +  \mathcal{O}(\lgb^2)\,,
\end{eqnarray}
for the second, and 
\begin{eqnarray}
\label{crit-point3}
&\,& \qfr_{\rm c,3}^2 \approx 2.90684 \pm 1.66612 i+\lgb(-2.16892 \pm 68.0434 i) + \mathcal{O}(\lgb^2)\,,\\
&\,& \wfr_{\rm c,3} \approx \pm 2.38819 -2.13154 i+\lgb(\mp 3.28617 +19.0509 i) +  \mathcal{O}(\lgb^2)\,,
\end{eqnarray}
for the third.

\subsubsection{Non-perturbative calculation}
Eqs.~\eqref{corr-1}  and \eqref{fluct-eq-main-n} can be solved fully non-perturbatively in $\lgb$. Several examples of the shear channel Gauss-Bonnet quasinormal spectrum are shown in fig.~\ref{GB-shear-1} and fig.~\ref{GB-shear-2}. Their characteristic feature, explored in 
refs.~\cite{Grozdanov:2016fkt,Grozdanov:2016vgg}, is the presence of non-perturbative modes located (for real $\qfr^2$) on the imaginary axis in the complex $\wfr$-plane. At sufficiently small values of $|\lgb|$, these modes lead to new critical points only for large values 
of $|\qfr^2|$: the closest to the origin critical points setting the radius of convergence of the shear mode are not affected by them 
(see  fig.~\ref{GB-shear-1}, where the spectrum is shown for $\lgb = -0.01$ and various values of complex $\qfr^2$). At larger values of  $|\lgb|$, however, the situation changes qualitatively. Now the closest to the origin critical point and thus the radius of convergence are set by the non-perturbative mode (this is illustrated by  fig.~\ref{GB-shear-2}, where the spectrum is shown for $\lgb = -0.03$). The transition between the two regimes occurs at $\lgb \approx -0.0198$.  The dependence of the radius of convergence of the shear mode's dispersion relation on Gauss-Bonnet coupling is shown in  fig.~\ref{fig:qc-lambda-shear}.

\subsection{Sound channel}
Finally, we repeat the same analysis for the sound channel of the Einstein-Gauss-Bonnet theory using  eqs.~\eqref{corr-1} and the equation of motion \eqref{fluct-eq-main-n} with $i=3$.

\subsubsection{Perturbative calculation}
To linear order in the  perturbative expansion in $\lgb$, we  find the closest to the origin pair of critical points at
\begin{align}
\qfr_{\rm c}^2 &\approx  \pm 2i +\lgb(-10.8809   \pm 10.4314 i ) + \mathcal{O}(\lgb^2)\,,\\
\wfr_{\rm c} &\approx \pm 1 - i + \lgb (\mp 2.05394  + 3.49495i) +  \mathcal{O}(\lgb^2)\,.
\end{align}
Hence, the radius of convergence in the sound channel is given (perturbatively) by
\begin{equation}\label{RsoundGB}
R_{\rm sound} (\lgb) \approx 2  +   15. 073 5 \, \lgb + \CO(\lgb^2)\,. 
\end{equation}
For ``physical'' values of $\lgb$ ($\lgb<0$), this is a decreasing function of $|\lgb|$ (which is different from the $\CN = 4$ SYM theory).

\subsubsection{Non-perturbative calculation}
\begin{figure*}[t]
\centering
\includegraphics[width=0.45\textwidth]{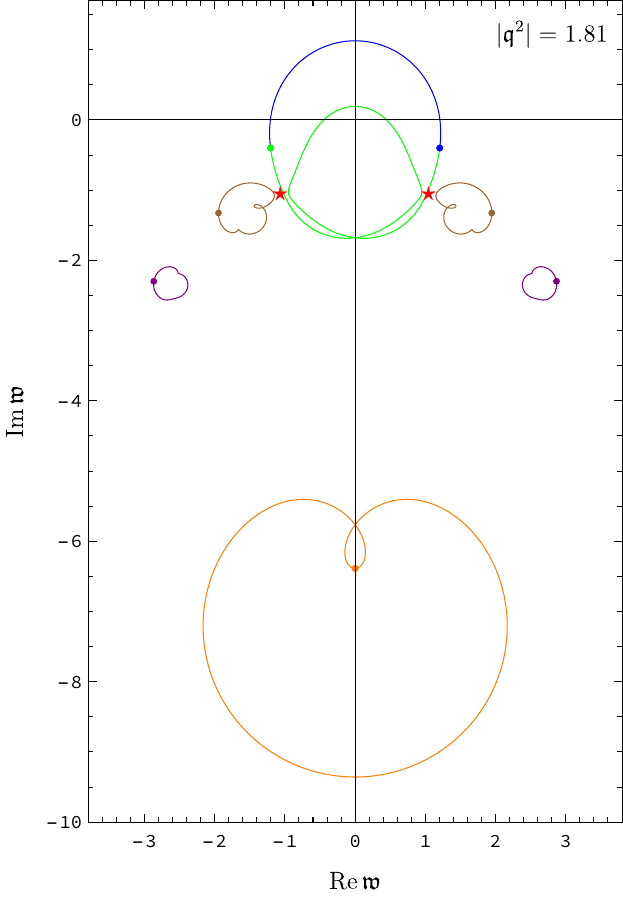}
\hspace{0.01\textwidth}
\includegraphics[width=0.45\textwidth]{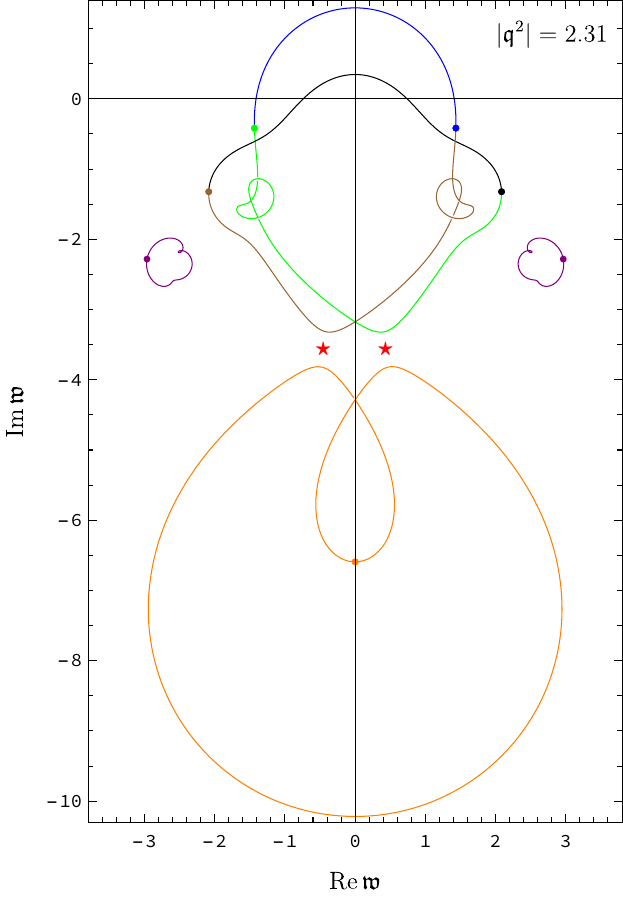}
\caption{
{\small  Quasinormal spectrum in the sound channel of Gauss-Bonnet theory, computed non-perturbatively in $\lgb$, at $\lgb=-0.02$. The trajectories are plotted for complex values of the spatial momentum squared, $\qfr^2= |\qfr^2|e^{i \varphi}$, where phase $\varphi$ is varied from 0 to $2\pi$.  The positions of quasinormal modes at $\varphi=0$ are shown by dots. The positions of the  critical points  are shown by red stars. 
}}
\label{GB-sound-1}
\end{figure*}

Computing critical points in the sound channel of the Einstein-Gauss-Bonnet theory non-perturbatively in $\lgb$, we find that for 
sufficiently small $|\lgb|$, the situation remains qualitatively the same as in the $\lgb=0$ case: the radius of convergence is determined by the level-crossing between sound modes and the top gapped modes in the complex $\wfr$-plane, as shown in fig.~\ref{GB-sound-1} (left panel). The non-perturbative mode also leads to critical points, but this happens at larger value of $|\qfr^2|$ (see  the right panel of 
fig.~\ref{GB-sound-1}). 

\begin{figure*}[t]
\centering
\includegraphics[width=0.45\textwidth]{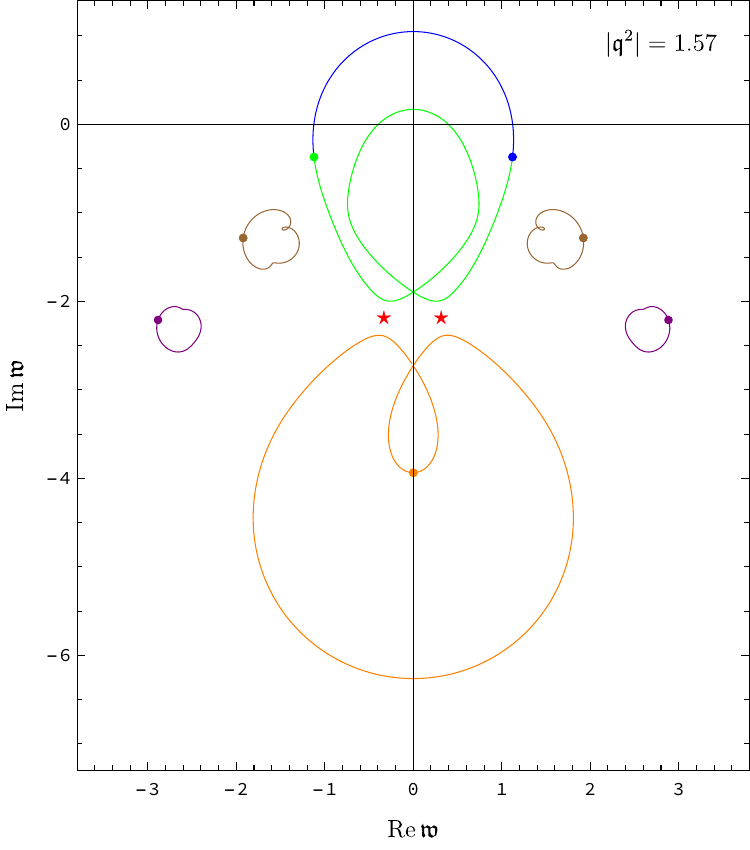}
\hspace{0.01\textwidth}
\includegraphics[width=0.45\textwidth]{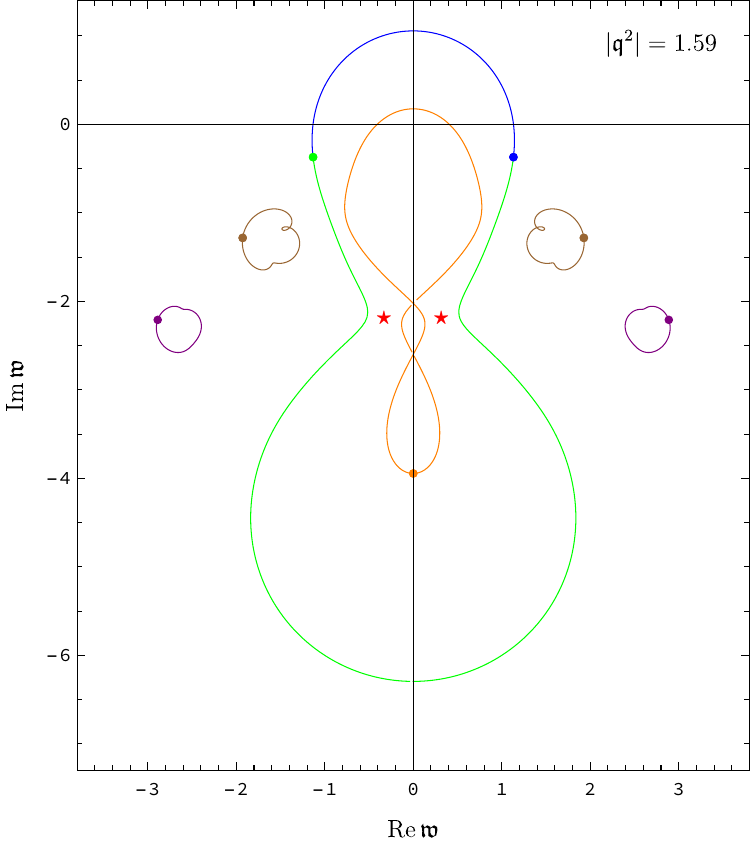}
\\
\includegraphics[width=0.45\textwidth]{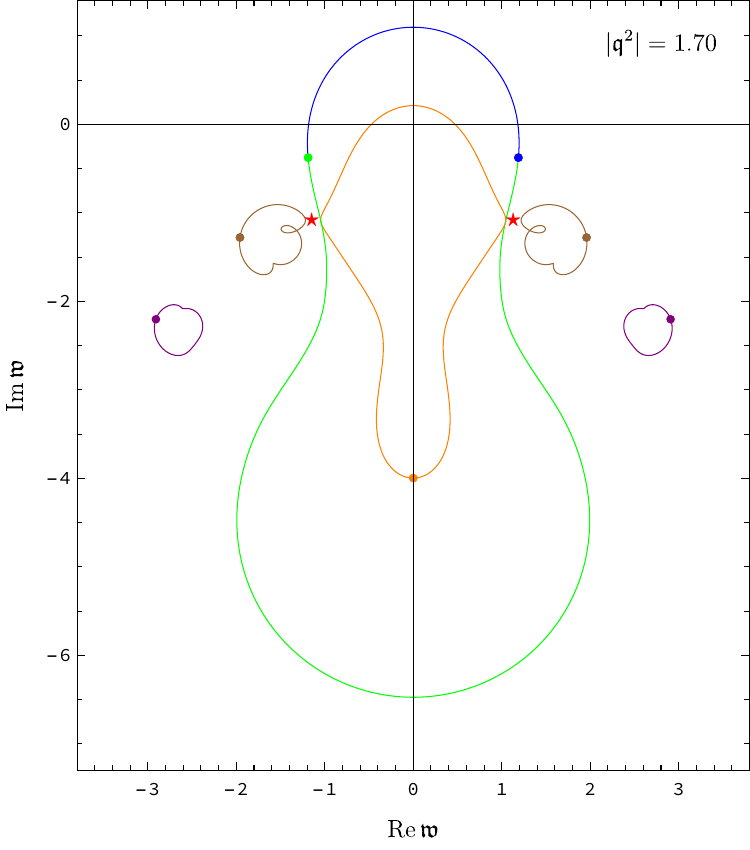}
\caption{
{\small  Quasinormal spectrum in the sound channel of Gauss-Bonnet theory, computed non-perturbatively in $\lgb$, at $\lgb=-0.04$. The trajectories are plotted for complex values of the spatial momentum squared, $\qfr^2= |\qfr^2|e^{i \varphi}$, where phase $\varphi$ is varied from 0 to $2\pi$.  The positions of  the quasinormal modes at $\varphi=0$ are shown by dots. The positions of the  critical points  are shown by red stars. 
}}
\label{GB-sound-2}
\end{figure*}

At larger value of the coupling $|\lgb|$, the picture changes qualitatively and the radius of convergence of the sound dispersion relation is now determined by the level-crossing with the non-perturbative quasinormal mode, as shown in the left panel of fig.~\ref{GB-sound-2}. 

\begin{figure*}[th]
\centering
\includegraphics[width=0.55\textwidth]{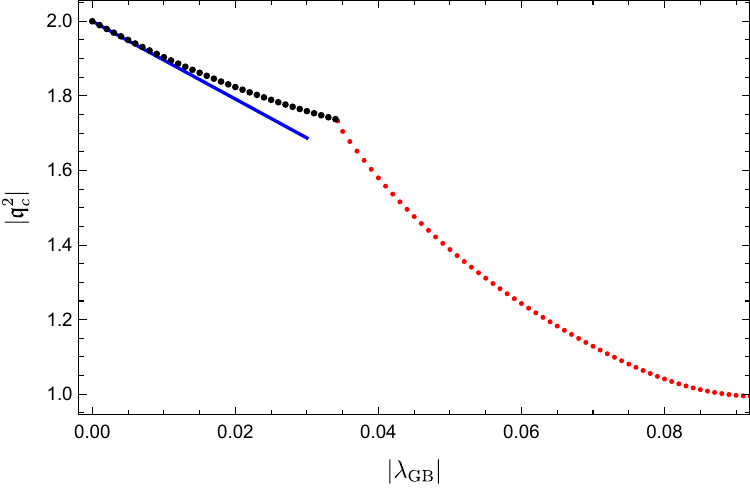}
\caption{ {\small Radius of convergence of the hydrodynamic sound mode in the Einstein-Gauss-Bonnet theory as a function of 
the magnitude of the higher-derivative coupling $\lgb$ ($\lgb <0$). The blue line is the linear approximation \eqref{RsoundGB}. Black dots are the non-perturbative result for the radius of convergence arising as shown in fig.~\ref{GB-sound-1}, left panel. Red dots are the radius of convergence arising as a result of the level-crossing between the sound mode and the non-perturbative mode (see fig.~\ref{GB-sound-2}, top left panel). The transition between the two regimes occurs at $ \lgb \approx -0.0338$.}}
\label{fig:radius_coupling_dependence_sound_GB}
\end{figure*}

The transition between the two regimes happens at $ \lgb \approx -0.0338$. The dependence of $R_{\rm sound}$ (including both perturbative and non-perturbative results) is shown in fig.~\ref{fig:radius_coupling_dependence_sound_GB}. 

\section{Non-perturbative quasinormal modes and singular perturbation theory}
\label{sec:validity}
In our analysis of the radii of convergence in the $\CN = 4$ SYM theory, as well as in previous works on higher-derivative holography \cite{Waeber:2015oka}, \cite{Grozdanov:2016vgg,Solana:2018pbk,Grozdanov:2018fic,Grozdanov:2018gfx}, the non-perturbative ``resummation'' and the appearance of the non-perturbative quasinormal modes played a rather prominent role. In particular, the existence of these non-perturbative features seems to be fully consistent with the requirement of a physically reasonable interpolation between  strongly coupled (holographic) regime and  weakly coupled (e.g. kinetic) regime in the same theory.  This  interpolation, even for simplest theories such as CFTs considered at finite temperature, is not fully understood \cite{Blaizot:2006tk,Grozdanov:2016vgg,Moore:2018mma,Grozdanov:2018gfx,Solana:2018pbk,Romatschke:2019gck,Romatschke:2019qbx,Du:2020odw}. Admittedly, relying --- even only qualitatively --- on the non-perturbative treatment in models arising as truncations of a perturbative expansion may seem to be unwarranted \cite{Buchel:2018eax}. However, before dismissing such an approach as ineffable nonsense, one may wish to consider examples where it is known to be successful. 
 
In this section, we first outline the problem as we see it, and then discuss in detail a simple example of an algebraic equation containing a small parameter, where similar issues arise.  
 
Consider the full set of solutions to an equation (algebraic or differential), which we schematically write as
\begin{equation}\label{Eq-0}
\CL [x, \epsilon] = 0\,.
\end{equation} 
We denote this (possibly infinite) set of solutions by $X = \{x_1 (\epsilon), x_2 (\epsilon), \ldots\}$. Here $\epsilon$ is a small parameter such that $\CL [x, \epsilon] $ can be formally expanded in a series,
 \begin{equation}\label{Eq-eps}
\CL [x, \epsilon] = \CL^{(0)} [x] + \epsilon \CL^{(1)} [x] + \epsilon^2 \CL^{(2)}[x] + \ldots = 0\,.
\end{equation} 
Truncating the series \eqref{Eq-0} at order $\epsilon^0$, $\epsilon^1$, $\epsilon^2$, $\dots$, and solving the corresponding equations, we obtain sets of solutions $X^{(0)} = \{x_1^{(0)}, x_2^{(0)}  , \ldots\}$, $X^{(1)} = \{x_1^{(1)}  (\epsilon), x_2^{(1)}  (\epsilon), \ldots\}$,  $X^{(2)} = \{x_1^{(2)}  (\epsilon), x_2^{(2)}  (\epsilon), \ldots\}$ and so on. A natural question to ask  is in what sense the solutions $X^{(0)}$, $X^{(1)}$, $X^{(2)},\ldots$ approximate the true solution $X$. Note that even the number of roots in each ``truncated'' set $X^{(n)}$ depends on $n$. For example, the equation
\begin{equation}\label{Eq-eps-ex-1}
 \CL [x]  = \CL^{(0)} [x]  = 1- x  = 0\,
\end{equation} 
has a single solution $X^{(0)} = \{ x_1^{(0)}=1\}$, whereas extending $\CL [x]$ by adding a term $\epsilon x^2$,
\begin{equation}\label{Eq-eps-ex-2}
 \CL [x, \epsilon] = \CL^{(0)} [x]  + \epsilon \CL^{(1)} [x] = 1- x  +\epsilon x^2 = 0\,,
\end{equation} 
leads to two solutions, $X^{(1)} = \{x_1^{(1)}  (\epsilon), x_2^{(1)}  (\epsilon)\}$, one of which is perturbative and another one is non-perturbative in $\epsilon$:
\begin{eqnarray}
&\,& x_1^{(1)}  (\epsilon) = \frac{1}{2 \epsilon} \left( 1 - \sqrt{1-4 \epsilon}\right) = 1 + \epsilon + O(\epsilon^2)\,, \label{psol}\\
&\,& x_2^{(1)}  (\epsilon) = \frac{1}{2 \epsilon} \left( 1 + \sqrt{1-4 \epsilon}\right) = \frac{1}{\epsilon} - 1 - \epsilon + O(\epsilon^2)\,.\label{npsol}
\end{eqnarray}
The solution $x_1^{(1)}  (\epsilon)$ can be constructed order by order in $\epsilon$ via standard perturbation theory (i.e., assuming a {\it perturbative ansatz} for the solution), whereas the $x_2^{(1)}  (\epsilon)$ is ``invisible'' in the standard perturbative approach yet it can be built consistently using singular perturbation theory  \cite{hinch-book}. Now imagine that eq.~\eqref{Eq-eps-ex-2} is a truncation of the equation 
  \begin{equation}\label{ex-2-2-x}
   \CL^{(N)} [x, \epsilon]  = x-1 +\sum\limits_{n=1}^N \epsilon^n x^{n+1} = 1-x +\epsilon x^2 + O(\epsilon^2) =0\,.
 \end{equation}
 At any finite $N$, among the $N+1$ roots of the equation \eqref{ex-2-2-x}, one is perturbative in $\epsilon$ (it is a regular perturbation 
 of the solution $x=1$ of the equation  \eqref{Eq-eps-ex-1}) and the remaining $N$ roots are non-perturbative: they disappear to infinity in the limit $\epsilon \to 0$. These extra 
 $N$ roots are located (approximately) along the circle $|x|=1/\epsilon$ in the complex $x$-plane. Finally, we note that 
 eq.~\eqref{ex-2-2-x} can be regarded as a truncation at order $\epsilon^N$ of an exact function 
  \begin{equation}\label{ex-2-4-x}
   \CL^{(\infty)} [x, \epsilon] = x-1 + \frac{\epsilon x^2}{1 - \epsilon x} =0\,.
 \end{equation}
 Note that eq.~\eqref{ex-2-4-x}  has only {\it one} solution,
 \begin{equation}\label{ex-2-5}
 x = \frac{1}{1+\epsilon}\,,
 \end{equation}
 whose small $\epsilon$ expansion coincides (for $|\epsilon|<1$)  with the perturbative solution
  of the equation \eqref{ex-2-2-x} at the appropriate order. Thus, in this example, truncating the  small $\epsilon$ expansion  of eq.~\eqref{ex-2-4-x} 
  at order $N$ produces one correct and  $N$ spurious roots located approximately at the boundary of analyticity of the function $ \CL^{(\infty)} [x, \epsilon]$,  i.e. at $|x|=1/\epsilon$.
 
This simple example seems to reinforce the ``conservative'' approach to the quasinormal spectrum in higher-derivative gravity suggesting that only perturbative solutions can be trusted. Non-perturbative solutions exist (and can be constructed via singular perturbation theory) at each given order of the expansion in a small parameter, but these solutions appear to be artefacts of the expansion and disappear when the full function is considered. However, such a verdict might be too quick, as the example in the next subsection shows.

\subsection{An algebraic equation example}
\label{sec:alg}
\begin{figure*}[t]
\centering
\includegraphics[width=0.45\textwidth]{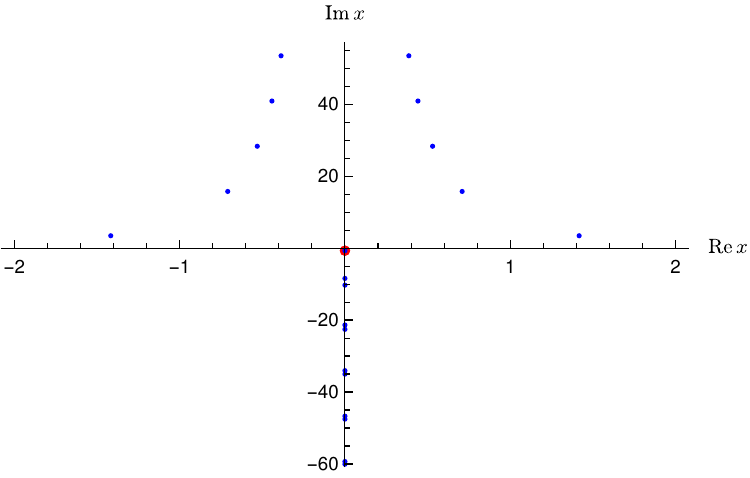}
\hspace{0.05\textwidth}
\includegraphics[width=0.45\textwidth]{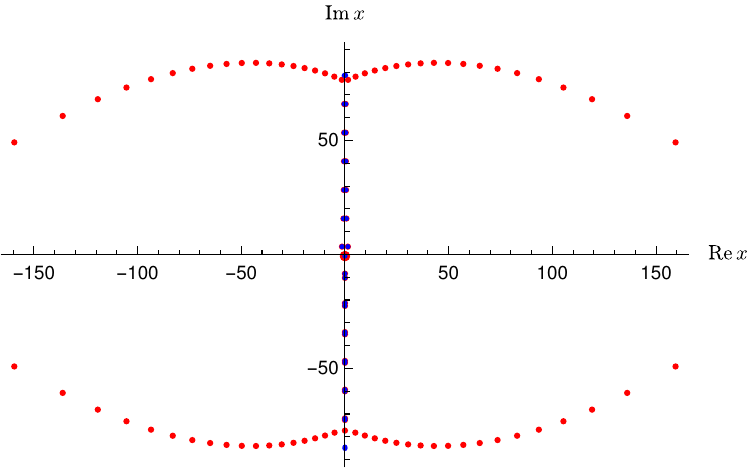}
\caption{ \small The roots of  eq.~\eqref{ex-3-1} (at $\epsilon=0.5$) closest to the origin in the complex $x$-plane (left panel). The large red dot corresponds to the perturbative root at $x\approx -0.735544 i$. (Note that the perturbative series \eqref{ex-3-2-p} fails to converge for this value of $\epsilon$.) The other roots on the imaginary axis are at  $x\approx -8.4414 i$, $x\approx -10.3128 i$, $x\approx -21.3770 i$, $x\approx -22.5885 i$, $x\approx -34.0718 i$, $x\approx -35.0365 i$, 
$x\approx - 46.7093 i$, $x\approx - 47.5349 i$. In the right panel, the same set of roots as in the left panel is approximated by the first $100$ roots of the polynomial  \eqref{ex-3-1-p}. The blue dots accurately reproduce the actual values from the left panel. The purpose of the zoomed-out plot is to show the spurious, unphysical red dots which move to infinity as the order of approximation is increased.} 
\label{fig:example-3}
\end{figure*}

Consider the algebraic equation
\begin{equation}\label{ex-3-1}
\CL[x,\epsilon] = i x - 1 - x \sinh{\epsilon x}  =0\,,
\end{equation}
where $\epsilon$ is a  parameter. At $\epsilon =0$, eq.~\eqref{ex-3-1} has a single root, $x=x_0=-i$. For  $\epsilon >0$, however, there are infinitely many solutions, parametrised by $\epsilon$ (see fig.~\ref{fig:example-3}, left panel, where  the roots of eq.~\eqref{ex-3-1}  closest to 
the origin in the complex $x$-plane are shown for $\epsilon=0.5$). Note that all these solutions but one are non-perturbative in $\epsilon$, since they must 
disappear from the set of solutions in the limit $\epsilon\to 0$ leaving the single root $x=x_0$ at  $\epsilon =0$.

The solutions to eq.~\eqref{ex-3-1} can be constructed as series in $\epsilon \ll 1$.  One such solution is the finite $\epsilon$ correction to the solution $x_0=-i$ of the equation at $\epsilon =0$. Using the standard perturbation theory, we find
\begin{equation}\label{ex-3-2-p}
 x_0 (\epsilon) = -i \left[ 1 -\epsilon + 2 \epsilon^2  - \frac{29}{6} \epsilon^3 + 13 \epsilon^4 - \frac{4481}{120} \epsilon^5 + \frac{5048}{45} \epsilon^6  + O(\epsilon^7) \right]\,.
 \end{equation}
The non-perturbative roots can be found analytically by using the methods of singular perturbation theory \cite{hinch-book}. Introducing a new variable $x = \bar x / \epsilon$ and taking the limit of $\epsilon \to 0$ in eq.~\eqref{ex-3-1} while keeping $\bar x$ fixed, we find the equation 
\begin{equation}
i - \sinh \bar x = 0\,.
\end{equation} 
The infinite set of solutions to this equation, $\bar x_n = i \pi (1 + 4 n) / 2$, where $n\in\mathbb{Z}$, then gives all  the non-perturbative roots of the original eq.~\eqref{ex-3-1} as 
\begin{equation}
x^\pm_n (\epsilon) = \frac{1}{\epsilon} \left[ \frac{ i \pi \left(1 + 4 n\right)}{2} \pm \frac{2}{ \sqrt{ \pi \left(1+4 n\right)   } }  \epsilon^{1/2}  + \frac{4 i}{\left[\pi \left(1+4n\right)\right]^2} \epsilon \pm \CO(\epsilon^{3/2}) \right] \,.
\label{ex-s}
\end{equation}
The series in eqs.~\eqref{ex-3-2-p} and \eqref{ex-s} converge\footnote{The radius of convergence is determined by the closest to the origin critical point of the function \eqref{ex-3-1}. \label{fn5}} for $\epsilon < |\epsilon_c| \approx 0.2625$. 

We now replace $\CL[x,\epsilon]$ by a finite polynomial (a truncated Taylor expansion of eq.~\eqref{ex-3-1}), 
\begin{equation}\label{ex-3-1-p}
\CL^{(2N+2)}[x,\epsilon]=   i x - 1 - \epsilon x^2 \sum\limits_{n=0}^N \frac{\epsilon^{2 n} x^{2n}}{(2n+1)!}   =0\,,
\end{equation}
and ask whether the $2N+2$ roots of  eq.~\eqref{ex-3-1-p} approximate the exact solutions of eq.~\eqref{ex-3-1} as we increase $N$. Our previous discussion implies that at least one such approximation, a regular perturbative correction to the zeroth-order root $x_0=-i$, should exist. Indeed, solving the corresponding equations $\CL^{(2)}[x,\epsilon]=0$, $\CL^{(4)}[x,\epsilon]=0$, etc.,  perturbatively in $\epsilon$, we 
 reproduce the series \eqref{ex-3-2-p} term by term. All others roots are non-perturbative in $\epsilon$, 
disappearing to complex infinity in the limit $\epsilon \to 0$.

To mimic our approach to the quasinormal spectra in higher-derivative gravity, we now solve  eq.~\eqref{ex-3-1-p} numerically at each order of $N$, without assuming $\epsilon$ to be small. A generic result is illustrated  in the right panel of fig.~\ref{fig:example-3}, where $\epsilon = 0.5$, and both the exact solutions to eq.~\eqref{ex-3-1} and the roots of the polynomial \eqref{ex-3-1-p} with $N=50$ are shown. It is clear that in a (large) region of the complex plane containing the origin, the solutions of eq.~\eqref{ex-3-1} (both perturbative and non-perturbative) are well approximated by some of the roots of  the polynomial \eqref{ex-3-1-p}. There are also ``spurious roots'', forming  the top and the bottom red ``arcs'' in fig.~\ref{fig:example-3}, right panel: they do not approximate any solution. The exact solutions of eq.~\eqref{ex-3-1} located outside of the domain bounded by the red arcs are not approximated by any of the roots of the polynomial \eqref{ex-3-1-p} and remain ``invisible''. The domain bounded by the red arcs of spurious roots increases with $N$ increasing. Our main conclusion is that it is possible in principle to approximate at least some of the exact non-perturbative solutions by the non-perturbative roots of the perturbative truncation of the exact equation. Of course, this example is not a justification of our ``resummation'' of quasinormal spectra but we hope it shows that such an approach has the right to exist.

\begin{figure*}[t]
\centering
\includegraphics[width=0.45\textwidth]{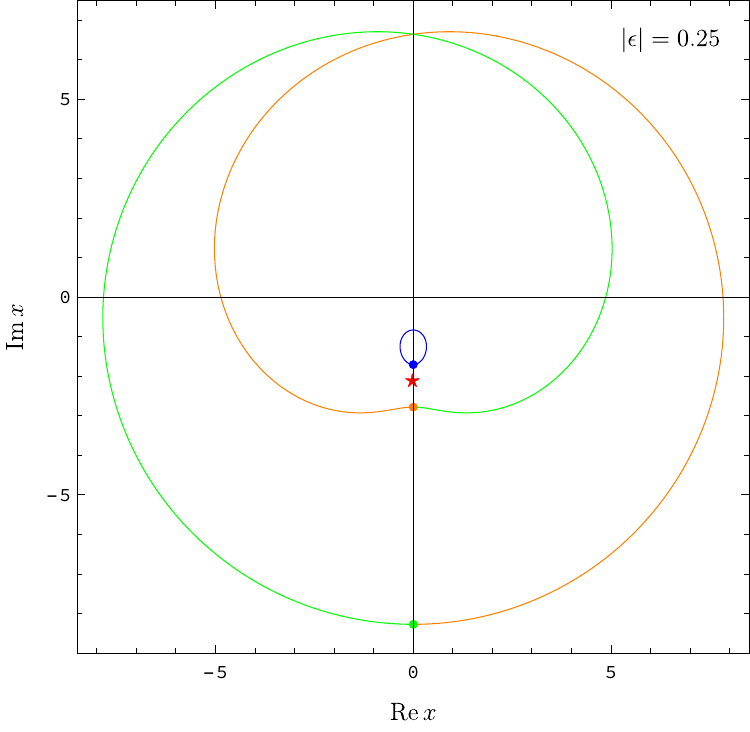}
\hspace{0.05\textwidth}
\includegraphics[width=0.45\textwidth]{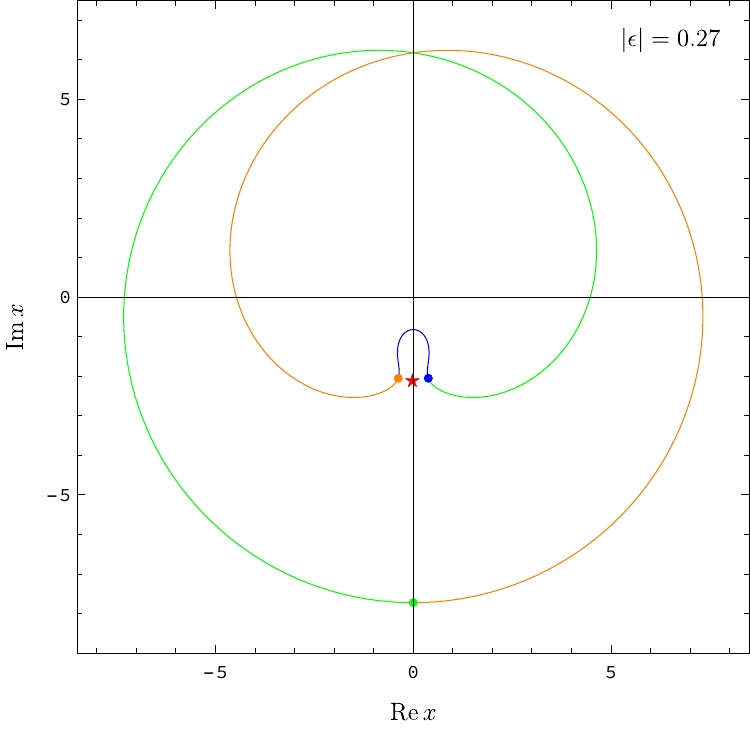}
\caption{ \small The two closest to the origin solutions $x_0$ and $x_1$  to eq.~\eqref{ex-3-1} in the complex $x$-plane as functions of 
the complexified  parameter $\epsilon= -|\epsilon|e^{i\varphi}$, where $\varphi \in [0,2\pi]$  for $|\epsilon| = 0.25$ (left panel) and $|\epsilon |=  0.27$ (right panel). Coloured dots correspond to $\varphi =0$. The critical point setting the radius of convergence of the 
series  \eqref{ex-3-2-p} is located at $x \approx -2.1176 i$ (it is shown by the red star).} 
\label{fig:example-4}
\end{figure*}

\begin{figure*}[t]
\centering
\includegraphics[width=0.43\textwidth]{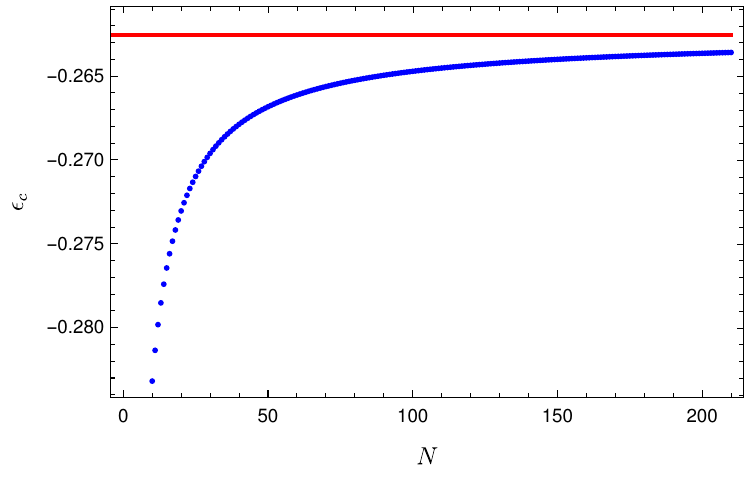}
\hspace{0.05\textwidth}
\includegraphics[width=0.465\textwidth]{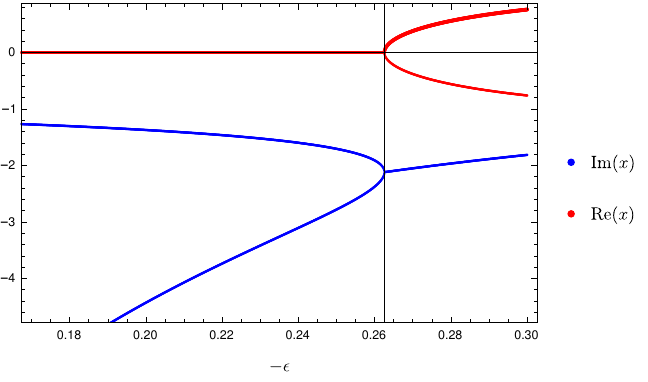}
\caption{{\small The critical points $\epsilon_c(N)$ computed perturbatively using the polynomial approximations \eqref{ex-3-1-p} of order $2N+2$ (blue line) converges to the ``exact'' critical point $\epsilon_{\rm c}\approx -0.2625$ (flat red line) of the full equation  \eqref{ex-3-1} (left panel).  
The dependence of  real and imaginary parts of the root $x$ on $\epsilon$ (right panel). The level-crossing occurs at $\epsilon_{\rm c} \approx -0.2625$. The same transition in the complex $x$-plane is shown in fig.~\ref{fig:example-4}. 
}
\label{fig:algeb-conv}}
\end{figure*}

We now ask a different question. As noted in footnote \ref{fn5}, the radius of convergence of the series \eqref{ex-3-2-p} representing the perturbative solution $x_0=x_0(\epsilon)$ of eq.~\eqref{ex-3-1} is set by the closest to the origin (in the complex $\epsilon$-plane) critical point of the complex curve \eqref{ex-3-1} determined by the conditions 
\begin{equation}
\label{aco}
\CL[x,\epsilon] = 0\,, \qquad \partial_x \CL[x,\epsilon] = 0\,.
\end{equation}
The solution of eqs.~\eqref{aco} with the smallest $|\epsilon|$ is  $x \approx -2.1176 i$, $\epsilon \approx - 0.2625$. In the complex $x$-plane, the critical point corresponds to the level-crossing between the perturbative mode $x_0$ and the nearest {\it non-perturbative} solution of eq.~\eqref{ex-3-1}, as shown in fig.~\ref{fig:example-4}. Now suppose that we only have access to the 
successive polynomial approximations \eqref{ex-3-1-p} to the full equation \eqref{ex-3-1}. Can we detect the existence of the correct critical point and hence the non-perturbative solution by using only these polynomial approximations and the regular perturbation theory? The answer is affirmative: solving perturbatively  eqs.~\eqref{aco} with  $\CL[x,\epsilon]$ replaced by the polynomial approximation \eqref{ex-3-1-p}  of order $2N+2$, we find that the (smallest in magnitude) critical value $\epsilon_c(N)$ converges to the ``exact'' result $\epsilon_{\rm c}\approx - 0.2625$ with $N$ increasing (see  fig.~\ref{fig:algeb-conv}, left panel).
Since the  critical point corresponds to the level-crossing (see  fig.~\ref{fig:algeb-conv}, right panel), i.e., a collision between perturbative and non-perturbative roots in the complex $x$-plane (see fig.~\ref{fig:example-4}), the above observation suggests  that  the perturbative calculation 
``knows'' about this collision, even though the non-perturbative root responsible for it is inaccessible in perturbation theory and remains ``invisible''. 
This raises a question of whether the non-perturbative quasinormal modes arising in higher-derivative gravity can be indirectly detected by a perturbative analysis of the relevant critical points. We investigate this question  in the next section.

\subsection{A signature of non-perturbative modes from the perturbative analysis of the Einstein-Gauss-Bonnet critical points}
\label{sign-egb}
\begin{figure*}[t]
\centering
\includegraphics[width=0.32\textwidth]{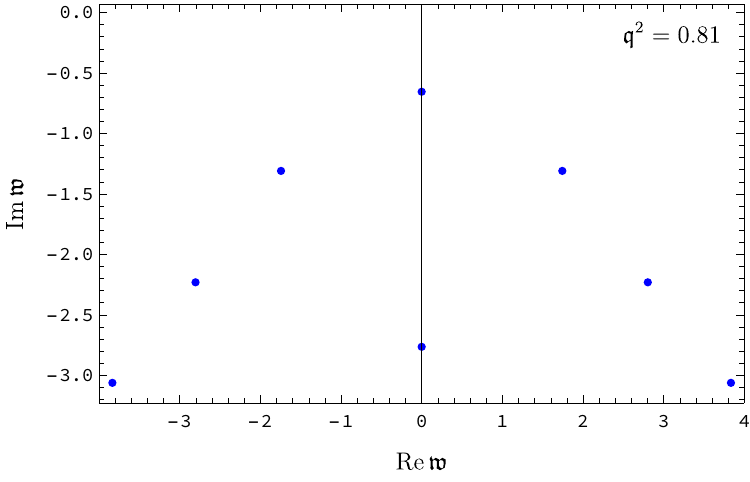}
\includegraphics[width=0.32\textwidth]{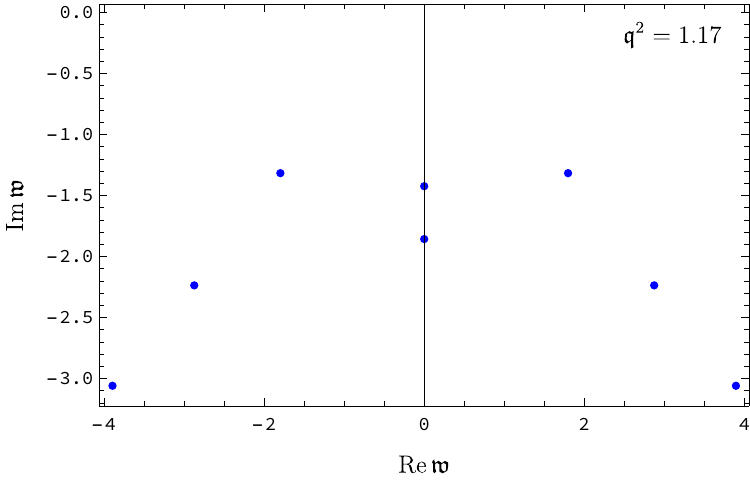}
\\
\includegraphics[width=0.32\textwidth]{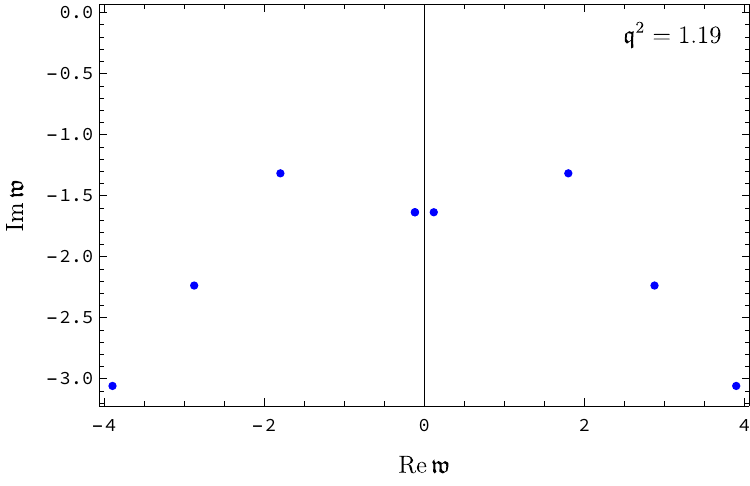}
\includegraphics[width=0.32\textwidth]{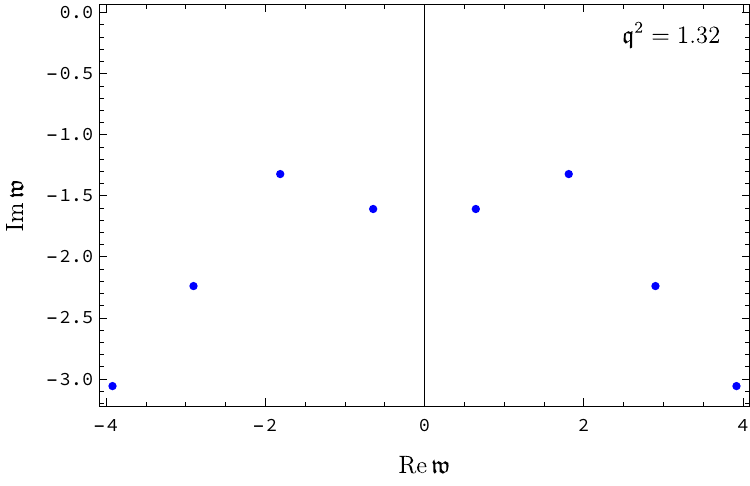}
\includegraphics[width=0.32\textwidth]{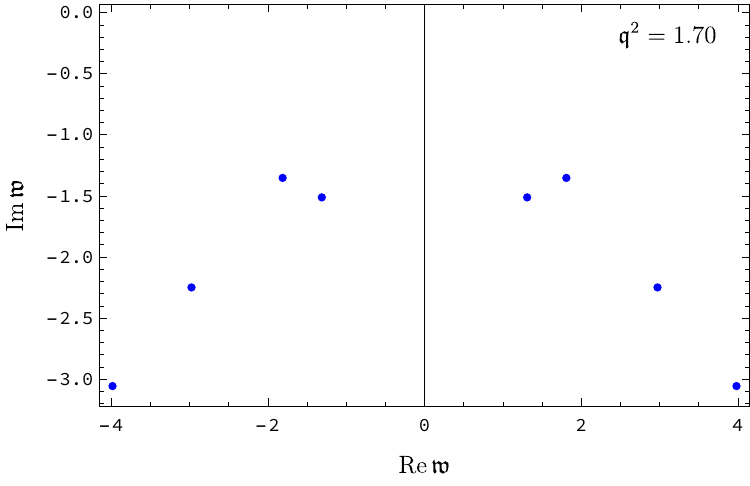}
\\
\includegraphics[width=0.32\textwidth]{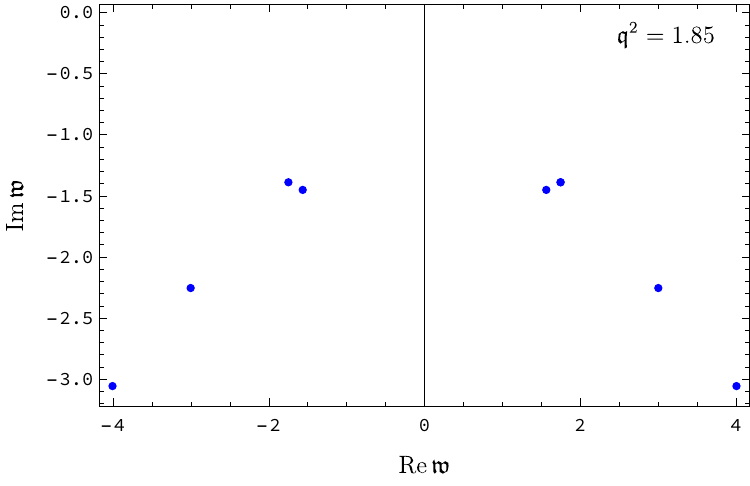}
\includegraphics[width=0.32\textwidth]{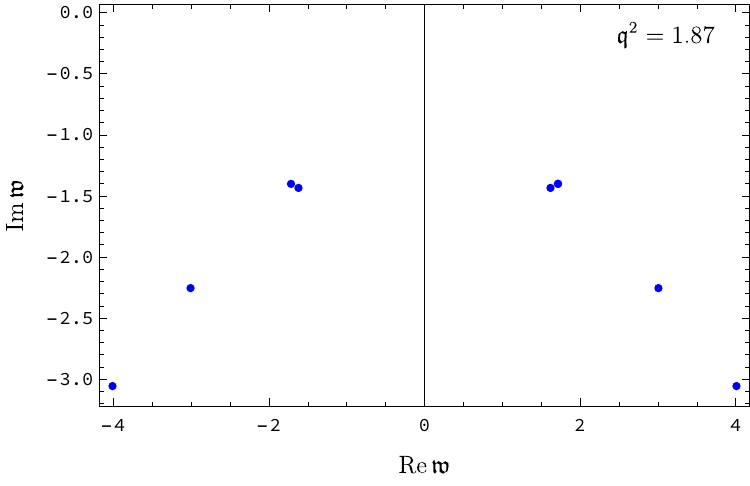}
\includegraphics[width=0.32\textwidth]{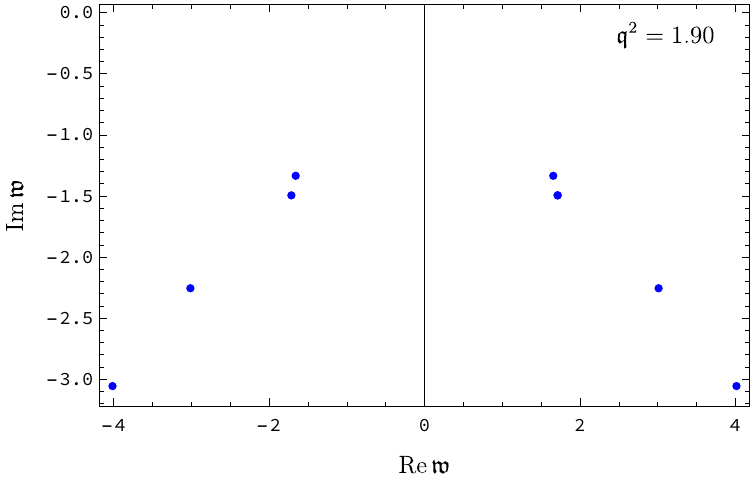}
\caption{
\label{fig:shar-ch-modesssss}
{\small The non-perturbative quasinormal spectrum in the shear channel of the Einstein-Gauss-Bonnet theory  at $\lgb \approx -0.04956$, 
plotted for several  (increasing) values of $\qfr^2 \in \mathbb{R}$. At $\qfr^2=0.81$, the shear mode and the non-perturbative mode are present on the imaginary axis; with $\qfr^2$ increasing, the two modes approach each other on the imaginary axis, collide at the critical point \eqref{crit-point-spec-nonpert1x}, move off the axis as the pair of propagating modes (shown in plots with $\qfr^2=1.19; 1.32; 1.70; 1.85; 1.87$) and then collide with the pair of the ``Christmas tree'' gapped modes at the second critical point \eqref{crit-point-spec-nonpertx}.
}}
\end{figure*}

The quasinormal spectrum in the shear channel of Einstein-Gauss-Bonnet gravity contains modes located on the imaginary axis which are non-perturbative in the Gauss-Bonnet coupling $\lgb$ (see fig.~\ref{GB-shear-1} and  refs.~\cite{Grozdanov:2016vgg,Grozdanov:2016fkt}). As discussed in section \ref{sec:GB}, for some range of parameters, the top (closest to the origin) non-perturbative mode collides with the hydrodynamic shear mode at real $\qfr_c^2$ and purely imaginary $\wfr_c$, as illustrated in fig.~\ref{GB-shear-2}. This collision is followed by another collision between the emergent pair of propagating modes  with the pair of gapped modes from the standard ``Christmas tree'' sequence.  For a fixed value of $\lgb$, the second collision again occurs at real  $\qfr^2_c$. We illustrate this in fig.~\ref{fig:shar-ch-modesssss}, where the shear channel Einstein-Gauss-Bonnet spectrum is shown  at $\lgb \approx -0.04956$ for several values of  real $\qfr^2$: as $\qfr^2$ is increased from $\qfr^2 =0.81$ to $\qfr^2=1.17$, the hydrodynamic mode and the non-perturbative mode on the imaginary axis approach each other, colliding at $\qfr_{\rm c}^2\approx 1.18544$ and forming a pair of propagating modes, which move off the axis and at $\qfr_{\rm c}^2\approx 1.87785$ collide again, now with the pair of the ``Christmas tree'' gapped modes. The two (sets of) critical points resulting from these collisions are  given by
\begin{align}
&\lgb \approx -0.04956\,:  \label{onj}\\
&\qfr_{\rm c}^2\approx 1.18544\,, \quad \wfr_{\rm c}\approx -1.63793i\,, \label{crit-point-spec-nonpert1x}\\
&\qfr_{\rm c}^2\approx 1.87785\,,\quad \wfr_{\rm c}\approx\pm 1.67253 -1.41811i \, .\label{crit-point-spec-nonpertx} 
\end{align}
We emphasise that the existence of the critical points \eqref{crit-point-spec-nonpert1x} and \eqref{crit-point-spec-nonpertx} is the direct consequence of the existence of the non-perturbative mode on the imaginary axis, absent at  $\lgb = 0$. Indeed, for $\lgb = 0$ (which corresponds to  the $\CN =4$ SYM theory at infinite `t Hooft 
coupling), the hydrodynamic shear mode  travels unobstructedly along the imaginary $\wfr$ axis towards negative infinity as the real $\qfr^2$ is increased: no critical point exists in the theory for purely imaginary $\wfr_{\rm c}$ at purely real and positive $\qfr_{\rm c}^2$.  In contrast, at finite $\lgb$, the shear mode collides with the non-perturbative mode on the imaginary axis at real  and positive $\qfr_{\rm c}^2$ given by eq.~\eqref{crit-point-spec-nonpert1x}, and then the resulting two propagating modes give rise to the {\it pair of critical points} at another real value of $\qfr^2_c$ (eq.~\eqref{crit-point-spec-nonpertx}). The pair of the propagating modes leading to the pair of critical points in eq.~\eqref{crit-point-spec-nonpertx} does not exist in the perturbative spectrum: it is created by the collision of the shear mode and the non-perturbative mode on the imaginary axis.

Can the existence of  non-perturbative critical points such as the ones in eq.~\eqref{crit-point-spec-nonpertx} be inferred from the perturbative data 
(i.e., from eqs.~\eqref{CriticalPointsX}, solved perturbatively in $\lgb$),  in analogy 
with what has been observed in section~\ref{sec:alg}? The goal is to find a perturbative approximation to the pair of critical points \eqref{crit-point-spec-nonpertx} with real $\qfr^2$.

\begin{figure*}[t]
\centering
\includegraphics[width=0.31\textwidth]{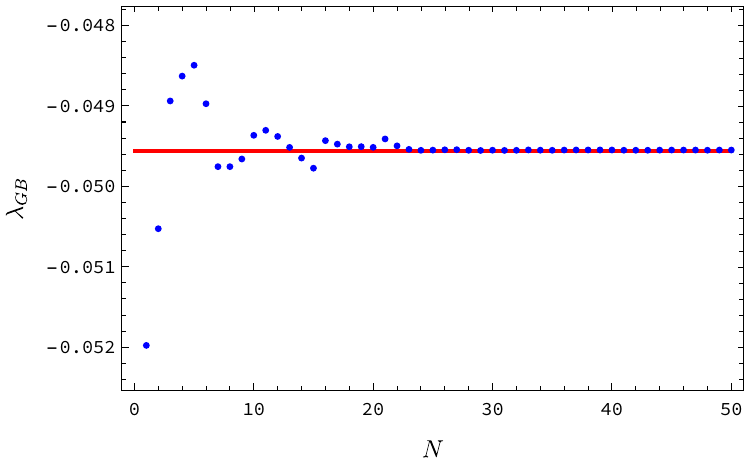}
\hspace{0.01\textwidth}
\includegraphics[width=0.31\textwidth]{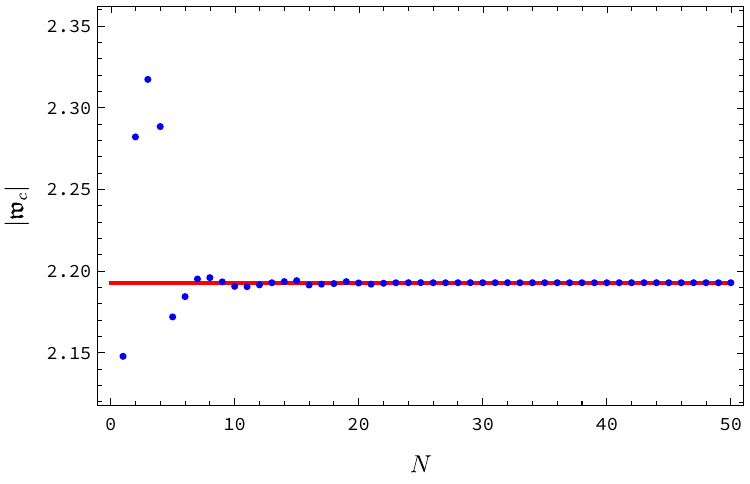}
\hspace{0.01\textwidth}
\includegraphics[width=0.31\textwidth]{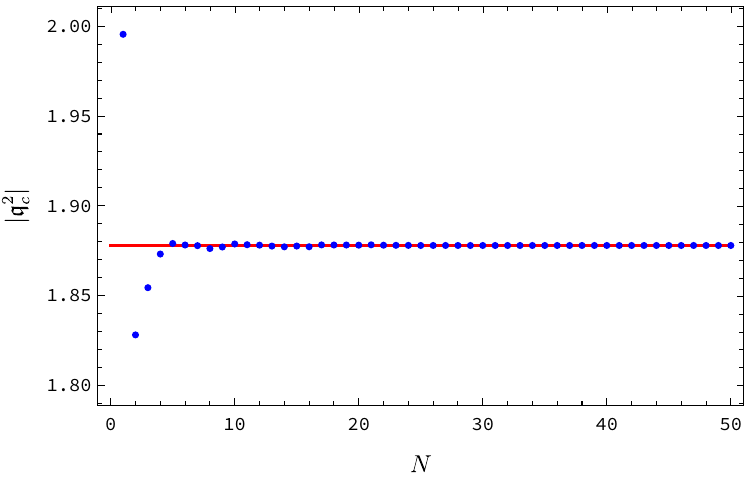}
\caption{
\label{fig:shar-ch-modes11www}
{\small Parameters of the critical point $\lgb (N)$, $|\wfr_{\rm c}(N)|$, $\qfr_{\rm c}^2(N)$ obtained from the perturbative analysis of eqs.~\eqref{CriticalPointsX} supplemented by a Pad\' e resummation (blue dots). The orders of the Pad\' e approximant $[a/b]$ were chosen as follows: 
$[N/N]$ for $1\leq N < 5$; $[N-3/2]$ for $5\leq N < 15$; $[N-11/10]$ for $15\leq N \leq 50$. The red lines are the non-perturbative results given 
by eqs.~\eqref{onj},  \eqref{crit-point-spec-nonpert1x} and \eqref{crit-point-spec-nonpertx}. }}
\end{figure*}

Perturbatively, to leading  order in $\lgb$, the closest to the origin  critical point is given by 
eq.~\eqref{crit-point}, reproduced here for convenience:
\begin{eqnarray} 
&\,& \qfr_{\rm c}^2 \approx 1.89065 \pm 1.17115 i + \lgb (-2.01742 \pm 22.5317i)  + \mathcal{O}(\lgb^2)\,, \label{crit-point-recall-1} \\
&\,& \wfr_{\rm c} \approx\pm 1.44364 - 1.06923 i + \lgb (\mp 1.69340  + 8.39996 i)  +  \mathcal{O}(\lgb^2)\,.
\label{crit-point-recall-2}
\end{eqnarray}
Eq.~\eqref{crit-point-recall-1} implies that the critical value of $\qfr^2$ is purely real for $\lgb\approx-0.0519780$, and is given by
\begin{align}
\label{crit-point-spec-x}
&\lgb\approx-0.0519780\, : \\
&\qfr_{\rm c}^2\approx 1.9955 + \mathcal{O}(\lgb^2)\,, \qquad \wfr_{\rm c}\approx \pm 1.53166 -1.50584 i + \mathcal{O}(\lgb^2)\,.
\label{crit-point-spec-xx}
\end{align}
We conjecture that the critical point \eqref{crit-point-spec-xx}  at $\lgb\approx-0.0519780$ is the perturbative approximation to the ``exact'' critical point \eqref{crit-point-spec-nonpertx} at $\lgb \approx -0.04956$. To see that this is indeed the case, we extend the perturbative expansion in 
eqs.~\eqref{crit-point-recall-1}, \eqref{crit-point-recall-2}  to order $\CO(\lgb^N)$. At each $N$, we then numerically compute the value of 
$\lgb$ (denoted by $\lgb(N)$) from the algebraic condition that sets $\im \, \qfr^2_{\rm c} (\lgb) = 0$ (among multiple roots $\lgb(N)$, we choose the solution  which is real and closest to the non-perturbative value $\lgb\approx-0.04956$). Then, we use thus obtained $\lgb(N)$ to evaluate  
 $\wfr_{\rm c}(N)$ and $\qfr_{\rm c}^2(N)$ from the perturbative series in $\lgb$. The numerical sequences $\wfr_{\rm c}(N)$, $\qfr_{\rm c}^2(N)$ do not converge and have to be Pad\'{e}-resummed. The Pad\'{e}-resummed values then converge to  the corresponding values for the non-perturbative critical point \eqref{crit-point-spec-nonpertx} with $N$ increasing (see fig.~\ref{fig:shar-ch-modes11www}). 
 
Hence, a perturbative analysis of eqs.~\eqref{CriticalPointsX} (admittedly, aided by  a Pad\'{e} resummation) seems to be capable of reproducing at least one  of the non-perturbative critical points of the full quasinormal spectrum. Since such a point can only occur due to the presence of a non-perturbative mode in the spectrum, we conclude that the calculation indirectly confirms the existence of the non-perturbative mode itself.


Another example of an inherently non-perturbative critical point that can be reproduced this way is the critical point located at
\begin{align}\label{nonpert-crit-point-spec1}
&\lgb\approx-0.02327\, :\nn
&\qfr_{\rm c}^2\approx 2.8462\,, \quad \wfr_{\rm c}\approx \pm 2.6329-2.4987i\,.
\end{align}
The appropriate perturbative critical point is given by eq.~\eqref{crit-point3}. Requiring $\im \, \qfr^2_{\rm c} (\lgb) = 0$, to first order in $\lgb$ we have
\begin{align}\label{crit-point-spec1}
&\lgb\approx-0.024486\, : \nn
&\qfr_{\rm c}^2= 2.95995 + \mathcal{O}(\lgb^2)\,, \quad \wfr_{\rm c}= \pm 2.46865 -2.59802 i + \mathcal{O}(\lgb^2)\,.
\end{align}
Continuing the expansion \eqref{crit-point-spec1} to higher orders of $\lgb$ as described above, we find that the sequence of order-$N$ 
critical points 
$\lgb (N)$, $\wfr_{\rm c}(N)$, $\qfr_{\rm c}^2(N)$ converges to the non-perturbative result \eqref{nonpert-crit-point-spec1} with $N$ increasing.

\subsection{A signature of non-perturbative modes from the perturbative analysis of critical points in the ${\cal N}=4$ SYM theory}

We now turn to the perturbative analysis of critical points in  the ${\cal N}=4$ SYM theory. Can we detect the existence of non-perturbative modes in the quasinormal spectrum by using the approach of the previous section? The recipe is to look for a pair of critical points with real value of $\qfr^2$. Considering the lowest perturbative critical point \eqref{crit-pointn4-1x} 
and finding $\gamma$ from the condition $\im \, \qfr^2_{\rm c} (\gamma) = 0$, we find to leading order in $\gamma$:
\begin{align}
\label{spec-crit-point-n4}
&\gamma\approx-0.0006290\, : \nn
&\qfr_{\rm c}^2\approx -0.67674\,, \quad \wfr_{\rm c}\approx \pm 0.86542-1.92493 i\,.
\end{align}
Such a pair of critical points at a single real $\qfr^2_c$ does not exist in an infinitely strongly coupled theory (at $\gamma=0$). At finite $\gamma$, it can arise as a result of the collision on the imaginary axis between the shear mode and the non-perturbative mode, followed by the second collision between the pair of the resulting propagating modes and the two gapped ``Christmas tree'' modes, similarly to what happens in the Einstein-Gauss-Bonnet case as described in section \ref{sign-egb}. We conjecture that eqs.~\eqref{spec-crit-point-n4}  are approximations to the non-perturbative critical point appearing in the right panel of fig.~\ref{shear-qnm-spectrum-example-3}. Of course, unlike in the  Einstein-Gauss-Bonnet theory, here we have no access to $\gamma^2$ and higher terms in the action and thus cannot verify this conjecture. Nevertheless, we can interpret the existence of the point \eqref{spec-crit-point-n4} as an  indirect evidence for the existence of the non-perturbative modes in the full quasinormal spectrum of the gravitational background dual to the $\CN =4$ SYM theory.

\section{Discussion}
\label{sec:disc}
In this paper, we have determined the dependence on the coupling of the radii of convergence of the shear and sound hydrodynamic dispersion relations in the strongly coupled $\CN =4$ SYM theory. This dependence is shown in fig.~\ref{fig:radius_coupling_dependence_n4_main_intro}. Limiting ourselves to the results obtained using the standard perturbation theory only, the coupling constant dependence of the radii  in the shear and sound channels is given by eqs.~\eqref{RN4Shear}, \eqref{RN4Sound} (the blue lines in  fig.~\ref{fig:radius_coupling_dependence_n4_main_intro}), respectively. These perturbative results suggest that the radii of convergence increase with the `t Hooft coupling decreasing from its infinite value. 

However, we argued that the presence of non-perturbative modes in the quasinormal spectrum of the dual black brane background at large but finite coupling will modify the perturbative result by making the radii's dependence on the coupling piecewise continuous, as shown in fig.~\ref{fig:radius_coupling_dependence_n4_main_intro}. This type of dependence is familiar from the finite density examples of holographic theories at infinite coupling \cite{aiks,Jansen:2020hfd} and from the Sachdev-Ye-Kitaev (SYK) chain at finite coupling \cite{Choi:2020tdj}. The origin of this similarity is not entirely clear to us.\footnote{We would like to thank the anonymous referee for raising this issue.} In the SYK case, it is the coupling dependence that gives rise to the piecewise behaviour, in parallel to what we observe in this paper. The strong-weak dependence in that case is, however, reversed. In the charged case, the chemical potential normalised by temperature plays the role of the coupling in the present paper being responsible both for the extra modes on the imaginary axis and the piecewise continuous dependence of the radius of convergence. One plausible guess is the emergence of approximate symmetries in all these cases in line with what was discussed in ref.~\cite{Grozdanov:2018fic}. 

Curiously, in the regime where the  non-perturbative mode becomes relevant, the radius of convergence of the shear mode  dispersion relation coincides with 
the endpoint $\qfr^2_c (\lambda)$ of the hydrodynamic regime introduced earlier in ref.~\cite{Grozdanov:2016vgg}. We then repeated our analysis for the case of the Einstein-Gauss-Bonnet theory using it as a theoretical laboratory to test our methods. Again, due to the presence of non-perturbative quasinormal modes in the spectrum, we  found the  piecewise dependence of the radii of convergence on the coupling (see figs.~\ref{fig:qc-lambda-shear} and fig.~\ref{fig:radius_coupling_dependence_sound_GB} for the shear and sound channel results, respectively). The role of the non-perturbative modes in a quasinormal spectrum thus appears to be significant. Their properties are similar to the properties of
 non-perturbative roots of algebraic equations with a small parameter, where singular perturbation theory is often useful. We have discussed in detail a simple example of an algebraic equation whose perturbative and non-perturbative roots can be consistently found using singular perturbation theory. Although at present  we cannot offer a generalisation of such a discussion to differential operators, we believe a qualitatively similar picture should hold there, too, and therefore the non-perturbative quasinormal modes can represent the qualitatively correct feature of the full theory. Finally, we show that the presence of the non-perturbative modes in the spectrum can be indirectly inferred by the analysis of critical points using the perturbative data only.

From the work in refs.~\cite{Grozdanov:2019kge,Grozdanov:2019uhi} it is clear that thermal two-point functions of the energy-momentum tensor in the  $\CN =4$ SYM theory at infinite `t Hooft coupling contain multiple branch point singularities in the complex plane of $\qfr^2$. At large but finite coupling, the location of these singularities acquires a dependence on the coupling. Our analysis in the present paper suggests that the property of being a singularity closest to the origin (and thus setting the radius of convergence) may change discretely as a function of the coupling, leading to the piecewise nature of the dependence of the radii of convergence on coupling, and that such a change is induced by the presence of non-perturbative quasinormal modes in the spectrum of a dual gravitational theory. It would be interesting to investigate whether a similar phenomenon is observed at small but finite coupling. Of special interest are the studies of the hydrodynamic series convergence in strongly coupled theories with non-zero chemical potential 
\cite{Jansen:2020hfd,Abbasi:2020ykq}, \cite{Soltanpanahi:2021mys}, where the coupling dependence has not yet been explored.

\acknowledgments{\small The work of S.G. was supported by the research programme P1-0402 of Slovenian Research Agency (ARRS). The work of P.T. is supported by an Ussher Fellowship from Trinity College Dublin. P.T. would like to thank Robin Karlsson for discussions. We also thank Jorge Noronha for comments on the draft of the paper.}

\appendix

\section{Critical points and the radius of convergence in holography}
\label{app-critical-points-radius}
Here, we briefly review the main points of the method introduced in refs.~\cite{Grozdanov:2019kge,Grozdanov:2019uhi}  to determine the radii of convergence of hydrodynamic dispersion relations in holography. In the limit where the dual gravity description of a QFT is valid (e.g. in the $N_c\to\infty$ limit of the ${\cal N} =4$ $SU(N_c)$ SYM theory), information about the hydrodynamic and other dispersion relations is contained in the quasinormal spectrum of the bulk equations of motion. In terms of a gauge-invariant variable $Z$, a typical bulk equation of motion is a second-order ODE
\begin{equation}
\partial^2_u Z + A(u,\wfr,\qfr^2) \partial_u Z +  B(u, \wfr,\qfr^2) Z =0\,,
\end{equation}
where $u$ is the bulk radial coordinate with $u=0$ corresponding to the boundary of the dual gravity background. 
The dependence of the coefficients of the equation on $\qfr^2$ reflects the rotation invariance of the theory, whereas the dependence on $\wfr$ is a consequence of the choice of the boundary condition at the horizon. The quasinormal spectrum $\wfr_i=\wfr_i(\qfr^2)$  is determined by the equation 
\begin{equation}
P(\qfr^2,\wfr)\equiv Z(u=0,\qfr^2,\wfr)=0\,,
\end{equation}
which also defines the spectral curve. The critical point  condition \eqref{CriticalPointsX} means that $p\geq 2$ branches $\wfr_i=\wfr_i(\qfr^2)$  
collide at $(\wfr_c,\qfr^2_c)\in \mathbb{C}^2$. The branches are locally represented by Puiseux series whose analytic structure is determined by the coefficients of the spectral curve via, e.g., the Newton polygon method. The situations when $\wfr_i \sim (\qfr^2 -\qfr_c^2)^{-\nu}$, where $\nu$ is a fractional power, e.g. $\nu=-1/2$, are common. The closest to the origin (in the complex $\qfr^2$-plane)  critical point with such a branch point singularity limits the convergence of the series $\wfr_i=\wfr_i(\qfr^2)$ centered at $\qfr^2=0$  and thus sets its radius of convergence, $R=|\qfr_c^2|$. This is the phenomenon of quasinormal level-crossing, analogous to the quantum-mechanical level-crossing \cite{Grozdanov:2019kge,Grozdanov:2019uhi}. When $\nu$ is a negative integer or zero, the branches are locally analytic, and we have ``level-touching'' rather than ``level-crossing'' \cite{Grozdanov:2019uhi}. 

 In practice, the bulk ODEs are sufficiently complicated  and have to be solved numerically. With such a solution in hand, one first solves eqs.~\eqref{CriticalPointsX} to find the critical points in the complex $\qfr^2$-plane, and then determines the degree of  the singularity at the critical points by considering the quasinormal mode behaviour in the complex $\wfr$-plane under the monodromy $\qfr^2= |\qfr^2| e^{i\varphi}$, 
where $\varphi \in [0,2\pi]$. We illustrate the difference between ``level-crossing''  and  ``level-touching''  by the following simple example.

Consider the complex curves
\begin{eqnarray}
&\,& P_1(x^2,y)=a^2 - b^2 + 2 b c x^2 - c^2 x^4 - 2 a y + y^2=0\,, \label{cu-1}\\
&\,& P_2(x^2,y)= a^2 - b + c x^2 - 2 a y + y^2 =0\,, \label{cu-2}
\end{eqnarray}
where $x^2\in \mathbb{C}$, $y\in \mathbb{C}$, and the coefficients $a,b,c$ are some fixed complex numbers. Applying the criterium  \eqref{CriticalPointsX} to $P_1$ and $P_2$, we find that both curves have the $p=2$ type crtitical point at  $(x^2_c,y_c)=(b/c,a)$. The two branches of the curve $P_1$ are given by 
\begin{eqnarray}
&\,& y_1^{(1)}(x^2) =  a+ b - c x^2 = y_c -c (x^2 -x^2_c)\,, \\
&\,& y_1^{(2)}(x^2) = a - (b-c x^2) = y_c + c(x^2-x^2_c)\,,
\end{eqnarray}
whereas the two branches of $P_2$ are
\begin{eqnarray}
&\,& y_2^{(1)}(x^2) =  a - \sqrt{b - c x^2} = y_c -i \sqrt{c} (x^2 - x^2_c)^{1/2}\,, \\
&\,& y_2^{(2)}(x^2) =  a + \sqrt{b - c x^2} = y_c +i \sqrt{c} (x^2 - x^2_c)^{1/2}\,.
\end{eqnarray}
\begin{figure*}[t]
\centering
\includegraphics[width=0.4\textwidth]{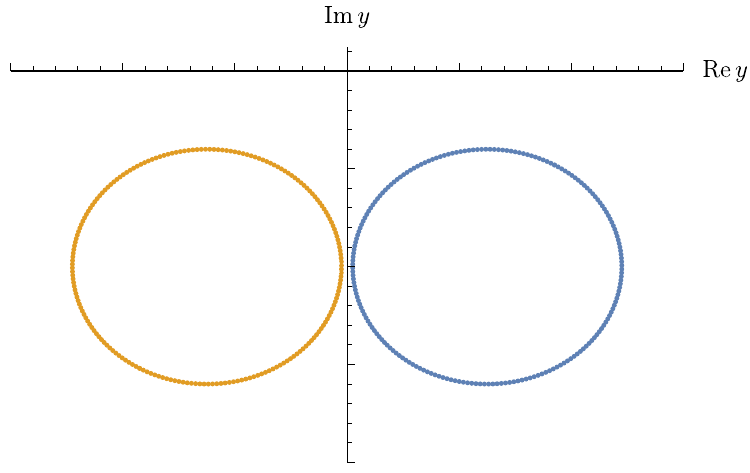}
\hspace{0.01\textwidth}
\includegraphics[width=0.4\textwidth]{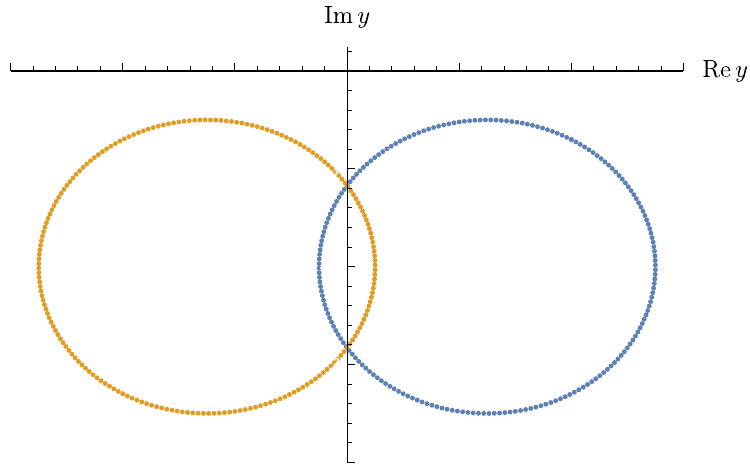}
\caption{
{\small  Level-touching: the two branches $y_1^{(1,2)}(x^2)$ of the spectral curve $P_1$ at complex $x^2 = |x^2| e^{i\varphi}$, with $\varphi$ varying from $0$ to $2\pi$, at fixed $|x_1^2| <|x_c^2|$ (left panel) and  $|x_2^2|>|x_c^2|$ (right panel).
}}
\label{level-touching-ex}
\end{figure*}
\begin{figure*}[t]
\centering
\includegraphics[width=0.45\textwidth]{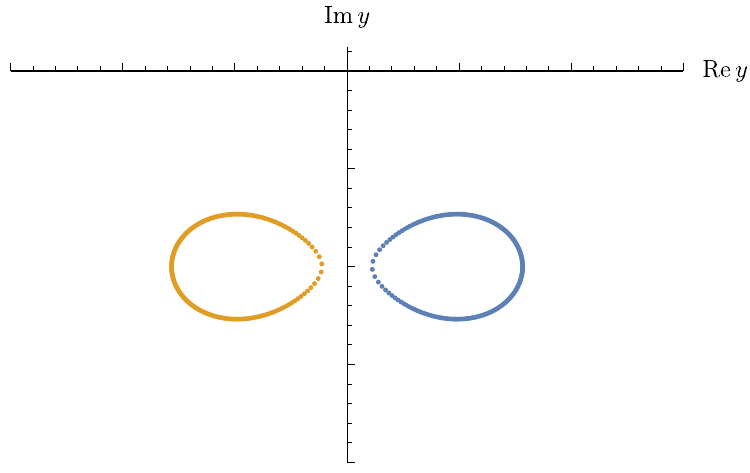}
\hspace{0.01\textwidth}
\includegraphics[width=0.45\textwidth]{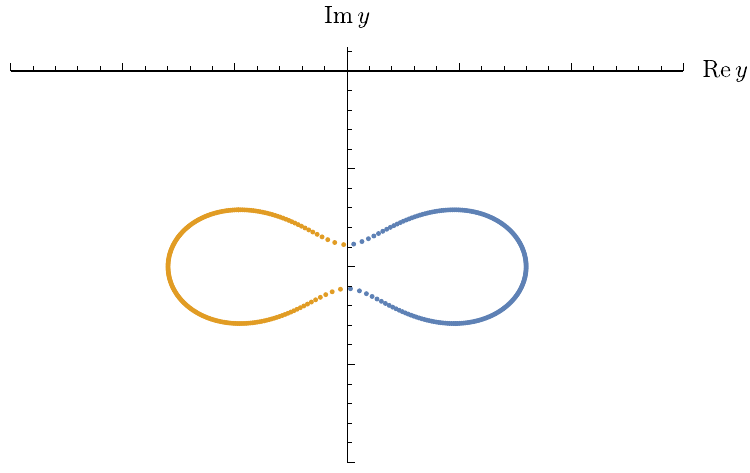}
\caption{
{\small  Level-crossing: the two branches $y_2^{(1,2)}(x^2)$ of the spectral curve $P_2$ at complex $x^2 = |x^2| e^{i\varphi}$, with $\varphi$ varying from $0$ to $2\pi$, at fixed $|x_1^2| <|x_c^2|$ (left panel) and  $|x_2^2|>|x_c^2|$ (right panel).
}}
\label{level-crossing-ex}
\end{figure*}

\noindent For the curve $P_1$, the two branches at the critical point are analytic functions of $x^2$. For the curve $P_2$, the critical point is a branch 
point singularity. Numerically, one can distinguish between the regular and the singular behaviour by finding local solutions $y_i=y_i (x^2)$ to eqs.~\eqref{cu-1}, \eqref{cu-2} at complex $x^2 = |x^2| e^{i\varphi}$, with fixed $|x^2|$ and $\varphi \in [0,2\pi]$. The ``trajectories'' traced by the branches as the phase $\varphi$ varies from $0$ to $2\pi$ are shown in fig.~\ref{level-touching-ex} for the curve $P_1$ 
 and in  fig.~\ref{level-crossing-ex} for the curve $P_2$ at fixed $|x_1^2| <|x_c^2| < |x_2^2|$. The regular branches of $P_1$ approach and  touch each other at $x^2=x^2_c$, with individual trajectories remaining closed under monodromy at $|x^2|>|x^2_c|$, as shown in fig.~\ref{level-touching-ex}. This is ``level-touching'', observed e.g. in the case of the BTZ quasinormal spectrum \cite{Grozdanov:2019uhi}. In case of a branch point singularity at the critical point, the two branches merge into a single trajectory for $|x^2|>|x^2_c|$, as they are mapped into each other under the monodromy (fig.~\ref{level-crossing-ex}). This is the level-crossing phenomenon \cite{Grozdanov:2019kge,Grozdanov:2019uhi}.

\section{How to reconstruct the Puiseux exponent from the power series}
\label{darboux}
Given the hydrodynamic expansion  (e.g. \eqref{PuiseuxShear} or \eqref{PuiseuxSound}) of the dispersion relation
 $\wfr = \wfr (\qfr^2)$ around $\qfr^2=0$, one can determine, at least in principle, the exponent of the Puiseux expansion $\wfr \sim (\qfr^2 - \qfr_c^2)^{-\nu}$ at the nearest to the origin critical point $\qfr^2_c$, given by eq.~\eqref{CriticalPointsX}. This can be done by using the Darboux  theorem (see e.g. \cite{henrici-book}, Theorem 11.10b) which states that if a function $p(t)$ has a singularity at $t=t_0$  of the form 
 \begin{equation}\label{1-1app}
p(t)\sim r(t) \left(1 - \frac{t}{t_0}\right)^{-\nu} \,, \qquad t\to t_0\,,
\end{equation}
where $r(t)$ is an analytic function and $\nu$ is not a negative integer or zero, then the coefficients $a_n$ of the Taylor expansion of $p(t)$ at the origin, $p(t) = \sum\limits_{n=0}^\infty a_n t^n$, have the following asymptotic form as $n\to \infty$
\begin{equation}\label{1-2app}
a_n \sim \frac{\Gamma (n+
\nu)}{\Gamma (\nu)  t_0^n n!} \left[ r(t_0) - \frac{(\nu - 1) t_0 r^\prime(t_0)}{(n+\nu -1)} +  
 \frac{(\nu - 1) (\nu -2) t_0^2 r^{\prime\prime}(t_0)}{2!(n+\nu -1)(n+\nu -2)} +\cdots  \right]\,.
\end{equation}
Keeping only the leading term in eq.~\eqref{1-2app}, we find 
\begin{equation}\label{1-3app}
\nu = \lim_{n\to \infty} \left(  t_0 (n+1) \frac{a_{n+1}}{a_n} - n \right)\,.
\end{equation}
Then, using the subdominant terms in eq.~\eqref{1-2app}, one can in principle reconstruct the function $r (t)$ by recovering its derivatives. In practice, 
a finite (and often relatively small) number of the coefficients $a_n$  (e.g. computed numerically) is sufficient to determine $\nu$ with a good precision. Thus, if $t_0$ is the critical point closest to the origin,  one can determine (or, strictly speaking, conjecture) whether  this point is a singularity of $p(t)$ (i.e., whether $\nu$ is a fractional or positive integer  number) and therefore sets the radius of convergence  and reconstruct the Puiseux expansion around $t_0$ by analysing the coefficients $a_n$.

A complication arises when there are two or more critical points located at the same distance from the origin. This is the case for the hydrodynamic dispersion relations in the  ${\cal N} =4$ SYM theory at infinite `t Hooft coupling, where for the shear and sound modes we have a pair of complex conjugate  critical points in the complex $\qfr^2$-plane \cite{Grozdanov:2019kge}. Such cases were studied for example in ref.~\cite{hunter_deducing_1980}. Instead, we find it more convenient to reduce the case with two critical points to the previous one with the help of a conformal transformation. Indeed, let $t=t_1$ and $t=t_2$, where $|t_1|=|t_2|$, be the critical points  located at the same distance from the origin $t=0$. By performing a M\"{o}bius transformation
\begin{equation}\label{1-1a}
t \to z = \frac{a t +b}{c t+d}
\end{equation}
and requiring that under the map, $0\to 0$, $t_1\to 1$ and $t_2\to z_2$, where $|z_2|>1$, we reduce the situation to the one of the Darboux theorem: e.g., the exponent $\nu_1$ of the branch point $t=t_1$ is determined by applying the Darboux procedure to the point $z=1$. The inverse transformation is 
\begin{equation}\label{1-1b}
t = \frac{d z - b}{a-c z}\,.
\end{equation}
We must also require that the singularity at $z\equiv z_s=a/c$ stays outside of the unit circle in the complex $z$-plane, so that any analytic part of $p(t)$ remains analytic after the transformation. This implies the requirement $|a/c|>1$.  Moreover, removing $z_s$ sufficiently far from the unit circle is advisable from the  following technical point of view: analytic functions such as $e^t$ acquire an essential singularity $\sim \exp[1/(z-z_s)]$ at $z = z_s$, and this would complicate numerics if $z_s$  were too close to the unit circle. 
Explicitly, we have
\begin{equation}\label{1-1c}
t \to z = \frac{t z_2 (t_1-t_2)}{t (t_1 z_2 - t_2) +t_1 t_2 (1-z_2) }\,,
\end{equation}
whereas the inverse transformation is given by
\begin{equation}\label{1-1d}
t = \frac{z t_1 t_2 (z_2-1)}{(t_1 z_2 -t_2) z + (t_2 -t_1) z_2}\,.
\end{equation}
The singularity $z_s$ is at
$$
z_s = \frac{(t_1-t_2) z_2}{t_1 z_2 - t_2} = \frac{(\alpha -1) z_2}{\alpha z_2 -1}\,,
$$
where $\alpha \equiv t_1/t_2$, with $|\alpha|=1$. We need to choose $|z_2|>1$. Let $z_2 = x + i y$. Then,
\begin{equation}\label{1-1f}
|z_s|^2 = \frac{|\alpha-1|^2 (x^2+y^2)}{x^2+y^2 +1 -2\, x\, \re \, \alpha \, + 2\, y\, \im\,  \alpha\, }\,.
\end{equation}
The zero of the denominator of \eqref{1-1f} is at $x= \re \, \alpha $, $y=- \im\,  \alpha$, with $|z_2|^2=x^2+y^2=1$, since $|\alpha|=1$. By choosing $x=\re \, \alpha  +\epsilon$ and $y=-\im \, \alpha  -\epsilon$, where $\epsilon >0$, we can satisfy the requirements $|z_2|>1$ and $|z_s|\gg1$. Indeed,
\begin{equation}\label{1-1g}
|z_s|^2 = \frac{|\alpha-1|^2}{2 \epsilon^2}    \left( 1 + 2\, \epsilon\, \re \, \alpha \, + 2\, \epsilon\, \im\,  \alpha\,+2 \epsilon^2 \right)\,,
\end{equation}
and so $|z_s| \sim |\alpha -1|/\sqrt{2}\epsilon$ for small $\epsilon$ (in numerical calculations, $\epsilon$ should not be too small, otherwise we need a large number of the coefficients $a_n$ to achieve a satisfactory convergence in  eq.~\eqref{1-3app}).

Applying this procedure to the hydrodynamic series of the  ${\cal N} =4$ SYM theory using the data of ref.~\cite{Grozdanov:2019kge}, we find $\nu =-1/2$ to a good precision, confirming that  the dispersion relations have branch point singularities of the square root type at the closest to the origin critical points. This is fully consistent with the characteristic quasinormal mode behaviour described in Appendix \ref{app-critical-points-radius}.

\section{Coefficients ${\cal A}_{(i)}$ and ${\cal B}_{(i)}$ of eq.~\eqref{fluct-eq-main-x-n4} in the ${\cal N} =4$ SYM theory}
\label{sec:appendix-N=4}
The coefficients can be written in the form
\begin{eqnarray}
{\cal A}_{(i)} (u, \wfr, \qfr, \gamma) 
&=& {\cal A}_{(i)}^{(0)}(u, \wfr, \qfr) + \gamma {\cal A}_{(i)}^{(1)}(u, \wfr, \qfr)\,,  \\
{\cal B}_{(i)} (u, \wfr, \qfr, \gamma) 
&=& {\cal B}_{(i)}^{(0)}(u, \wfr, \qfr) + \gamma {\cal B}_{(i)}^{(1)}(u, \wfr, \qfr)\,, 
\end{eqnarray}
where $i=1,2,3$ for the scalar, shear and sound channels, respectively. We have for the scalar channel:
\begin{align}
&{\cal A}_{(1)}^{(0)}(u, \wfr, \qfr) = - \frac{1+u^2}{u\left(1-u^2\right)}\,,   \\
&{\cal B}_{(1)}^{(0)}(u, \wfr, \qfr) =  \frac{\wfr^2 - \qfr^2 \left(1-u^2 \right)}{u \left(1-u^2\right)^2} \,,  \\
& {\cal A}_{(1)}^{(1)}(u, \wfr, \qfr) = 6  u \left(160 \qfr^2 u^3+129 u^4+94 u^2-25\right)\,,   \\
& {\cal B}_{(1)}^{(1)}(u, \wfr, \qfr) = \frac{192 \qfr^4 u^5-\qfr^2 \left(851 u^6-789 u^4+75 u^2+30\right)+6 \left(-89 u^4+30 u^2+5\right) \wfr^2}{u \left(1- u^2\right)}\,. 
\end{align}
For the shear channel:
\begin{align}
{\cal A}_{(2)}^{(0)}(u, \wfr, \qfr) = &\, - \frac{\left(1+u^2\right) \wfr ^2-\qfr^2 \left(1-u^2\right)^2}{u \left(1-u^2\right) \left(\wfr^2-\qfr^2 \left(1-u^2\right)\right)}\,, \\
{\cal B}_{(2)}^{(0)}(u, \wfr, \qfr) = &\, \frac{\wfr ^2 - \qfr^2 \left(1-u^2\right)}{u \left(1-u^2\right)^2}\,, 
\end{align}
\begin{align}
{\cal A}_{(2)}^{(1)}(u, \wfr, \qfr) = &\, \frac{2u}{\left(\wfr^2 - \qfr^2\left(1-u^2\right)\right)^2 } \biggr[ 640 \qfr^6 u^3 \left(u^2-1\right)^2  \nonumber  \\
&- 4 \qfr^4 u^2 \left(135 u^6-450 u^4-248 u^3 \wfr^2+495 u^2+200 u \wfr^2-180\right) \nonumber   \\
&+\qfr^2 \wfr^2 \left(-462 u^6+1374 u^4+160 u^3 \wfr^2-1002 u^2+75\right) \nonumber \\
&+ 3\left(129 u^4+94 u^2-25\right) \wfr^4 \biggr] \,,  \\
{\cal B}_{(2)}^{(1)}(u, \wfr, \qfr) =  &- \frac{3}{u \left(1-u^2\right) \left(\wfr^2 - \qfr^2\left(1-u^2\right)\right) } \biggr[ -64 \qfr^6 u^5 \left(u^2-1\right)\nn
&+\qfr^4 \left(425 u^8-880 u^6-64 u^5 \wfr^2+480 u^4-15 u^2-10\right)\nn
&+\qfr^2 \left(699 u^6-693 u^4+75 u^2+20\right) \wfr^2+2 \left(89 u^4-30 u^2-5\right) \wfr^4  \biggr]  \,.
\end{align}
For the sound channel:
\begin{align}
{\cal A}_{(3)}^{(0)}(u, \wfr, \qfr)  =&\, - \frac{3 \left(1+u^2\right) \wfr ^2 -  \qfr^2 \left(3-2 u^2 +3 u^4\right)}{u \left(1-u^2\right) \left(3 \wfr^2 - \qfr^2 \left(3-u^2\right)\right)}\,, \\
{\cal B}_{(3)}^{(0)}(u, \wfr, \qfr)  =&\,\frac{3 \wfr ^4 - 2  \left(3-2 u^2\right) \wfr ^2 \qfr^2  -  \qfr^2 \left(1-u^2\right) \left(4 u^3+\qfr^2 \left(u^2-3\right)\right)}{u \left(1-u^2\right)^2 \left(3 \wfr ^2 -  \qfr^2 \left(3-u^2\right)\right)}   \,,
\end{align}
 \begin{align}
{\cal A}_{(3)}^{(1)}(u, \wfr, \qfr)  =&\,  \frac{2u}{\left(3 \wfr ^2 -\qfr^2 \left(3-u^2\right) \right)^3}    \biggr[  32 \qfr^8 u^3 \left(35 u^6-291 u^4+753 u^2-585\right) \nn
&-3 \qfr^6 \left(3741 u^{10}-27911 u^8-2720 u^7 \wfr^2+60804 u^6+12992 u^5 \wfr^2-50112 u^4 \right.\nn
&\left. -12960 u^3 \wfr^2+16887 u^2-225\right)+3 \qfr^4 \wfr^2 \left(-19401 u^8+59832 u^6+4960 u^5 \wfr^2 \right. \nn
&\left. -53892 u^4-7200 u^3 \wfr^2+26094 u^2-1125\right)+9 \qfr^2 \wfr^4 \left(-1263 u^6+99 u^4 \right.\nn
&\left. +160 u^3 \wfr^2-3915 u^2+525\right)+81 \left(129 u^4+94 u^2-25\right) \wfr^6 \biggr]\,,
\end{align}
 \begin{align}
{\cal B}_{(3)}^{(1)}(u, \wfr, \qfr)  =&\,  \frac{1}{u \left(1-u^2\right) \left(3 \wfr ^2 - \qfr^2 \left(3-u^2\right)\right)^3} \biggr[192 \qfr^{10} u^5 \left(u^2-3\right)^3 \nn
&-\qfr^8 \left(u^2-3\right) \left(5811 u^{10}-41287 u^8-1728 u^7 \wfr^2+74004 u^6+5184 u^5 \wfr^2 \right.\nn
&\left. -35169 u^4+495 u^2+270\right)-3 \qfr^6 \left(11184 u^{13}-90072 u^{11}+17099 u^{10} \wfr^2 \right.\nn
&\left.+223952 u^9-106323 u^8 \wfr^2-16 u^7 \left(108 \wfr^4+12971\right)+185876 u^6 \wfr^2\right. \nn
&\left. +1728 u^5 \left(3 \wfr^4+34\right)-91107 u^4 \wfr^2+1800 u^3+2835 u^2 \wfr^2+1080 \wfr^2\right) \nn
&+3 \qfr^4 \wfr^2 \left(-68316 u^{11}+279504 u^9-40333 u^8 \wfr^2-319056 u^7+121158 u^6 \wfr^2\right.\nn
&\left. +36 u^5 \left(48 \wfr^4+2713\right)-81018 u^4 \wfr^2+3600 u^3+6075 u^2 \wfr^2+1620 \wfr^2\right)\nn
&-9 \qfr^2 \wfr^4 \left(21708 u^9-37140 u^7+7003 u^6 \wfr^2+12972 u^5-10017 u^4 \wfr^2 \right.\nn
&\left.+600 u^3+1755 u^2 \wfr^2+360 \wfr^2\right)-162 \left(89 u^4-30 u^2-5\right) \wfr^8  \biggr]\,.
\end{align}
See ref.~\cite{Grozdanov:2016vgg} for details.

\section{Coefficients ${\cal A}_{(i)}$ and ${\cal B}_{(i)}$ of eq.~\eqref{fluct-eq-main-n} in the
Einstein-Gauss-Bonnet theory}
\label{sec:appendix-GB}
In the scalar, shear and sound channels ($i=1,2,3$) we have, correspondingly,
\begin{align}
{\cal A}_{(1)} = \, & -\frac{1}{u}   - u \left[ \frac{1}{\left(\ggb ^2-1\right) \left(1-u^2\right)^2 + 1 - u^2} + \frac{1}{\left(1-u^2\right) \sqrt{\ggb ^2-\left(\ggb^2-1\right) u^2}}\right]\,, \\
{\cal B}_{(1)} = \, &\frac{(\ggb -1) (\ggb +1)^2 \left(3 \left(\ggb ^2-1\right) u^2-\ggb ^2\right)\left(-\ggb ^2+\left(\ggb ^2-1\right) u^2+U\right)}{4 u \left(\ggb^2-\left(\ggb ^2-1\right) u^2\right)^{3/2} \left(-\ggb ^2+\left(\ggb
   ^2-1\right) u^2+2 U-1\right)} \qfr^2 \nn
\, & 
+ \frac{\left(\ggb ^2-1\right)^2 \left(-\ggb ^2+\left(\ggb ^2-1\right)u^2+U\right)}{4 u (U-1) \sqrt{\ggb ^2-\left(\ggb ^2-1\right) u^2} \left(-\ggb^2+\left(\ggb ^2-1\right) u^2+2 U-1\right)}\wfr^2\,,
\end{align}
\begin{align}
 {\cal A}_{(2)} = \,&-\frac{2 \ggb ^4 (\ggb +1)  \left[\frac{1}{2} \left(1-\ggb ^2\right) \left(u^2-1\right) (U-2)+U-1\right]}{u (U-1) U^3 \left[\ggb ^2 (\ggb +1) (U-1)  \qfr^2  -\left(\ggb ^2-1\right) U^2 \wfr ^2\right]}  \qfr^2 \nn
\,&-\frac{\left(1-\ggb ^2\right) \left(\ggb ^4+\left(1-\ggb^2\right)^2 u^4- 2 \left(1-\ggb ^2\right) u^2 \left(U-\ggb^2\right)-\ggb ^2 U\right)}{u (U-1) U \left[\ggb ^2 (\ggb +1)(U-1)  \qfr^2  -\left(\ggb ^2-1\right) U^2 \wfr ^2\right]}   \wfr^2\,, \\
 {\cal B}_{(2)} = \,&  \frac{\ggb ^2 (\ggb +1) (U+1)}{4 u \left(u^2-1\right) U^2} \qfr^2 + \frac{\left(U^2+2 U+1\right) }{4 u \left(u^2-1\right)^2}  \wfr^2 \,,
\end{align}
\begin{align}
{\cal A}_{(3)} =\,& \frac{3}{2 u} + \frac{3 (\ggb -1)  \left[\left(\ggb ^2-1\right) u^2-\ggb^2\right] \left[\left(\ggb ^2-1\right) u^2 (5 U-7)-5 \ggb ^2(U-1)\right]}{2 u (U-1) U^2 D_1} \wfr^2 \nn
\,&+ \frac{ \left(\ggb ^2-1\right)^2 u^4 \left(-3 \ggb ^2+5 U-7\right)+\ggb ^2 \left(\ggb ^2-1\right) u^2 \left(18 \ggb ^2-13 U+10\right) }{2 u (U-1) U^2 D_1} \qfr^2 \nn
\,&- \frac{ 15 \ggb ^4 \left(\ggb ^2-2 U+1\right) }{2 u (U-1) U^2 D_1} \qfr^2\,, \\
{\cal B}_{(3)} =&~ \frac{\left(\ggb ^2-1\right)^2}{D_{0}} \biggr\{ \,12 (\ggb -1)^2 \ggb ^2 (\ggb +1) \qfr^2 u^5-4 (\ggb -1) \ggb ^2 \qfr^2 u^3 \left(3 \ggb ^2-7 U+4\right) \nn
&+ \left(\ggb ^2-1\right)^3 \qfr^2 u^6 \left(3 (\ggb -1) \wfr ^2+\qfr^2\right) \nn
& -u^2 \ggb ^2 \left(\ggb ^2-1\right)  \left[\qfr^4 \left(\ggb ^2+2 U\right)+(\ggb -1) \qfr^2 \wfr ^2 \left(9 \ggb ^2-4 U\right)-6 (\ggb-1)^2 U \wfr ^4\right] \nn
&  + \left(\ggb ^2-1\right)^2 u^4 \left[\qfr^4 \left(3 \ggb ^2 (U-2)+U\right)+2(\ggb -1) \qfr^2 U \wfr ^2-3 (\ggb -1)^2 U \wfr ^4\right]\nn
& - 3 \ggb ^4 \left[\qfr^4 \left(\ggb ^2 (U-2)+U\right)+2 (\ggb -1) \qfr^2 \wfr^2 \left(U-\ggb ^2\right)+(\ggb -1)^2 U \wfr ^4\right]  \,\, \biggr\}\,,
\end{align}
where we have defined
\begin{align}
D_1 &\equiv \left(\ggb ^2-1\right) u^2 \left(3(\ggb -1) \wfr^2+\qfr^2\right)+3 \ggb ^2 \left(\qfr^2 (U-1)-(\ggb -1)\wfr^2\right)\,, \\
D_0 &\equiv 4 (\ggb -1) u (U-1)^2 U^3 D_1\,.
\end{align}
In the above expressions, we also used $U^2 = u^2 + \ggb^2 - u^2 \ggb^2$, as well as the dimensionless 
frequency and momentum $\wfr = \omega/2\pi T$, $\qfr = q/2\pi T$,  where $T$ is the Hawking temperature of the black brane background. See ref.~\cite{Grozdanov:2016fkt} for details. 

\bibliographystyle{JHEP}
\bibliography{refs-instability.bib}{}
\end{document}